\documentclass[3p,twocolumn]{elsarticle}
\usepackage{amsmath,bm,epsfig,subfigure}
\usepackage{epstopdf}
\usepackage{indentfirst}
\usepackage{amssymb}
\usepackage{graphicx}
\usepackage{float}
\usepackage{color}
\usepackage[normalem]{ulem}
\usepackage{lineno,hyperref}
\modulolinenumbers[5]
\usepackage[toc,page]{appendix}
\usepackage{natbib}

\DeclareMathAlphabet{\mathpzc}{OT1}{pzc}{m}{it}

\journal{Physica D}

\begin{document}

\begin{frontmatter}

\title{Dynamics and stability of a discrete breather in a harmonically excited chain with vibro-impact on-site potential}

\author{Nathan Perchikov}

\author[]{O.V. Gendelman\corref{mycorrespondingauthor}}
\cortext[mycorrespondingauthor]{Corresponding author}
\ead{ovgend@tx.technion.ac.il}

\address{Faculty of Mechanical Engineering, Technion, Haifa 32000, Israel}

\begin{abstract}
We investigate the existence and stability of discrete breathers in a chain of masses connected by linear springs and subjected to vibro-impact on-site potentials. The latter are comprised of harmonic springs and rigid constraints limiting the possible motion of the masses. Local dissipation is introduced through a non-unit restitution coefficient characterizing the impacts. The system is excited by uniform time-periodic forcing. The present work is aimed to study the existence and stability of similar breathers in the space of parameters, if additional harmonic potentials are introduced. Existence--stability patterns of the breathers in the parameter space and possible bifurcation scenarios are investigated analytically and numerically. In particular, it is shown that the addition of harmonic on-site potential can substantially extend the stability domain, at least close to the anti-continuum limit. This result can be treated as an increase in the robustness of the breather from the perspective of possible practical applications.
\end{abstract}


\end{frontmatter}


\section{Introduction}
\label{sect1}

Discrete breathers (DBs) have long been a subject of both theoretical analysis and experimental studies \cite{Ovchinnikov, Flach1, Flach2}. In \cite{Marin2001}, DBs of the sine-Gordon type are analyzed for various coupling strength values, including the no-coupling limit, showing noteworthy features in terms of existence, stability, bifurcation types, mobility and interaction, and exhibiting properties similar to some of those exhibited by Hamiltonian systems. In \cite{Lazarides2006}, magnetic meta-material breathers are analyzed with special emphasis on the weak coupling limit and with stability and mobility investigated for both energy-conserving and dissipative systems. In \cite{Cubero2009}, a model with quartic nonlinearity is analyzed from the perspective of spontaneous creation and annihilation of DBs due to thermal fluctuations, exhibiting the features of stochastic resonance, such as, for example, non-monotonic dependence on noise. A seminal experimental work \cite{Cuevas2009}, investigated stability exchange between different localized modes in forced-damped coupled pendula and their possible relation to dislocation dynamics.

In the majority of theoretical studies related to DBs, the considered models are Hamiltonian. Still, in many applications the damping cannot be neglected, and in order to maintain the DB, one should compensate it by some kind of direct or parametric external forcing \cite{Flach2}. Many of the DBs observed in experiments exist in damped systems and should be maintained by some external forcing.

Lack of Hamiltonian structure radically changes the properties of the DBs. To name just one point, instead of a continuous family of localized solutions, one expects to obtain a discrete set of attractors. Accordingly, many of the methods devised for computation and analysis of Hamiltonian DBs are not applicable in forced-damped systems. Recently, it was demonstrated that one can derive exact solutions for DBs in vibro-impact chain models. Such lattices have been investigated analytically both for the Hamiltonian case \cite{GendelmanManevitch2008} and for the forced-damped case \cite{Gendelman2013}.

In both cases, representation of the nonlinearity, responsible for the localization effect, with the help of  the impact conditions, turned out to be advantageous, both for the derivation of an analytic solution and for the stability analysis.  To simplify the numeric simulations, in \cite{Gendelman2006}, a method is suggested for modeling impact conditions by smooth potentials for both symmetric and single-impacts scenarios. By application of group theory techniques, one can derive smoothened potential and dissipation terms, which rigorously mimic the non-elastic impacts in a limit of large smoothening exponent.  The obvious advantage of smoothened impact conditions is the ability to incorporate them into an explicit, stable integration scheme, such as the backward Euler scheme, for example. An inevitable (although, perhaps, acceptable) shortcoming of the method is in that it makes the equations stiff in finite intervals, in finite proximity of the impact constraints.   Another shortcoming is the relative complexity of a linear stability analysis of a solution obtained for the smooth problem, relatively to the case where genuine impact conditions are imposed. A beneficial approach could be the use of a combination of impact and smooth potentials $-$ performing a linear stability analysis with respect to the impact-potential representation, as argued to be advantageous in \cite{Champneys2008} and \cite{Zhusubaliyev2003}, integrating with the impact scheme when it converges and with a smoothened scheme where there seems to be an instability and it is to be determined whether it arises from the physics or from the integration algorithm.

The present work is in a sense a continuation of \cite{Gendelman2013}, where exact expressions for the displacements of the masses in an infinite chain were obtained in the form of a convergent Fourier series. Moreover,  for every value of the (dimensionless) link stiffness, a range of amplitudes of the time-harmonic excitation was found, for which a localized breather exists. Interestingly, it was found that no solution corresponding to a phonon-emitting breather could exist. Linear stability analysis based on Floquet theory was performed, utilizing the method of \cite{FredrikssonNordmark2000}. Three noteworthy features were revealed. First, it was found that for a large-enough value of the dimensionless link stiffness (smaller than the maximum value corresponding to breather existence), one observes loss of stability by delocalization. Second, for low enough link stiffness, there exists a critical value of the excitation amplitude smaller than the critical value corresponding to the limit of existence of the breather, at and above which loss of stability by symmetry breaking takes place. Third, it was found that the delocalization instability sub-domain boundary is non-monotonous with respect to the link stiffness (or the excitation amplitude).

The motivation for the present investigation is two-fold. First, we would like to explore an additional feature of the system that may be reflecting a state of affairs more commonly encountered in practice. For instance, the harmonic part of a uniform on-site potential may represent the effect of weak, non-dissipative coupling to the environment.  Second, the extension of the parameter space could supply more information about generic bifurcations and stability of the DBs.

The structure of the present paper is as follows. In Section \ref{sect2}, the model system is discussed and exact expressions for DBs are derived. In Section \ref{sect3}, detailed characteristics of the solution are derived for the case of single-harmonic excitation. In Section \ref{sect4}, the problem of the existence of localized breathers is explored and existence charts in the parameter-space are presented and discussed. In Section \ref{sect5}, linear stability analysis is performed. In Section \ref{sect6}, the equations of motion of the system are integrated numerically for periodic boundary conditions, in order to validate the analytic solution, to check the effect of the boundary conditions and to verify the stability picture. Section \ref{sect7} is devoted to concluding remarks.

\section{Description of the model and analytic treatment}
\label{sect2}

We consider an infinite system of \emph{N} identical masses, connected by linear elastic springs, each having dimensionless rigidity ${\gamma}$, subjected to harmonic  on-site potentials with dimensionless rigidity ${\kappa}$ and excited by a time-periodic spatially uniform external loading force $F(t)$, having a period of $2\pi$. The equation of motion for the displacements ${u_n(t)}$ in this case takes the following form:
\begin{equation}
\begin{split}
\ddot{u}_n+(2\gamma+\kappa)u_n-\gamma u_{n+1}-\gamma u_{n-1}  = F(t)  ,\\    \lvert u_n \rvert  \leq 1  ,   \forall     n \in \mathbb{Z}
\label{eq1}
\end{split}
\end{equation}

We suggest that each oscillator is subjected to rigid symmetric vibro-impact non-elastic constraints at distances $\pm 1$ from the equilibrium positions of the oscillators. Each impact results in an abrupt change of the velocity of the impacting particle. The formal general expression for this can be written as follows:
\begin{equation}
\begin{split}
\dot{u}_n|{t=\phi^++\pi n}= \\
 U\left ( \dot{u}_n|_{t=\phi^-+\pi N},u_n|_{t=\phi^-+\pi N}\right ) \ , \ \forall \ n,N \in \mathbb{Z}
\label{eq2}
\end{split}
\end{equation}
where $\phi$ represents the time phase lag between the external forcing and the impacts, and the impact function $U$ is to be specified later.
At this point we limit ourselves by seeking only those solutions that correspond to strongly localized breathers, when only one particle experiences impact. Hence, we assume that the impact conditions are fulfilled only for the zeroth mass, namely ${ \lvert u_n \rvert \leq 1}$ is replaced by $ \lvert u_n \rvert < 1  \ \forall \ n \in \mathbb{N} , \ \lvert u_0 \rvert \leq 1$. We then eliminate the nonsmooth bounding condition by representing it as an external loading force, following \cite{Gendelman2013}:
\begin{equation}
\begin{split}
\ddot{u}_n+(2\gamma+\kappa)u_n-\gamma u_{n+1}-\gamma u_{n-1}  = F(t)  + \\  2p \delta  _{n0} \sum _{j=- \infty}^{j=\infty} \delta(t-\phi+\pi(2j+1)) \\ - \delta(t-\phi+2\pi j)
\label{eq3}
\end{split}
\end{equation}
where $2p$ stands for the change in the linear momentum of the zeroth mass due to a single impact incident.

As the external forcing is spatially uniform, the solution may be decomposed into a uniform and a non-uniform part:
\begin{equation}
 u_n=v_n+f(t)
\label{eq4}
\end{equation}
where the uniform part satisfies the equation:
\begin{equation}
\ddot{f}(t)+\kappa f(t) = F(t)
\label{eq5}
\end{equation}
the general solution of which is:
\begin{equation}
f(t) = \kappa^{-1/2}\int\limits^{t}{\sin[\kappa^{1/2}(t-\tau)]F(\tau)~d\tau}
\label{eq6}
\end{equation}

Substitution of (\ref{eq4}) and (\ref{eq6})  into (\ref{eq3}) gives an equation for $v_n(t)$:
\begin{equation}
 \begin{split}
 \ddot{v}_n+(2\gamma+\kappa)v_n-\gamma v_{n+1}-\gamma v_{n-1}  =\\  2p \delta  _{n0} \sum _{j=- \infty}^{j=\infty} \delta(t-\phi+\pi(2j+1)) \\ - \delta(t-\phi+2\pi j)
 \label{eq7}
 \end{split}
 \end{equation}
Expanding the right-hand side of (\ref{eq7}) into a cosine Fourier series yields:
\begin{equation}
\begin{split}
\ddot{v}_n+(2\gamma+\kappa)v_n-\gamma v_{n+1}-\gamma v_{n-1}  =\\ -4p\pi^{-1} \delta  _{n0} \sum _{l= 0}^{\infty} \cos[(2l+1)(t-\phi)]
\label{eq8}
 \end{split}
 \end{equation}

The equation of motion in (\ref{eq8}) leads to the following dispersion relation for $v_n(t)$:
\begin{equation}
\omega(\zeta)=\sqrt{\kappa+2\gamma(1-\cos{\zeta})}
\label{eq9}
\end{equation}
where $\zeta$ is a wavenumber. Hence, a solution may in general be phonon-emitting and contain harmonics corresponding to propagating frequencies in the strip: $\sqrt{\kappa} \le  2l+1 \le \sqrt{\kappa+4\gamma}$.

Consequently, we decompose Eq. (\ref{eq8}) into two localized and one propagating part, producing the following equations:
\begin{equation}
\begin{split}
 \ddot{\hat{v}}_n+(2\gamma+\kappa)\hat{v}_n-\gamma \hat{v}_{n+1}-\gamma \hat{v}_{n-1}  =\\-\frac{4p}{\pi}  \delta  _{n0} \sum _{l= 0}^{\left \lfloor \sqrt{\kappa}/2-1/2 \right \rfloor} \cos[(2l+1)(t-\phi)]
 \label{eq10}
 \end{split}
 \end{equation}
\begin{equation}
\begin{split}
 \ddot{\tilde{v}}_n+(2\gamma+\kappa)\tilde{v}_n-\gamma \tilde{v}_{n+1}-\gamma \tilde{v}_{n-1}  =\\-\frac{4p}{\pi} \delta  _{n0} \sum _{l= \left \lceil \sqrt{\kappa}/2-1/2 \right \rceil}^{\left \lfloor \sqrt{\kappa+4\gamma}/2-1/2 \right \rfloor} \cos[(2l+1)(t-\phi)]
 \label{eq11}
 \end{split}
 \end{equation}
\begin{equation}
\begin{split}
\ddot{\bar{v}}_n+(2\gamma+\kappa)\bar{v}_n-\gamma \bar{v}_{n+1}-\gamma \bar{v}_{n-1}  =\\-\frac{4p}{\pi} \delta  _{n0} \sum _{l= \left \lceil \sqrt{\kappa+4\gamma}/2-1/2 \right \rceil}^{\infty} \cos[(2l+1)(t-\phi)]
\label{eq12}
\end{split}
\end{equation}
\begin{equation}
 v_n=\hat{v}_n+\tilde{v}_n+\bar{v}_n
\label{eq13}
 \end{equation}

Expression (\ref{eq10}) corresponds to the lower attenuation zone, and  expression  (\ref{eq12}) - to the upper one. The following expansions are assumed for the unknown functions:
\begin{equation}
 \hat{v}_n(t)  =\sum _{l= 0}^{\left \lfloor \sqrt{\kappa}/2-1/2 \right \rfloor} \hat{V}_n \cos[(2l+1)(t-\phi)]
 \label{eq14}
 \end{equation}
\begin{equation}
\begin{split}
 \tilde{v}_n(t)  =\\  \sum _{l= \left \lceil \sqrt{\kappa}/2-1/2 \right \rceil}^{ \left \lfloor \sqrt{\kappa+4\gamma}/2-1/2 \right \rfloor}  \left \lbrace A_n \cos[(2l+1)(t-\phi)] \right. \\ +\left. B_n \sin[(2l+1)(t-\phi)] \right \rbrace
 \label{eq15}
 \end{split}
 \end{equation}
\begin{equation}
  \bar{v}_n(t)  = \sum _{l= \left \lceil \sqrt{\kappa+4\gamma}/2-1/2 \right \rceil}^{\infty}  \bar{V}_n\cos[(2l+1)(t-\phi)]
  \label{eq16}
  \end{equation}

Obviously the phonon-emitting part creates additional phase lag, giving rise to the additional sine series in Eq. (\ref{eq15}). For $N \to \infty$, the $n$-dependent coefficients, being powers of $n$, can either diverge or vanish at $|n| \to \infty$. For an infinite system, only the solution vanishing at infinity can be valid, and no other boundary condition is required.  Substituting Eqs. (\ref{eq14}-\ref{eq16}) into Eqs. (\ref{eq10}-\ref{eq12}), solving the resulting second order linear recursion equations for the amplitudes, choosing the spatially non-diverging solutions and assuming spatial symmetry with respect to $n=0$, (thus permitting only outward phonon emission), we get:
\begin{equation}
\begin{split}
 u_n(t)=  \int\limits^{t}{\frac{\sin[\kappa^{1/2}(t-\tau)]}{\sqrt{\kappa}}F(\tau)~d\tau} \\
-\frac{4p}{\pi} \left(\frac{1}{2\gamma}\right)^{\lvert n \rvert}
\sum _{l= 0}^{\left \lfloor \sqrt{\kappa}/2-1/2 \right \rfloor}  \frac{\cos[(2l+1)(t-\phi)]}{\sqrt{Q_l^2-(2\gamma)^2}}\\ \times \left \lbrace \vphantom{\sqrt{Q_l^2}} Q_l \right.\left. -\sqrt{Q_l^2-(2\gamma)^2}\right\rbrace^{\lvert n \rvert} \\
  +\frac{4p}{\pi}   \sum _{l= \left \lceil \sqrt{\kappa}/2-1/2 \right \rceil}^{ \left \lfloor \sqrt{\kappa+4\gamma}/2-1/2 \right \rfloor}   \frac{1}{\sqrt{(2\gamma)^2-Q_l^2}}  \\  \times \sin \left \lbrace \lvert n \rvert \arcsin{\frac{\sqrt{(2\gamma)^2-Q_l^2}}{2\gamma}}   \right.  \\ \left.   -(2l+1)(t-\phi) \vphantom{\frac{\sqrt{1^2}}{2}} \right \rbrace
 +\frac{4p}{\pi} \left(-\frac{1}{2\gamma}\right)^{\lvert n \rvert} \times \\
\sum _{l= \left \lceil \sqrt{\kappa+4\gamma}/2 -1/2 \right \rceil}^{\infty} \frac{\cos[(2l+1)(t-\phi)]}{\sqrt{Q_l^2-(2\gamma)^2}}
 \times \\  \left\lbrace \vphantom{\sqrt{[Q_l^2}} Q_l \right. \left. -\sqrt{Q_l^2-(2\gamma)^2}\right\rbrace^{\lvert n \rvert}
 \label{eq17}
 \end{split}
 \end{equation}
where
\begin{equation}
Q_l \triangleq|(2\gamma+\kappa)-(2l+1)^2|
\label{eq17b}
\end{equation}

\section{Exact solutions for the DB\lowercase{s} in case of single-harmonic excitation}
\label{sect3}

To derive a specific form of the DB solution, we consider a single-harmonic uniform harmonic excitation. Without loss of generality, we can express single-harmonic excitation as:
\begin{equation}
F(t)=a\cos{t}  \ , \ a>0
\label{eq18}
\end{equation}

Consequently, we have:
\begin{equation}
 u_0(\phi+2\pi N)=\frac{a\cos\phi}{\kappa-1}+\frac{4S_0}{\pi}p  \ \ \ , \ \  \forall     N \in \mathbb{Z}
 \label{eq19}
 \end{equation}
where
\begin{equation}
 \begin{split}
 S_0  =   \sum _{l= \left \lceil \sqrt{\kappa+4\gamma}/2-1/2 \right \rceil}^{\infty}  \frac{1}{\sqrt{Q_l^2-(2\gamma)^2}} \\ -\sum _{l= 0}^{\max{(0,\left \lfloor \sqrt{\kappa}/2-1/2 \right \rfloor)}}  \frac{1}{\sqrt{Q_l^2-(2\gamma)^2}}
 \label{eq20}
 \end{split}
 \end{equation}

Consistently with Eqs. (\ref{eq1}-\ref{eq3}) and with no loss of generality we can write the following equation for ${\phi}$:
\begin{equation}
  u_0(\phi+2\pi N) =1
 \label{eq21}
 \end{equation}

Having two eigenvalues in this nonlinear eigenvalue problem we obviously need two equations to describe the impact: one for the impact phase, ${\phi}$, given by Eq. (\ref{eq21}), and one for the momentum jump at an impact instance, $2p$.

In order to write an equation for the momentum jump at the moment of impact we need a constitutive model for the impact function, $U$. Following \cite{GendelmanManevitch2008} and \cite{Gendelman2013}, we employ here perhaps the simplest possible jump condition, based on the assumption of a constant restitution coefficient, $k$, namely:
\begin{equation}
\lim_{h\to 0}\frac{\dot{u}_0(\phi+2\pi N+h)}{\dot{u}_0(\phi+2\pi N-h)}=-k
\label{eq22}
\end{equation}

Substituting (\ref{eq4}) into (\ref{eq22}), recalling that from (\ref{eq17}) and (\ref{eq18}) we know that ${u_0(t)}$ is ${2\pi}$-periodic, and knowing from (\ref{eq6}) and (\ref{eq18}) that ${f(t)}$ is smooth, we get:
\begin{equation}
\lim_{h\to 0}\frac{\dot{v}_0(\phi+h)+\dot{f}(\phi)}{\dot{v}_0(\phi-h)+\dot{f}(\phi)}=-k
\label{eq23}
\end{equation}

Next, integrating Eq. (\ref{eq7}) for ${n=0}$ over the domain $t \in{[\phi-h,\phi+h]}$ for ${h \to 0}$ and finding from Eqs. (\ref{eq13}-\ref{eq16}) that ${v_0(t)}$ and ${v_1(t)=v_{-1}(t)}$ are continuous and hence do not contribute to the momentum change integral, we find that:
\begin{equation}
 \lim_{h\to 0} [\dot{v}_0(\phi+h)-\dot{v}_0(\phi-h)]=-2p
 \label{eq24}
 \end{equation}

So far, we have assumed that our solution is spatially symmetric with respect to ${n=0}$ and temporally-periodic with a period of ${2\pi}$. Henceforth, we assume that the solution is also temporally symmetric with respect to ${t=\phi}$, implying ${v_n(\phi+t)=v_n(\phi-t) \ , \  \forall  \ t \in [0,2\pi]}$ and thus, naturally, we have:
\begin{equation}
 \lim_{h\to 0} \frac{\dot{v}_0(\phi+h)}{\dot{v}_0(\phi-h)}=-1
 \label{eq25}
 \end{equation}

Solving system (\ref{eq24}-\ref{eq25}), rearranging and substituting the result into (\ref{eq23}) yields:
\begin{equation}
\dot{f}(\phi)=\frac{1-k}{1+k}p
\label{eq26}
\end{equation}

Substituting (\ref{eq18}) into (\ref{eq6}), taking a time derivative and substituting ${t=\phi}$ in the result, gives:
\begin{equation}
 \dot{f}(\phi)=\frac{a\sin\phi}{1-\kappa}
 \label{eq27}
 \end{equation}

Combining (\ref{eq26}) and (\ref{eq27}) yields an expression for the impact time lag explicitly, in terms of $p$:
\begin{equation}
 \phi= \arcsin{ \begin{bmatrix}\frac{1-k}{1+k}(1-\kappa)p/a \end{bmatrix}}
 \label{eq28}
  \end{equation}

Similarly, substituting (\ref{eq28}) into (\ref{eq19}) and relying on the $2\pi$ temporal periodicity of the solution results in an explicit expression for $p$:
\begin{equation}
 p=\frac{\chi_0+\sqrt{(q^2+\chi_0^2)\hat{a}^2-q^2}}{q^2+\chi_0^2} .
 \label{eq29}
 \end{equation}
were we define:
\begin{equation}
 q \triangleq \frac{1-k}{1+k} \ \ , \ \ \chi_0 \triangleq \frac{4S_0}{\pi} \ \ , \ \ \hat{a} \triangleq \frac{a}{1-\kappa}
\label{eq30}
 \end{equation}
and the choice of the positive sign in front of the square root in (\ref{eq29}) is based on physical reasoning in the ${k \to 1}$ limit, which is:
\begin{equation}
 \lim_{k \to 1}p=\frac{1+\hat{a}}{\chi_0} \ \  ,  \ \  \lim_{k \to 1}\phi=0
 \label{eq31}
  \end{equation}

It follows from (\ref{eq31}) that for perfect restitution, the impact instances coincide with the temporal extrema of the excitation.

Next, without damaging the argumentation, one can assume infinitesimally weak links between the masses and vanishingly small foundation stiffness, namely, ${\gamma \to 0 \ , \kappa \to 0}$. This results in:
\begin{equation}
\lim_{\kappa \to 0}{\lim_{\gamma \to 0}\lim_{k \to 1}p}=\frac{1+a}{\pi/2}   >0
\label{eq32}
\end{equation}

Obviously a positive momentum jump would only increase with the amplitude of coherent single-harmonic excitation. Thus taking a negative sign in front of the square root in (\ref{eq29}) would have given an unphysical result.

Eq. (\ref{eq31}) shows linear relation between $p$ and $a$ in the limit of perfect restitution. This does not imply, of course, that the whole system becomes linear in the sense that the amplitudes of the displacements of the masses are linear in the amplitude of the applied excitation, since the displacement amplitudes are always non-smoothly bounded. Furthermore, the response to single-harmonic excitation stays multi-harmonic even in the limit case explored in (\ref{eq31}).

A corollary of (\ref{eq32}) is that the series in (\ref{eq17}) are at least conditionally convergent, since their supremum is a finite number, as represented by the denominator in the right-hand side of (\ref{eq32}).

\section{Existence of localized DB\lowercase{s} for single-harmonic excitation}
\label{sect4}

It is obvious that for impact to take place in steady-state, the amplitude of the excitation has to be large enough to compensate for the energy lost during the imperfect impacts. This understanding is illustrated by expression (\ref{eq29}). The requirement that $p$ be real, corresponding to actual impact preserving time-periodicity, results in a lower bound for the excitation amplitude:
\begin{equation}
 \hat{a}=\frac{a}{1-\kappa}\ge\frac{q}{\sqrt{q^2+\chi_0^2}} >0
 \label{eq33}
 \end{equation}

Hence, necessary conditions for the existence of a breather are:
\begin{equation}
\kappa<1 \ \ , \ \ a\ge \frac{q(1-\kappa)}{\sqrt{q^2+\chi_0^2}}
\label{eq34}
\end{equation}

Consequently, the displacements take the following form:
\begin{equation}
 \begin{split}
  u_n(t)= \frac{a\cos t}{\kappa - 1} \\
+\frac{4p}{\pi} \sum _{l= 0}^{ \left \lfloor \sqrt{\kappa+4\gamma}/2-1/2 \right \rfloor}   \frac{1}{\sqrt{(2\gamma)^2-Q_l^2}}  \\ \times \sin\left \lbrace \lvert n \rvert \arcsin{\frac{\sqrt{(2\gamma)^2-Q_l^2}}{2\gamma}} \right. \\ \left.-(2l+1)(t-\phi) \vphantom{\frac{\sqrt{1^2}}{2}}\right \rbrace
 +\frac{4p}{\pi} \left(-\frac{1}{2\gamma}\right)^{\lvert n \rvert} \times  \\
\sum _{l= \left \lceil \sqrt{\kappa+4\gamma}/2-1/2 \right \rceil}^{\infty} \frac{\cos[(2l+1)(t-\phi)]}{\sqrt{Q_l^2-(2\gamma)^2}} \\
 \times \left  \lbrace \vphantom{\sqrt{Q_l^2}} Q_l \right.  \left.  -\sqrt{Q_l^2-(2\gamma)^2} \right \rbrace^{\lvert n \rvert}
 \label{eq35}
 \end{split}
 \end{equation}
and one now has:
\begin{equation}
\begin{split}
S_0  = \sum _{l= \left \lceil \sqrt{\kappa+4\gamma}/2-1/2 \right \rceil}^{\infty}  \frac{1}{\sqrt{Q_l^2-(2\gamma)^2}}
\label{eq36}
\end{split}
 \end{equation}
with account of definitions (\ref{eq30}). Taking the limit, one obtains the upper bound for the lowest value of $a$ for which a localized breather exists:
\begin{equation}
\begin{split}
 a_{min} \le  \lim_{\kappa \to 0}{\lim_{\gamma \to 0}\lim_{k \to 0}a_{min}} = \\ \frac{1}{\sqrt{1+(\pi/2)^2}} \approx \ 0.537
\end{split}
 \label{eq37}
 \end{equation}

An upper bound on the amplitude emerges from a consistency requirement, related to the assumption that impact is experienced only by the central mass. In other words, we should require:
\begin{equation}
\begin{split}
a \le \bar{a} \triangleq \min_{n,t}\max_{a}L(a) \ , \ L(a) = a \ \\  s.t.: \lvert u_n(t,a) \rvert < 1 \  \ \forall \ |n| \in \mathbb{N}
\label{eq38}
\end{split}
\end{equation}
where ${u_n(t,a)}$ is given by Eq. (\ref{eq35}) with Def. (\ref{eq17b}) and $\phi$ and $p$ are calculated from Eqs. (\ref{eq28}-\ref{eq30}, \ref{eq36}).

The general analytic solution of (\ref{eq38}) is not within reach. Hence, we employ here a combination of analytic approximations and bounds and numerical calculation, carried-out on a Matlab platform, for a typical set of parameters.

First, we show that for single-harmonic excitation there are no phonon-emitting localized solutions. We begin by showing it analytically, in an approximate manner, and later verify it numerically for a typical set of parameters. The simplest approximate analytic way to show that phonon-emitting localized solutions cannot exist in the examined case is to show that the localization condition is violated already for an amplitude as small as the lower bound, corresponding to the limit where impact first takes place at $n=0$. One can show that for the smallest feasible amplitude, infinitely distant masses have maximal displacements larger than unity for almost the entire phonon-emitting parameter sub-domain. This can be done in several steps. First, it can be shown that the third term in (\ref{eq35}) vanishes for $n \gg 1$. We note that for every two real numbers $A$ and $B$ satisfying $A \ge B>0$, the following algebraic property holds:
\begin{equation}
\begin{split}
 A \ge B > 0 \Rightarrow B(2A-B) \ge B^2 \Rightarrow \\ A^2 - B^2 \ge  A^2 - B(2A-B) \\ = (A-B)^2
 \Rightarrow A \ge  \sqrt{A^2 - B^2} \ge \\ A-B \ge 0 \Rightarrow  0 \le A -\sqrt{A^2 - B^2} \\ \le B
 \Rightarrow 0 \le \frac{A -\sqrt{A^2 - B^2}}{B} \le 1 \\ \Rightarrow \lim_{n \to \infty}{{\left (\frac{A -\sqrt{A^2 - B^2}}{B} \right )}^n}=0
  \label{eq39}
  \end{split}
  \end{equation}
Then, if we set:
\begin{equation}
 A \triangleq (2l+1)^2-(2\gamma+\kappa)  \ \ ; \ \ \ B \triangleq 2\gamma
 \label{eq40}
  \end{equation}
we would have $A \ge B \ge 0$ whenever $(2l+1)^2>4\gamma+\kappa$ and the third term in (\ref{eq35}) would vanish for $n \to \infty$.

Next, we look at the expression for the displacement of infinitely distant masses at time $t=2\pi N +\pi /2$, with $N \in \mathbb{N}$ and N being possibly very large, even infinite, for causality to hold:
\begin{equation}
\begin{split}
\bar{u}(y) \triangleq u_{n\to \infty}(\pi /2)= \\
\frac{4p}{\pi} \sum _{l= 0}^{ \left \lfloor \sqrt{\kappa+4\gamma}/2-1/2 \right \rfloor}   \frac{\sin(2\pi y)}{\sqrt{(2\gamma)^2-Q_l^2}} \\
\label{eq41}
 \end{split}
 \end{equation}
where
\begin{equation}
\begin{split}
 y \triangleq \lim_{n \to \infty}\left \lbrace\frac{ \lvert n \rvert \arcsin{\frac{\sqrt{(2\gamma)^2-Q_l^2}}{2\gamma}}}{2\pi} \right. \\ \left. -\frac{(2l+1)(\pi/2-\phi)}{2\pi} \vphantom{\frac{\frac{\sqrt{[1]^2}}{1}}{2}} \right \rbrace _{frac}
 \label{eq42}
 \end{split}
 \end{equation}
and $\left \lbrace \cdot{} \right \rbrace _{frac}$ denotes the fractional part of a real number, i.e. $\lbrace r \rbrace _{frac}= r-\lfloor{r}\rfloor$.

Now, obviously the consistency condition, which allowed us to write Eq. (\ref{eq7}) with $\delta_{n0}$ at the right-hand side, is violated if $\underset{y}{\operatorname{max}} \ \bar{u}(y)>1$.
Defining,
\begin{equation}
 \begin{split}
  A_l \triangleq 0  \ , \  \text{if} \ \  l  \ > \  \lfloor{\sqrt{\kappa+4\gamma}/2-1/2} \rfloor \\
A_l \triangleq \frac{4p}{\pi} \frac{1}{\sqrt{(2\gamma)^2-Q_l^2}} \ , \\  \text{if} \ \  l \  \le \  \lfloor{\sqrt{\kappa+4\gamma}/2-1/2} \rfloor
 \label{eq43}
 \end{split}
 \end{equation}
and acknowledging that $\bar{u}(y)$ is a real odd function, we can express it as a Fourier sine series,
\begin{equation}
\bar{u}(y) =  \sum _{l= 0}^{\infty} A_l \sin{(2\pi y)}
\label{eq44}
\end{equation}

Next, the following obviously holds, due to Parseval's theorem,
\begin{equation}
\begin{split}
 [\underset{y}{\operatorname{max}} \ \bar{u}(y) ]^2 = \int_{0}\limits^{1}{ [\underset{y}{\operatorname{max}} \ \bar{u}(y) ]^2 dy} \ge \\ \int_{0}\limits^{1}{ \bar{u}^2 (y)  dy} = \sum_{l=0}^{\infty} A_l^2
 \label{eq45}
\end{split}
 \end{equation}

And thus, a sufficient condition for inconsistency is:
\begin{equation}
 \frac{4p_*}{\pi} \sqrt{S_1} > 1
 \label{eq46}
 \end{equation}
where $p_*$ is the minimum with respect to $a$ of the expression given in (\ref{eq29}) and $S_1$ follows from (\ref{eq43}) as:
\begin{equation}
 S_1 \triangleq \sum_{l=0}^{\lfloor{\sqrt{\kappa+4\gamma}/2-1/2}\rfloor} \frac{1}{{(2\gamma)^2-Q_l^2}}
 \label{eq47}
  \end{equation}

The condition for inconsistency is thus fulfilled if:
\begin{equation}
 S_0 (\sqrt{S_1} - S_0) > (\pi q/4)^2
 \label{eq48}
 \end{equation}
where $S_0$ is given by (\ref{eq36}).

Now, in order to illustrate analytically why phonon emission is not taking place in the examined scenario, we shall perform several approximations. First, since for practical purposes one would most probably consider at least moderately high restitution coefficients, we shall assume $q \ll 1$ and hence analyze a simplified form of the inconsistency condition, namely: $\sqrt{S_1}>S_0$. Also, in order to make the analytic reasoning clearer, we narrow down the argumentation to show that the lowest frequency phonon is inconsistent with the assumption of localization. In order to check whether the lowest frequency in the propagation band violates the model's assumptions, we retain only the first frequency in the sum in (\ref{eq47}), which corresponds to the parameter range: $1/4 < \gamma <  9/4$. Moreover, in order to eliminate the $\kappa$-dependence, we replace the aforementioned condition by a stricter one, namely: $\underset{\kappa}{\operatorname{min}} \ \sqrt{S_1} > \underset{\kappa}{\operatorname{max}} \ S_0$.

Consequently, the inconsistency condition becomes:
\begin{equation}
 \begin{split}
    \frac{1}{\sqrt{4\gamma-1}} >\\  \sum _{l= 1}^{\infty}  \frac{1}{\sqrt{(2l+1)^2-1}\sqrt{(2l+1)^2-1-4\gamma}} \ , \\ \ \text{if} \ 1/2 \le \gamma < 9/4 \\
 \frac{1}{{2\gamma}} > \\ \sum _{l= 1}^{\infty}  \frac{1}{\sqrt{(2l+1)^2-1}\sqrt{(2l+1)^2-1-4\gamma}} \ , \\ \ \text{if} \ 1/4 < \gamma< 1/2
 \label{eq49}
 \end{split}
  \end{equation}

Clearly, condition (\ref{eq49}) is satisfied whenever a stricter, and somewhat simpler, condition is satisfied:
\begin{equation}
 \begin{split}
   \Gamma_1(\gamma) \triangleq  \frac{1}{\sqrt{4\gamma-1}} -\frac{1}{4\sqrt{2}\sqrt{2-\gamma}}>   \sigma_1  \ , \\ \text{if} \ 1/2 \le \gamma < 9/4 \\
 \Gamma_2(\gamma) \triangleq \frac{1}{{2\gamma}} -\frac{1}{4\sqrt{2}\sqrt{2-\gamma}} >    \sigma_2  \ , \\ \text{if} \ 1/4 < \gamma< 1/2
  \label{eq50}
  \end{split}
  \end{equation}
where,
\begin{equation}
\begin{split}
  \sigma_1 \triangleq  \sum _{l= 2}^{\infty}  \frac{[(2l+1)^2-1]^{-1/2}}{\sqrt{(2l+1)^2-10}} \approx 0.1399   \\ \sigma_2 \triangleq  \sum _{l= 2}^{\infty}  \frac{[(2l+1)^2-1]^{-1/2}}{\sqrt{(2l+1)^2-3}} \approx 0.1276
  \label{eq51}
  \end{split}
  \end{equation}

Obviously, $\Gamma_2(0) \to \infty$ and $\Gamma_2(2) \to -\infty$ and thus, given that  $\Gamma_2(\gamma)$ is smooth in the domain (0,2), the critical value for the fulfillment of the second condition in (\ref{eq50}) is a real number within the range $(0,2)$. Plotting $\Gamma_2(\gamma)$ reveals that the second condition in (\ref{eq50}) is satisfied for: $0<\gamma<1.403$ and as the second condition in (\ref{eq50}) is only relevant in the range (1/4,1/2), which is a sub-domain of (0,1.403), evidently the solution given in (\ref{eq35}) is inexistent for: $1/4<\gamma<1/2$.

Similarly, for the first condition in (\ref{eq50}) we have: $\Gamma_1(1/4) \to \infty$ and $\Gamma_1(2) \to -\infty$ and thus, given that  $\Gamma_1(\gamma)$ is smooth in the domain (1/4,2), the critical value for the fulfillment of the first condition in (\ref{eq50}) is a real number within the range $(1/4,2)$. Neglecting $\sigma_1$ in the first inequality in (\ref{eq50}) gives an approximate critical value of $\gamma_{cr}=65/36 \approx 1.8056$, below which there exists no localized breather. Plotting $\Gamma_1(\gamma)$ reveals that the first condition in (\ref{eq50}) is exactly satisfied for: $1/4<\gamma<1.621$ and as it is relevant only within (1/2,9/4), it is only meaningful in (1/2,1.621).

Finally, combining the two results, we get a semi-analytic evidence that for $q \ll 1$, there is no solution for: $1/4<\gamma<1.621$, for any value of $\kappa$.

We have thus shown that in the majority of the parameter range corresponding to the smallest propagating frequency, there exists no solution. Numerical analysis shows that for the remaining part of the first frequency band, $\gamma \in [1.621,9/4)$ there is no solution either, due to violation of the consistency condition by masses lying in the proximity of the symmetry point. The proof of non-existence of solutions in domain-bands corresponding to higher frequency phonons is expected to be similar to the one outlined here, and numerical analysis is performed instead.

\begin{figure}
\includegraphics[scale=0.53]{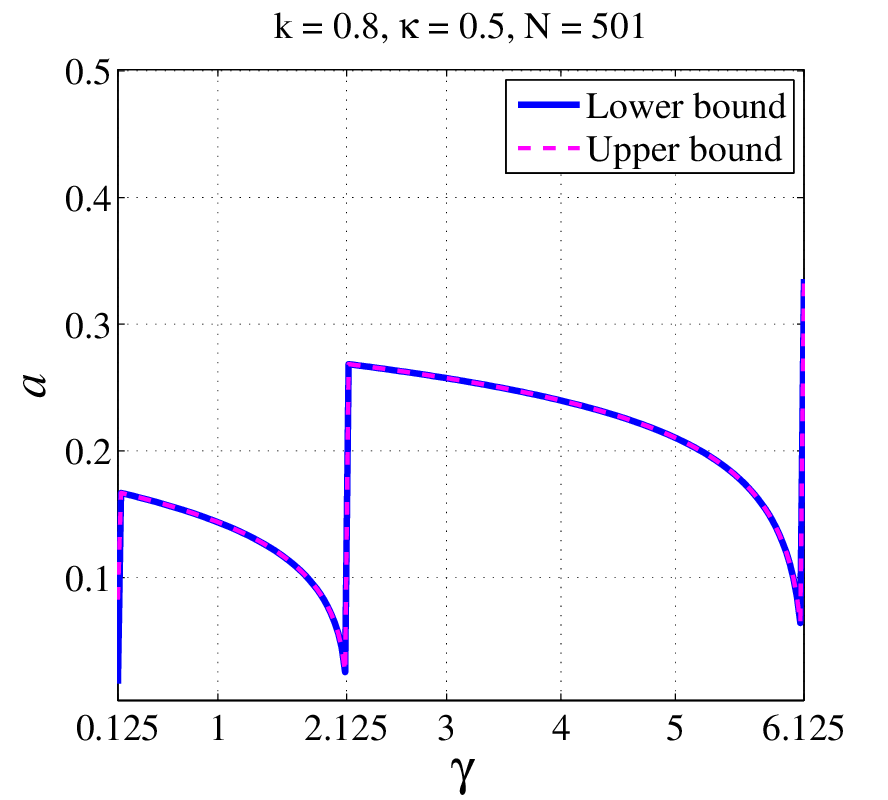}
\caption{\small The lower and upper bounds on the excitation amplitude for which the localized breather exists, versus the link stiffness.}
\label{Figure1}
\end{figure}

\begin{figure}
\includegraphics[scale=0.53]{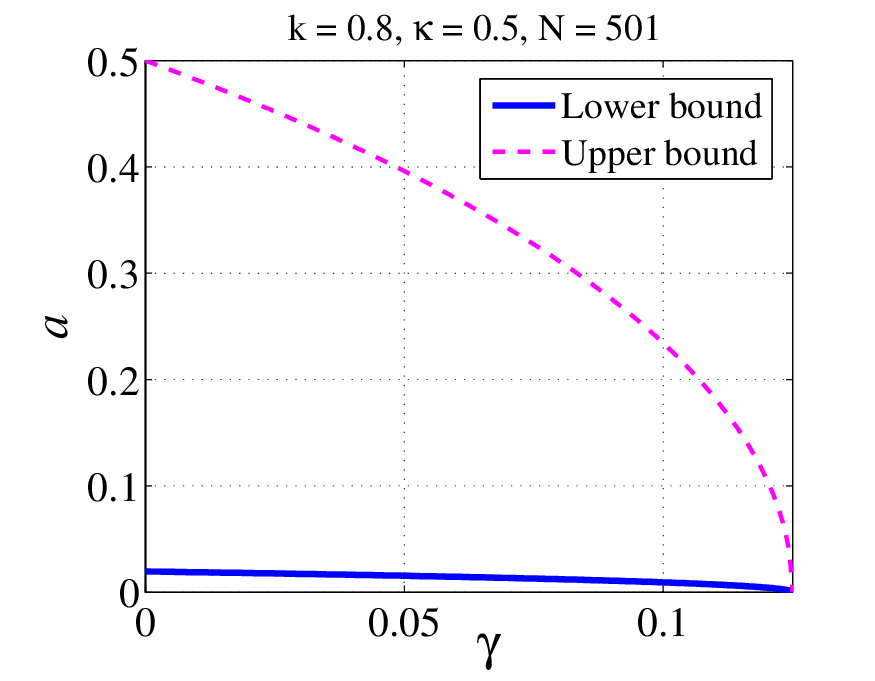}
\caption{\small The lower and upper bounds on the excitation amplitude for which the localized breather exists, versus the link stiffness - attenuation zone}
\label{Figure1b}
\end{figure}

In order to illustrate the non-existence of phonon-emitting localized breathers for the examined system, we present, below, the numerical solution of system (\ref{eq38}). The resulting upper bound, as well as the analytically calculated lower bound on the excitation amplitude $a$, versus $\gamma$, for $\kappa$ taken exactly from the middle of the feasible range and for typical value of $k$, are plotted in Figures \ref{Figure1}-\ref{Figure1b}.

Figure \ref{Figure1} shows that for values of $\gamma$ corresponding to propagation frequencies, the lower and upper bounds on the excitation amplitude coincide. A non-zero gap between the upper and lower existence-domain boundaries exists only for: $\gamma<(1-\kappa)/4$. This means that all of the frequencies of the DB should lie in the upper attenuation zone. This phenomenon is reproduced for parameter values chosen arbitrarily from the feasible domain. Together with the analytic illustration of the absence of self-consistency of phonon-emitting solutions, we conclude that the breather solution for single-harmonic excitation is not phonon-emitting and can thus be expressed in the following form:
\begin{equation}
 \begin{split}
 u_n(t)= -\frac{a\cos t}{1-\kappa} + \frac{4p}{\pi} \left(-\frac{1}{2\gamma}\right)^{\lvert n \rvert} \times \\
\sum _{l=0}^{\infty} \frac{\cos[(2l+1)(t-\phi)]}{\sqrt{[(2l+1)^2-(2\gamma+\kappa)]^2-(2\gamma)^2}}   \\ \times \left \lbrace \vphantom{\sqrt{[(1)^2]^2}} (2l+1)^2-(2\gamma+\kappa) \right. \\ \left. -\sqrt{[(2l+1)^2-(2\gamma+\kappa)]^2-(2\gamma)^2}\right \rbrace^{\lvert n \rvert}
\label{eq52}
\end{split}
  \end{equation}
with the necessary conditions for existence being: $\gamma<(1-\kappa)/4$ and $0<\kappa<1$.

Now that it has been shown that the solution can be expressed by a single time Fourier series as given above, three other points should be clarified in regard with the self-consistency of the DB solution. First, it is to be shown that the displacement of the central mass is indeed always smaller than unity between subsequent impacts, practically speaking, in the interval: $\phi<t<\phi+\pi$. Subsequently, (the evident) periodicity assures the absence of impacts between the nominal impacts for arbitrarily large values of $t$. Second, it should be shown that no other mass beside the central one experiences impact. Third, series convergence issues are to be addressed. The first and the third matters are addressed in detail in \ref{AppendixA}. The second matter is addressed in the context of the solution of problem (\ref{eq38}), which is carried out numerically and shown in Figures \ref{Figure1} and \ref{Figure1b}, and is further investigated in an approximate analytical manner in \ref{AppendixB}.

To conclude this section, we present, below, the second-order Taylor expansions of both the lower and the upper bound on the excitation amplitude guaranteeing the existing of the breather, the former being obtained by direct expansion of the function given in (\ref{eq34}), acknowledging (\ref{eq30}) and (\ref{eq36}) with a lower bound of $0$, and the latter $-$ from an approximate analytical treatment of problem (\ref{eq38}), as described in detail in \ref{AppendixB}:
\begin{equation}
\begin{split}
\underline{a}=\left[ \frac{2}{\pi}-\left(\frac{2}{\pi}+\frac{\pi}{6}\right)\kappa+\left(\frac{\pi}{6}-\frac{\pi^3}{360}\right)\kappa^2\right]q \\+ \left[-\frac{\pi}{3}+\left(\frac{\pi}{3}-\frac{\pi^3}{90}\right)\kappa +\left(\frac{\pi^3}{90}+\frac{11\pi^5}{1680} \right. \right. \\ \left. \left. -\frac{\pi^7}{1152}\right)\kappa^2 \right]q\gamma + \left[-\frac{2\pi^3}{45}+\left(\frac{2\pi^3}{45} \right. \right. \\ \left. \left. -\frac{3\pi^5}{560} \right)\kappa +\left(\frac{3\pi^5}{560}-\frac{41\pi^7}{75600} \right)\kappa^2 \right]q \gamma ^2
\end{split}
\label{eq88}
\end{equation}

\begin{equation}
\begin{split}
\bar{a}=1-\kappa -\left[ \frac{\pi^2}{6}+\frac{7\pi^2-60}{360}\pi^2\kappa \right. \\ \left.+\frac{31\pi^2-294}{15120}\pi^4\kappa^2-\left(\frac{2}{3}-\frac{20+\pi^2}{30}\kappa \right. \right. \\ \left.  \left.+\frac{168-10\pi^2}{5040}\pi^2\kappa^2\right)q^2\right]\gamma
-\left[ \frac{\pi^4}{40}\right. \\ \left. +\frac{25\pi^2-126}{5040}\pi^4\kappa +\frac{147\pi^2-1000}{201600}\pi^6\kappa^2\right. \\+ \left. \left(\frac{11\pi^2}{90}-\frac{1155-110\pi^2}{9450}\pi^2\kappa \right. \right.\\ \left. \left. -\frac{264-17\pi^2}{22680}\pi^4\kappa^2\right)q^2\right]\gamma^2
\end{split}
\label{eq89}
\end{equation}

These closed-form analytic expressions may be more convenient to use in the case of weak coupling in a relatively weak quadratic potential under single-harmonic excitation with the central mass impacting with nearly perfect restitution.

\section{Linear stability analysis}
\label{sect5}
\subsection{Theoretical background}
\label{sect5A}

The derivation presented in the previous section, along with the details given in the appendices, in fact proves the existence of a localized DB in the examined setting, by showing that there exists a nonempty sub-domain in the excitation amplitude-link stiffness space, in which all the conditions for the existence of the breather are satisfied. Some full plots of the bounds of the aforementioned sub-domain, including the lower bound for several representing values of the remaining problem parameters, are presented in this section.

The next question one naturally aims to address, after proving the presence of a non-empty existence domain in the parameter space, is the question of the linear stability of the solution. Once we have shown that a linear elastic system with an additional limiting constraint, specified by a restitution coefficient, has a \emph{periodic} solution when subjected to periodic loading, the question of linear stability can be examined using Floquet theory. According to the theory, unstable points in the parameter space correspond to Floquet multipliers lying outside the unit circle in the complex plane, Floquet multipliers being the eigenvalues of the monodromy matrix.

In dynamical systems' notation, $\textbf{\emph{x}}_p(t)=(\textbf{u}^\top,\dot{\textbf{u}}^\top)^\top$ is a linearly stable periodic solution if and only if the monodromy matrix $\textbf{M}$, specified by:
\begin{equation}\delta\textbf{\emph{x}}_p(t_0+T)=\textbf{M} \delta\textbf{\emph{x}}_p(t_0)
\label{eq90}
\end{equation}
where $T$ is the period, has no eigenvalues outside the unit circle. It should be noted that due to the $\delta$ appearing in Eq. (\ref{eq90}), when employing Floquet theory, one considers only linear stability and thus only small perturbations to the periodic solution. This is why eigenvalues lying exactly on the unit circle do not necessarily imply instability.  This is important because in systems like ours, there is usually a large number of eigenvalues lying \emph{on} the unit circle, which correspond to the displacements and velocities of masses that do not experience impact.

In our case, since the period is $2\pi$ and there are two impacts during a single period, at $t=\phi$ and at $t=\phi+\pi$, and the response is linear between impacts, the monodromy matrix takes the following form (see \cite{FredrikssonNordmark2000}, for example):
\begin{equation}
\textbf{M}=\textbf{L}\textbf{S}\textbf{L}\textbf{S}=(\textbf{L}\textbf{S})^2 \ , \ \textbf{L}=\exp{(\pi\textbf{A})}
\label{eq91}
\end{equation}
where $\textbf{A}$ is the matrix representing the linear system given by (\ref{eq1}) when the displacement-limiting condition is omitted, and $\textbf{S}$ is the saltation matrix \cite{Champneys2008}, representing the leap in the perturbation vector corresponding to a non-smooth obstacle encountered by the trajectory of the perturbation of a limit-cycle in the phase space. As a geometric object in the phase space, the saltation matrix is the mapping matrix that shifts and rotates the perturbation vector as the dynamical flow approaches the discontinuity surface in the state space.

The saltation matrix can be expressed by use of geometrical construction in the phase space. Once the evolution vector $\textbf{f}_p$, embodied in the relation $\dot{\textbf{x}}_p=\textbf{f}_p$, and the jump surface $h(\textbf{x},t)$, represented by, say, its normal $\textbf{n}$, in phase space coordinates, are known, the saltation matrix can be expressed as:
\begin{equation}
\textbf{S}=(\nabla \textbf{g}_p^{\top})^{\top}+\frac{\left[\textbf{f}_p^+-(\nabla \textbf{g}_p^{\top})^{\top}\textbf{f}_p^-\right]\textbf{n}^\top}{\textbf{n}^\top\textbf{f}_p^-+\frac{\partial{h}}{\partial{t}}|_{t=t_p}}
\label{eq92}
\end{equation}
where $\textbf{g}_p(\textbf{x}_p^-)=\textbf{x}_p^+$ is the jump mapping \cite{Champneys2008, Zhusubaliyev2003}.

For a system of the sort that we examine, where the impact condition is represented by an instantaneous change of the sign of the velocity, with a (positive) restitution coefficient lower than unity and linear behavior between impacts, and with an infinite number of masses symmetrically positioned with respect to the central mass, which is assumed to be the only mass to experience impact, the saltation matrix, derived as in \cite{FredrikssonNordmark2000}, can be expressed by blocks construction, as follows:
\begin{equation}
\textbf{S} = \begin{pmatrix}
       \textbf{K} & \textbf{0}         \\
       \textbf{C}      & \textbf{K}
     \end{pmatrix}
\label{eq93}
\end{equation}
where $\textbf{0}$ is an $N\times N$ zero matrix with all entries equal to zero ($N$ standing for the total number of masses in the system), and $\textbf{K}$ and $\textbf{C}$ are $N\times N$ matrices, given by:
\begin{equation}
\begin{split}
\textbf{K}\triangleq
 \begin{pmatrix}
  -k & 0 & \cdots & 0 \\
  0 & 1 & \cdots & 0 \\
  \vdots  & \vdots  & \ddots & \vdots  \\
  0 & 0 & \cdots & 1
 \end{pmatrix} ,\\  \textbf{C}\triangleq\begin{pmatrix}
  C & 0 & \cdots & 0 \\
  0 & 0 & \cdots & 0 \\
  \vdots  & \vdots  & \ddots & \vdots  \\
  0 & 0 & \cdots & 0
 \end{pmatrix}
\end{split}
\label{eq94}
\end{equation}
where for a localized breather one can show, following \cite{FredrikssonNordmark2000}, that the nonzero entry $C$ in the saltation matrix is the same as it is for a single mass in a vibro-impact potential (since smooth forces cannot change the velocity in a single instance, and the only non-smoothness is associated solely with the central mass $-$ we choose a numbering order for which the first equation is for the central mass, and then, consecutively with an increasing $n$ there come the equations for the  rest of the masses, with the equation for $n=(N-1)/2$ followed directly by the equation for $n=-(N-1)/2$. Consistent construction is applied for $\textbf{K}$ and $\textbf{C}$), which for a system without linear damping is proportional to the acceleration just before an impact, and can be shown to be given explicitly by:
\begin{equation}
C \triangleq (1+k)\ddot{u}_0^{-}/\dot{u}_0^{-}
\label{eq95}
\end{equation}
in which, making use of (\ref{eq21}-\ref{eq24}) and (\ref{eq26}), the velocity just before impact can be shown to be given by:
\begin{equation}
\dot{u}_0^{-} = (1+q)p
\label{eq96}
\end{equation}
and the acceleration of the central mass just before impact can be obtained from (\ref{eq1}) and may be shown to satisfy:
\begin{equation}
\ddot{u}_0^{-} = \hat{a}\cos{\phi}-(\kappa+2\gamma)\chi_0 p -2\gamma\chi_1 p
\label{eq97}
\end{equation}
where use is made of definitions (\ref{eq28}-\ref{eq30}), in which $\chi_0$ is proportional to $S_0$, which for a periodic localized solution corresponding to an immobile breather is given by:
\begin{equation}
S_0=\underset{l=0}{\overset{\infty}{\sum}}\frac{1}{\sqrt{[\kappa+2\gamma-(2l+1)^2]^2-(2\gamma)^2}}
\label{eq98}
\end{equation}

Also, $\chi_1\triangleq4S_1/\pi$, where $S_1$ is given by:
\begin{equation}
\begin {split}
S_1=\frac{1}{2\gamma}\underset{l=0}{\overset{\infty}{\sum}}\frac{1}{\sqrt{[\kappa+2\gamma-(2l+1)^2]^2-(2\gamma)^2}} \\ \times \left|(2l+1)^2-(\kappa+2\gamma) \vphantom{\sqrt{[1]^2}} \right. \\ \left. -\sqrt{[\kappa+2\gamma-(2l+1)^2]^2-(2\gamma)^2}\right|
\end{split}
\label{eq99}
\end{equation}

To conclude this subsection, we note that the linear-dynamics matrix $\textbf{A}$ can be presented as follows:
\begin{equation}
\textbf{A} = \begin{pmatrix}
       \textbf{0} & \textbf{I}         \\
       \textbf{B}      & \textbf{0}
     \end{pmatrix}
\label{eq100}
\end{equation}

Once again, the essence of $\textbf{A}$ is in block $\textbf{B}$ (as $\textbf{I}$ and $\textbf{0}$ are simply $N\times N$ unity and zero matrices, respectively), given by the following augmented tridiagonal $N\times N$ matrix (in which $\mathcal{G}\triangleq \kappa+2\gamma$):
\begin{equation}
\textbf{B}\triangleq
 \begin{bmatrix}
  -\mathcal{G} & \gamma & 0 & \cdots & 0 & \gamma \\
  \gamma & -\mathcal{G} & \gamma & \cdots & 0 & 0\\
  0 & \gamma & -\mathcal{G} & \gamma & 0  & 0\\
  \vdots  & \vdots  & \ddots & \vdots  \\
  0 & 0 & \cdots  &  \gamma & -\mathcal{G}  & \gamma\\
  \gamma & 0 & 0 & \cdots  &  \gamma & -\mathcal{G}
 \end{bmatrix}
\label{eq101}
\end{equation}
where the first row corresponds to $u_0$ and the last row corresponds to $u_{-1}$, in coincidence with the ordering within the saltation matrix, and where, in principle, in order for the resulting monodromy matrix to determine the stability of the solution obtained in the preceding sections, one has to take the $N \to \infty$ limit. However, since for a large, tridiagonal matrix, spectral analysis can only be performed numerically, $N$ has to be finite. For a finite $N$, however, to use the analytic solution for the extremal beads and check the stability of the intermediate $N-2$ beads would be inconsistent with the uniform excitation assumption, since the displacements of the extremal beads would act like force-terms in the remaining equations of motion. In this paper we use periodic boundary conditions, thus turning the chain into a ring. The choice of this alternative is what produces the off-band terms in expression (\ref{eq101}) and the special form of the integration scheme, as given in Section \ref{sect6}. The following subsections outline the obtained analytic and numerical results most noteworthy from the aforementioned perspective.

\subsection{Analytic results}
\label{sect5B}

Before going into the process of the numerical solution of the linear stability problem, one can take advantage of the fact that in the case that we examine, the monodromy matrix can be expressed analytically and see whether one could also solve the full linear stability problem analytically, at least for special values of the problem parameters.

Clearly, the simplest case for which one could try to look for an analytic solution would be the $\gamma \to 0$ limit, since in that case one effectively gets a $2\times 2$ monodromy matrix. In what follows, we indeed investigate this limit-case analytically.

First, one notes that: $\textbf{B}_{\gamma \to 0} \to -\kappa$, and thus:
\begin{equation}
\textbf{L}_{\gamma \to 0} \to \exp{\begin{pmatrix} 0 & \pi \\ -\pi\kappa & 0 \end{pmatrix}}
\label{eq102}
\end{equation}

In order to perform matrix exponentiation, one has to make the matrix diagonal. The eigenvalues of $\pi\textbf{A}_{\gamma \to 0}$ can be easily found to be: $\pm \pi i \sqrt{\kappa}$.

Realizing that the (non-normalized) eigenvectors are $(1,\pm i\sqrt{\kappa})^\top$ we can write the diagonalization decomposition equation as:
\begin{equation}
\begin{split}
\textbf{L}_{\gamma \to 0}  \begin{pmatrix} 1 & 1 \\i\sqrt{\kappa} & -i\sqrt{\kappa}\end{pmatrix}= \\ \begin{pmatrix} 1 & 1 \\i\sqrt{\kappa} & -i\sqrt{\kappa}\end{pmatrix} \exp{\begin{pmatrix} \pi i \sqrt{\kappa} & 0 \\ 0 & -\pi i \sqrt{\kappa}\end{pmatrix}}
\end{split}
\label{eq103}
\end{equation}
where the eigenvector normalization prefactor $1/\sqrt{1+\kappa}$ is first introduced in order for the diagonalization matrix not to be affected by the exponentiation, and then canceled-out due to the presence of both the eigenvector matrix and its inverse in the diagonalization equation.

Next, acknowledging that: $\exp{[\text{Diag}(\textbf{\emph{v}})]}=\text{Diag}(\textbf{\emph{w}})$, where: $\emph{w}_j=\exp{(\emph{v}}_j)$, and Diag$(\textbf{\emph{y}})$ is an operator that returns a diagonal matrix with the diagonal entries corresponding in an order-preserving manner to the components of its argument, the vector $\textbf{\emph{y}}$, and using the appropriate trigonometric relations, we can solve (\ref{eq103}) to get (after performing matrix inversion and multiplication operations analytically):
\begin{equation}
\textbf{L}_{\gamma \to 0} =\begin{bmatrix} \cos{(\pi \sqrt{\kappa})} &\sin{(\pi \sqrt{\kappa})}/\sqrt{\kappa} \\-\sqrt{\kappa} \sin{(\pi \sqrt{\kappa})} & \cos{(\pi \sqrt{\kappa})} \end{bmatrix}
\label{eq104}
\end{equation}
which satisfies: $\det{\textbf{L}_{\gamma \to 0}}=1$.

Now, since essentially the $\gamma \to 0$ case is equivalent to the $N=1$ case, one obtains from (\ref{eq93}-\ref{eq94}) that the saltation matrix for this case is:
\begin{equation}
\textbf{S}_{\gamma \to 0} =\begin{pmatrix} -k & 0 \\ C & -k \end{pmatrix}
\label{eq105}
\end{equation}

Following (\ref{eq91}), the monodromy matrix for the $\gamma \to 0$ case becomes:
\begin{equation}
\textbf{M}_{\gamma \to 0} =k^2 \begin{pmatrix} M_{11} & \hat{M}_{12}/\sqrt{\kappa} \\ -\hat{M}_{21}\sqrt{\kappa} & M_{22}\end{pmatrix}
\label{eq106}
\end{equation}
where
\begin{equation}
\begin{split}
M_{11} \triangleq \cos(2\pi\sqrt{\kappa})-(3/2)\hat{C}\sin(2\pi\sqrt{\kappa}) \\ +\hat{C}^2\sin^2{(\pi\sqrt{\kappa})} \\ \hat{M}_{12} \triangleq \sin(2\pi\sqrt{\kappa})-\hat{C}\sin^2(\pi\sqrt{\kappa})\\
\hat{M}_{21} \triangleq  \sin(2\pi\sqrt{\kappa})+ \\ \hat{C}[2\cos^2{(\pi\sqrt{\kappa})}-\sin^2(\pi\sqrt{\kappa})] \\ -(1/2)\hat{C}^2\sin(2\pi\sqrt{\kappa})\\
M_{22} \triangleq \cos(2\pi\sqrt{\kappa})-(1/2)\hat{C}\sin(2\pi\sqrt{\kappa})
\end{split}
\label{eq107}
\end{equation}
and $\hat{C} \triangleq C/(k\sqrt{\kappa})$.

At this point the question of linear stability can be addressed by examining the characteristic equation:
\begin{equation}
(M_{11}-\lambda/k^2)(M_{22}-\lambda/k^2)+\hat{M}_{12}\hat{M}_{21}=0
\label{eq108}
\end{equation}

According to Floquet theory, critical stability corresponds to: $|\lambda|=1$.

Substituting this critical condition, along with definitions (\ref{eq107}) and the definition of $\hat{C}$ given above into (\ref{eq108}) and solving the resulting quadratic equation with respect to $C$, produces an expression for a critical $C$ value, as follows (formally Eq. (\ref{eq108}) may seem cubic in $C$, but the cubic term cancels out):
\begin{equation}
\begin{split}
C_{cr} = \\ 2k\sqrt{\kappa}\frac{\cos{(\pi\sqrt{\kappa})}\pm\sqrt{[1+ (k^2+k^{-2})/2]/2}}{\sin{(\pi\sqrt{\kappa})}}
\label{eq109}
\end{split}
\end{equation}

For the limit $k \to 1, \ \kappa \to 0$, if one takes the negative sign in (\ref{eq109}), then the critical value of $C$ becomes zero, which means that for a single-degree-of-freedom system with a full-restitution vibro-impact potential of finite stiffness, any perturbation from a periodic solution is fully damped during a single impact. In the same time, for $\gamma \to 0,\kappa \to 0$, the linear part of the system is unconditionally linearly stable, and yet the criticality implies that an eigenvalue of the monodromy matrix is on the verge of becoming smaller than $-1$ while staying real. This is a contradiction. The resolution of this contradiction is that for feasible values of $k$ and $\kappa$, only the 'plus' sign corresponds to critical conditions.
Consequently,
\begin{equation}
\begin{split}
C^{\gamma \to 0}_{cr} = \\ 2k\sqrt{\kappa}\frac{\cos{(\pi\sqrt{\kappa})}+\sqrt{[1+ (k^2+k^{-2})/2]/2}}{\sin{(\pi\sqrt{\kappa})}}
\label{eq110}
\end{split}
\end{equation}

In the same time, using (\ref{eq95}-\ref{eq97}) along with (\ref{eq28}-\ref{eq30}) gives:
\begin{equation}
\begin{split}
C^{\gamma \to 0}_{cr}=  \frac{(1+k)^2}{2} \times \\ \left [\sqrt{(\hat{a}/p)^2_{cr,\gamma \to 0}-q^2}-\sqrt{\kappa}\tan{(\pi\sqrt{\kappa}/2)}\right ]
\end{split}
\label{eq111}
\end{equation}

The combination of (\ref{eq110}) and (\ref{eq111}) yields:
\begin{equation}
\begin{split}
p_{cr,\gamma \to 0}=\hat{a}_{cr,\gamma \to 0} \times \\ \left \lbrace \vphantom{{\left[\frac{\sqrt{\frac{k^2}/2}}{1}\right]}^2} q^2+ \kappa \left [\vphantom{\frac{\sqrt{\frac{k^2}{2}}}{\sqrt{\kappa}}}\tan{\left(\frac{\pi\sqrt{\kappa}}{2}\right)} \vphantom{\frac{\sqrt{[1]^2}}{1}}+\frac{4k}{(1+k)^2}\times \right. \right. \\ \left. \left. \frac{\cos{(\pi\sqrt{\kappa})}+\sqrt{[1+ \frac{k^2+k^{-2}}{2}]/2}}{\sin{(\pi\sqrt{\kappa})}}  \right ] ^2 \right \rbrace ^{-1/2}
\end{split}
\label{eq112}
\end{equation}

Next, from (\ref{eq29}) and (\ref{eq30}), and the second-order Taylor-series expansion followed by analytical summation of $S_0$ as given by Eq. (\ref{eq98}) (which detailed derivation can be found in \ref{AppendixB} in Eqs. (\ref{eq78}) and (\ref{eq84})), we get:
\begin{equation}
\begin{split}
p_{cr,\gamma \to 0}=\frac{\kappa^{-1/2}\tan{(\pi\sqrt{\kappa}/2)}}{q^2+\kappa^{-1}\tan^2{(\pi\sqrt{\kappa}/2)}} \\ +\frac{\sqrt{[q^2+\kappa^{-1}\tan^2{(\pi\sqrt{\kappa}/2)}]\hat{a}_{cr,\gamma \to 0}^2-q^2}}{q^2+\kappa^{-1}\tan^2{(\pi\sqrt{\kappa}/2)}}
\end{split}
\label{eq113}
\end{equation}

Combining (\ref{eq112}) and (\ref{eq113}), we obtain the excitation amplitude corresponding to a critically linearly stable localized DB with vanishingly weak link stiffness, as an exact and explicit function of the coefficient of restitution and the foundation stiffness:
\begin{equation}
\begin{split}
a_{\underset{\gamma  \to 0}{cr}}^{LS}(k,\kappa) = \\ (1-\kappa)\left\lbrace \vphantom{{\left[\frac{\left\lbrace \sqrt{[\frac{k^2}{2}]} \right\rbrace}{(1)^2}\right]}^2} \frac{(1-k)^2}{(1+k)^2}+\kappa \left [ \vphantom{{\frac{\left\lbrace \sqrt{[\frac{k^2}{2}]} \right\rbrace}{(k)^2}}} \tan{\left(\frac{\pi\sqrt{\kappa}}{2}\right)}+\right.\right. \\ 4k\left. \left. \frac{\cos{(\pi\sqrt{\kappa})}+\sqrt{\left(1+ \frac{k^2+k^{-2}}{2}\right)/2}}{(1+k)^2\sin{(\pi\sqrt{\kappa})}} \right]^2\right\rbrace^{1/2} \\ \times \left\lbrace \vphantom{{\left[\frac{\left\lbrace \sqrt{[(1)^2]} \right\rbrace}{(1)^2}\right]}}  \frac{\tan{(\pi\sqrt{\kappa}/2)}}{\sqrt{\kappa}}\mp \sqrt{\kappa}\left [ \vphantom{{\frac{\left\lbrace \sqrt{[\frac{k^2}{2}]} \right\rbrace}{(k)^2}}}  \tan{\left(\frac{\pi\sqrt{\kappa}}{2}\right)} \right. \right. + \\ 4k\left. \left. \frac{\cos{(\pi\sqrt{\kappa})}+\sqrt{\left(1+ \frac{k^2+k^{-2}}{2}\right)/2}}{(1+k)^2\sin{(\pi\sqrt{\kappa})}}\right ]\right\rbrace^{-1}
\end{split}
\label{eq114}
\end{equation}
where the $\mp$ sign in the denominator corresponds to multiple stable branches as will be shown in the numerical results section below, and where $\mp$ is used rather than $\pm$ due to the fact that the $"-"$ sign produces more practical a value for the exact $\gamma \to 0$ limit, than does the $"+"$ sign.

Now, before proceeding with the derivation of the corollaries of (\ref{eq114}), we should justify the assumption that the eigenvalues of the monodromy matrix are real, which leads to Eq. (\ref{eq109}). Substituting (\ref{eq107}) into (\ref{eq108}), we get a quadratic equation with real-valued coefficients for $\lambda$. In order for the solutions of this equations to be real, the discriminant of the equation has to be positive. The discriminant is given by:
\begin{equation}
\begin{split}
\Delta_{\lambda} = k^4[2\sin{(2\pi\sqrt{\kappa})}-\sin^2{(\pi\sqrt{\kappa})}\hat{C}]^2\hat{C}^2 \\ +4k^2[2\cos{(2\pi\sqrt{\kappa})}-k^2]
\end{split}
\label{eq115}
\end{equation}

Clearly, a sufficient condition for $\Delta>0$ is $\kappa\le1/36$, which covers the $\kappa \to 0$ case as well.

For $\kappa>1/36$, more sophisticated argumentation should be employed. Substituting (\ref{eq107}) into (\ref{eq108}) for an arbitrary value of $\lambda$, one obtains a quadratic equation for $\hat{C}$ (of which (\ref{eq109}) is a special case). The solution of this equation for the critically stable case is:
\begin{equation}
\begin{split}
\hat{C}_{cr}= \\ 2\frac{\cos{(\pi\sqrt{\kappa})}\pm\sqrt{[1+ (k^{-2}\lambda_{cr}+k^2\lambda_{cr}^{-1})/2]/2}}{\sin{(\pi\sqrt{\kappa})}}
\label{eq116}
\end{split}
\end{equation}

In order for $\hat{C}_{cr}$ to be real, one has to have:
\begin{equation}
\begin{split}
\Im{(\lambda_{cr}^{-1})}=-|\lambda_{cr}|^{-2}\Im{(\lambda)}\underset{(\ref{eq116})}= \\ -k^{-4}\Im{(\lambda_{cr})} \underset{|\lambda_{cr}|=1>k^2}\Rightarrow \Im{(\lambda_{cr})}=0
\end{split}
\label{eq117}
\end{equation}

Also, from (\ref{eq116}),
\begin{equation}
\begin{split}
\lambda_{cr}=-1 \Rightarrow 2+k^{-2}\lambda_{cr}+k^2\lambda_{cr}^{-1} \\ =2-k^{-2}-k^2< 0 \ \ \forall \ \ k \in (0,1)  \\ \Rightarrow \Im{(C_{cr})} \ne 0
\end{split}
\label{eq118}
\end{equation}

Hence, since by definition: $\Im{(C_{cr})}=0$, one has: $\lambda_{cr}=1$, and thus Eq. (\ref{eq109}) is correct.

Returning to the expression for the excitation amplitude corresponding to critical linear stability in the $\gamma \ll 1$ limit, we should first examine several special cases of (\ref{eq114}).

First, in the $k \to 1$ limit, (\ref{eq114}) yields:
\begin{equation}
a_{cr}^{LS}(\kappa) \underset{\gamma\to 0,k\to 1} \to \frac{2\kappa(1-\kappa)}{1-\cos{(\pi\sqrt{\kappa})}\mp 2\kappa}
\label{eq119}
\end{equation}

This expression can produce at least three noteworthy results. The first is obtained by choosing a $"-"$ sign in (\ref{eq119})  (for a maximum linear-stability-wise critical amplitude) and taking the $\kappa \to 0$ limit by applying L'H\^opital's rule twice, consecutively:
\begin{equation}
\begin{split}
a_{cr}^{LS}(\gamma\to 0,k\to 1,\kappa \to 0) \underset{z \triangleq \sqrt{\kappa}} = \\ \underset{z \to 0}  \lim \frac{2z^2}{1-\cos{(\pi z)}- 2z^2} =\frac{4}{\pi^2- 4}\approx 0.6815 \end{split}
\label{eq120}
\end{equation}

This is the largest amplitude in the described limit that corresponds to linear stability. The maximum breather existence amplitude in this regime is 1. This means that only slightly more than two thirds of the existence-wise feasible gain is exploited.

It is exactly in that point in the analysis where one might ask for stabilization of the system. If one assumes that an unstable breather is uninteresting for practical purposes, then one might ask themselves whether the linearly stable existence domain in the amplitude-link stiffness space can be extended by the addition of an elastic, harmonic foundation potential, at least in the limit of weak coupling. The answer is affirmative and indeed a value higher than 0.6815 can be obtained by increasing $\kappa$ from zero to some finite value. In the $k \to 1$ case   this would not even yield a change in the minimum allowed amplitude for breather existence. In fact, it turns out that by increasing $\kappa$ from 0 up to a critical value, at a certain point, the linear stability-wise critical amplitude becomes equal to the maximum breather existence-preserving amplitude. Increasing $\kappa$ further beyond this value adds nothing, at least in the weak coupling limit. Thus, a critical value of $\kappa$ exists, for which $\hat{a}=1$. Substituting the equivalent, $a=1-\kappa$ condition into (\ref{eq119}) and choosing the "minus" sign for the result to correspond to a maximum amplitude value, a simple equation for the critical value of $\kappa$ in the $\gamma \to 0, k \to 1$ limit is obtained:
\begin{equation}
\cos{(\pi\sqrt{\kappa_{cr}})}=1-4\kappa_{cr}
\label{eq121}
\end{equation}
where the left-hand side is higher than $-1$ and thus: $\kappa_{cr}<1/2$. Also, one immediately observes that $\kappa=0$ and $\kappa_{cr}=1/4$ are solutions. The $\kappa_{cr}=0$ case was already examined above, thus we would next limit ourselves to the domain $(0,1/2]$. Now, since the right-hand side in (\ref{eq121}) is a linear form, a sufficient condition for the absence of additional solutions of (\ref{eq121}), aside from $\kappa_{cr}=1/4$, in the domain $(0,1/2]$, would be convexity of the left-hand side of (\ref{eq121}) in this domain. Differentiation yields:
\begin{equation}
\begin{split}
\frac{d^2\cos{(\pi\sqrt{\kappa_{cr}})}}{d\kappa_{cr}^2}= \\ \frac{\pi\cos{(\pi\sqrt{\kappa_{cr}})} [\tan{(\pi\sqrt{\kappa_{cr}})}-\pi\sqrt{\kappa_{cr}}]}{4\kappa_{cr}^{3/2}}
\label{eq122}
\end{split}
\end{equation}

Generally, the following holds:
\begin{equation}
\frac{d^2\tan{y}}{dy^2}=2(1+\tan^2{y})\tan{y}
\label{eq123}
\end{equation}

Now, $\tan{y}$ is positive for $y \in (0,\pi/2)$ and negative for $y \in(\pi/2,\pi)$. Thus, from (\ref{eq123}), $\tan{y}$ is convex for $y \in (0,\pi/2)$ and concave for $y \in(\pi/2,\pi)$. Hence (since $\tan{(0)}=0$ and also $(\tan{y})'|_0=y'=1$), $\tan{(y)}-y>0$ for $y \in (0,\pi/2)$ and $\tan{(y)}-y<0$ for $y \in (\pi/2,\pi)$. As $\cos{y}>0$ for $y \in (0,\pi/2)$, clearly the right-hand side in (\ref{eq122}) is positive for $\kappa_{cr} \in (0,1/4)$, which means that $\cos{(\pi\sqrt{\kappa})}$ is convex for $\kappa_{cr} \in (0,1/4)$.

Furthermore, $\cos{(\pi\sqrt{0})}=1-4\cdot{0}$ and
\begin{equation}
\begin{split}
\left.\frac{d\cos{(\pi\sqrt{\kappa_{cr}})}}{d\kappa_{cr}}\right \rvert_{\kappa_{cr}=0}= \\ -\pi^2/2 \approx -4.9<-4 =\left.\frac{d(1-4\kappa_{cr})}{d\kappa_{cr}}\right \rvert_{\kappa_{cr}=0}
\end{split}
\label{eq124}
\end{equation}
and thus $1-4\kappa_{cr}$ is an upper bounding chord for the convex curve $\cos{(\pi\sqrt{\kappa_{cr}})}$ in the region $\kappa \in (0,1/4)$. A convex curve is always strictly lower than its upper bounding chord, and thus (\ref{eq121}) has no solutions for $\kappa \in (0,1/4)$.

Similarly, as both $\cos{y}$ and $\tan{(y)}-y$ are negative for $y \in (\pi/2,\pi)$, clearly, $\cos{(\pi\sqrt{\kappa_{cr}})}$ is still convex also for $\kappa_{cr} \in (1/4,1)$.

Moreover, since $\cos{(\pi\sqrt{1/4})}=1-4\cdot1/4$ and
\begin{equation}
\begin{split}
\left.\frac{d\cos{(\pi\sqrt{\kappa_{cr}})}}{d\kappa_{cr}}\right \rvert_{\kappa_{cr}=1/4}= -\pi >-4 = \\ \left.\frac{d(1-4\kappa_{cr})}{d\kappa_{cr}}\right \rvert_{\kappa_{cr}=1/4}
\end{split}
\label{eq125}
\end{equation}
clearly, the convex curve $\cos{(\pi\sqrt{\kappa_{cr}})}$ lies strictly above a straight line intersecting it at $\kappa_{cr}=1/4$ and having a lower initial slope in the range $\kappa_{cr} \in (1/4,1)$, and therefore (\ref{eq121}) has no solution in the range: $\kappa_{cr} \in (1/4,1)$. Thus the only solution of (\ref{eq121}) in the interesting range $\kappa_{cr} \in (0,1/2)$ is the solution we already know of, namely,
\begin{equation}
\kappa_{cr}|_{a_{max}^{LS,\gamma \ll 1}} \underset{k \to 1} =1/4
\label{eq126}
\end{equation}

This value of the foundation stiffness has interesting properties. Substituting (\ref{eq126}) into (\ref{eq119}), one obtains:
\begin{equation}
a_{cr,\gamma \to 0}^{LS,k \to 1}\underset{\kappa=\kappa_{cr}}=\frac{3/4}{2\mp1}\rightarrow \lbrace 3/4,1/4\rbrace
\label{eq127}
\end{equation}
which produces two values, the overall maximum possible linearly stable amplitude (in the considered limit), $3/4$, and another value, $1/4$. This value of $1/4$ is the third of the aforementioned noteworthy results of (\ref{eq119}). The physical meaning of this value is somewhat non-trivial.

For $\gamma \to 0$, the system is linearly stable from both sides of $a=1/4$ in the vicinity of this value. Nevertheless, this value is analytically found to be linear stability-wise critical in some sense. The only explanation one can think of for this matter is that the point $(0,1/4)$ in the $(\gamma,a)$ plane is a bifurcation point from which, generically, two branches in the $(\gamma,a)$ plane originate. The analytic investigation presented in this subsection can only tell us that there exists a bifurcation point at $(\gamma \to 0,a=1/4)$ for the parameter choice: $k \to 1,\kappa=1/4$, and that from this bifurcation point, at least two stability boundaries in the $(\gamma,a)$ plane should come out.

Employing a fully analytic approach in the $\gamma \to 0$ limit and obtaining qualitative information about the stability of a spatially extended breather with finite link stiffness indeed seems a noteworthy result. The proceeding subsection shows, numerically, that for the aforementioned choice of parameters, there indeed exist two branches in the $(\gamma,a)$ plane, originating from the point $(0,1/4)$, between which there exists an unstable region, as arises in the simplest feasible scenario according to the analytic investigation.

Based on the aforementioned understanding, one clearly sees that in the case $\gamma\ne 0$, the addition of a critically stiff foundation potential, although increasing the maximum linearly stable amplitude, may introduce an unstable band at intermediate amplitudes, as implied by the analysis of the $k \to 1,\gamma \to 0,\kappa =1/4$ case.

Thus if one wishes to have a smooth, simply connected amplitude range of maximum breadth, corresponding to a linearly stable breather with finite link stiffness, then the optimal foundation stiffness should satisfy the inequality: $\kappa_{opt}<\kappa_{cr}$. In order to obtain such optimal a value, one may take advantage of the numerically obtained stability map, presented in the subsequent subsection.

Before proceeding to the numerical results subsection, it is desirable to extend the analysis, at least approximately, to the more practical, $k<1$ case.

The general expression for the linear stability-wise critical amplitude in the vanishing link-strength limit is already given by (\ref{eq114}) for $k \in (0,1)$. What one may wish to do next is to obtain two specific versions of this expression, one for the case: $\kappa=0$, as a reference for estimating the effect of adding a harmonic foundation, and another one for the case: $\kappa=\kappa_{cr}$.

The referential critical amplitude for an arbitrary coefficient of restitution, as one can obtain from (\ref{eq114}) by setting: $\kappa \to 0$ in (\ref{eq114}) and taking the appropriate limit, much like in (\ref{eq120}), is as follows:
\begin{equation}
\begin{split}
a_{cr}^{LS}(k,\kappa=0)= \\ \frac{\sqrt{\frac{(1-k)^2}{(1+k)^2}+\left[\frac{4k}{(1+k)^2}\frac{1+\sqrt{[1+(k^2+k^{-2})/2]/2}}{\pi}\right]^2}}{\pi/2-\frac{4k}{(1+k)^2}\frac{1+\sqrt{[1+(k^2+k^{-2})/2]/2}}{\pi}}
\end{split}
\label{eq128}
\end{equation}
(where after taking the $k \to 1$ limit, (\ref{eq120}) is reproduced).

Next, in order to obtain an expression for the maximum possible linearly stable amplitude for an arbitrary value of $k$ in the $\gamma \to 0$ limit, one should substitute $1-\kappa$ \ for the existence amplitude in the $\gamma \to 0$ limit into the left-hand side of Eq. (\ref{eq114}). The resulting equation would be the generalization of (\ref{eq121}) for an arbitrary $k$, namely an equation for $\kappa_{cr}(k)$, which we present here in the following compact form:
\begin{equation}
(1-q^2\kappa_{cr})\cos{(\pi\sqrt{\kappa_{cr}})}=1-\eta\kappa_{cr}
\label{eq129}
\end{equation}
where $q$ is as defined in (\ref{eq30}) and:
\begin{equation}
\eta\triangleq \left(\frac{1-k}{1+k}\right)^2+2\left\lbrace 1 +\frac{4k\sqrt{\frac{1}{2}+\frac{k^2+k^{-2}}{4}}}{(1+k)^2}\right\rbrace
\label{eq130}
\end{equation}
(note that in the $k \to 1$ limit, $q \to 0$ and $\eta\to 4$, and Eq. (\ref{eq121}) is reproduced).

Eq. (\ref{eq129}) is transcendental and thus, unlike Eq. (\ref{eq121}), does not  seem to have a solution expressible in radicals. Owing to the fact that for practical applications one would be interested in a restitution coefficient at least as high as say, $0.8$, but with no loss of generality, one could seek a solution of (\ref{eq130}) in the form of a Taylor series expansion in powers of $1-k$. Since the solution of (\ref{eq129}) for $1-k \to 0$ is known and the derivatives of $\kappa_{cr}(k)$ at $1-k \to 0$ can be obtained from (\ref{eq129}) using the known solution of (\ref{eq121}) and employing the chain rule, one readily gets: $\kappa_{cr}'(k)|_{k \to1}=0$, which implies, by the way, that it is not necessary to get perfect restitution and that a relatively high value, such as, say, $k\simeq 0.9$ should be almost just as good as $k \to 1$, in the sense of maximum linearly stable gain increase.

Applying the chain rule to (\ref{eq129}) twice and using (\ref{eq126}) and the aforementioned result, we get:
\begin{equation}
\kappa_{cr}(k) = \frac{1}{4}-\frac{3}{16}\frac{(1-k)^2}{4-\pi}+O[(1-k)^3]
\label{eq131}
\end{equation}

One can use (\ref{eq129}) to obtain higher-order corrections to (\ref{eq131}). We shall settle with the second order expansion though, as it is sufficient for getting the necessary insights and is accurate enough in the range $k \in (0.8,1)$, which, as one may argue, is the appropriate range for practical applications.

Anyhow, having the critical value of the foundation, as given above, one can readily obtain the maximum amplitude corresponding to a linearly stable breather for infinitesimally small values of the breather link stiffness and relatively high restitution coefficient values, using the relations: $a_{cr,\gamma \to 0}^{LS}(k)=a^{ex}_{\gamma \to 0}(k)=1-\kappa^{\gamma \to 0}_{cr}(k)$, as follows:
\begin{equation}
a_{cr,\gamma \to 0}^{LS}(k) \underset{1-k \ll 1}\to \frac{3}{4}+\frac{3}{16}\frac{(1-k)^2}{4-\pi}
\label{eq132}
\end{equation}

Consequently, if one uses the critical foundation stiffness, as given by (\ref{eq132}), then they should get the relative gain in the maximum amplitude corresponding to a linearly stable weakly-linked breather, in the form of a ratio of the expressions given in (\ref{eq128}) and (\ref{eq132}). As the expression in (\ref{eq132}) is merely a quadratic expansion with respect to $1-k$, there would hardly be use in expressing the aforementioned amplitude gain as anything but a quadratic expansion as well. Hence, we get:
\begin{equation}
\begin{split}
\frac{a_{cr,\gamma \to 0}^{LS}(k,\kappa=\kappa_{cr})}{a_{cr,\gamma\to 0}^{LS}(k,\kappa=0)}\underset{1-k \ll 1}\to \frac{3}{16}(\pi^2-4) \\ -\frac{3}{512}(\pi^2-4)\left(\pi^2-\frac{8}{4-\pi}\right)(1-k)^2
\end{split}
\label{eq133}
\end{equation}
which has a $k \to 1$ limit of $10\%$ increase in the maximum linear stability-preserving amplitude, corresponding to the introduction of critically stiff foundation (with respect to a system having no foundation at all). Moreover, the absence of a linear term in the expansion in (\ref{eq133}) suggests weak dependence of the amplitude gain on the restitution coefficient as $k \to 1$, which means that for all practical values of $k$, say, for $k>0.8$, one should expect about $10\%$ amplitude gain when introducing linear elastic foundation with the right stiffness.

In the following subsection, approximations (\ref{eq131}-\ref{eq133}), as well as the exact result given in (\ref{eq128}), are compared to fully numerical calculation in the $\gamma>0$ case, when numerically small values of $\gamma$ are chosen (the monodromy matrix is still, however, constructed analytically).

Having established the optimal foundation stiffness as a function of the coefficient of restitution for weakly-linked breathers, we wish to conclude this section by determining the optimal value of the coefficient of restitution, $k$, for which the overall linearly stable and breather existence-wise feasible amplitude range is maximal.

Substituting (\ref{eq131}) into (\ref{eq88}) and taking the $\gamma \to 0$ limit, we obtain the total gain bandwidth for the optimal foundation stiffness, $\Delta{a}_{cr}$. In line with the previous discussion, we should be unable to obtain an expansion of order higher than $2$ with respect to $1-k$ without doing an additional extensive body of work. For this specific case, however, it appears that linear expansion with respect to $1-k$ is already good enough for our purposes. To linear order, then, one has:
\begin{equation}
\begin{split}
\Delta{a}_{cr,\gamma \to 0}\underset{1-k \ll 1}\to \\ \frac{3}{4}-\left(\frac{3}{4\pi}-\frac{\pi}{64}-\frac{\pi^3}{1152}\right)(1-k)
\label{eq134}
\end{split}
\end{equation}

Linear expansion is sufficient in this case first because unlike in the calculation of other expansions in the present subsection, the linear order does not vanish here, owing to its appearance in the expression for $\underline{a}$, and second, since linear expansion yields: $d\Delta{a_{cr,\gamma \to 0}}/dk|_{k \to 1}=\frac{3}{4\pi}-\frac{\pi}{64}-\frac{\pi^3}{1152} \approx 0.16>0$.

This implies that as $k \to 1$, the gain bandwidth increases with $k$, which means that its local maximum point is $k_0 \to 1$, and second-order corrections cannot change that.

Therefore, in terms of the maximum value of the amplitude corresponding to a linearly stable localized breather and also the minimum amplitude for breather existence, and in terms of the maximization of the difference between the two, the $k \to 1,\kappa \to 1/4$ choice is optimal in the $\gamma \to 0$ case. However, in order for a breather to exist, one obviously has to have $\gamma>0$. Now, for $\gamma>0$, the results given in (\ref{eq127}) imply the emergence of an unstable region corresponding to $\kappa=1/4$. The existence of such a region clearly conflicts with the existence of a continuous linearly stable $a$-$\gamma$ band. As shown in the following subsection, for $k<1$, and the corresponding $\kappa_{cr}(k)$, the unstable zone does not start exactly at $\gamma=0$ but rather there is a linearly stable band in the $(\gamma,a)$ plane, starting at $\gamma =0$ and having finite width.

In line with this observation, one should expect to obtain the largest continuous linearly stable amplitude range by choosing: $0<\gamma \ll 1$, $k<1$, but still $0<1-k \ll1$ and $\kappa <1/4$, but still $0<1/4-\kappa \ll 1$.

Specific numerical values that correspond to the largest stable band, as well as other insights regarding the existence--stability picture, are given in the numerical results section below.

\subsection{Numerical Results}
\label{sect5C}

\begin{figure}[H]
\includegraphics[scale=0.45]{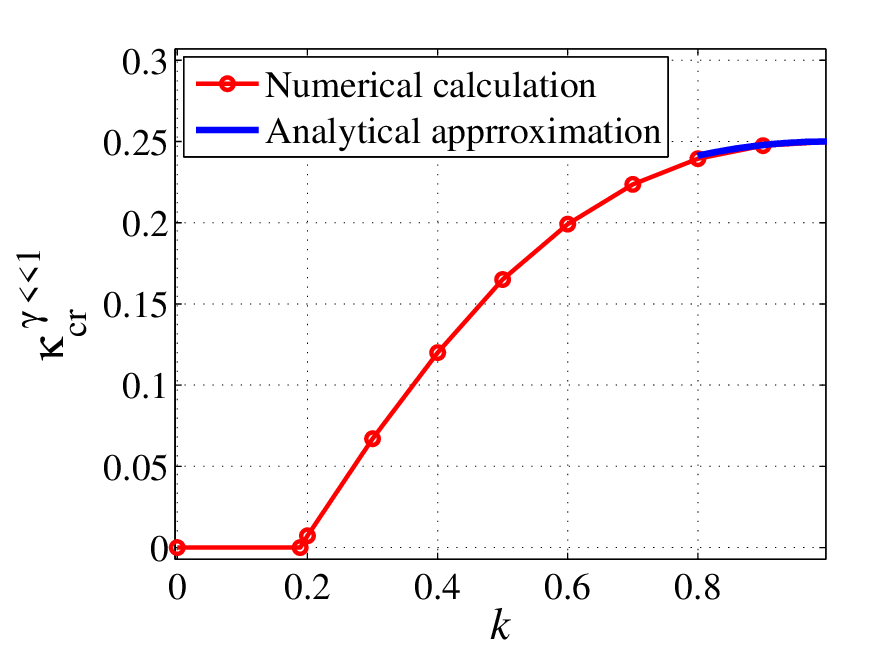}
\caption{\small Optimal foundation stiffness in the zero link stiffness limit calculated numerically (in red) and approximated analytically (in blue).}
\label{Figure9}
\end{figure}

\begin{figure}[H]
\includegraphics[scale=0.32]{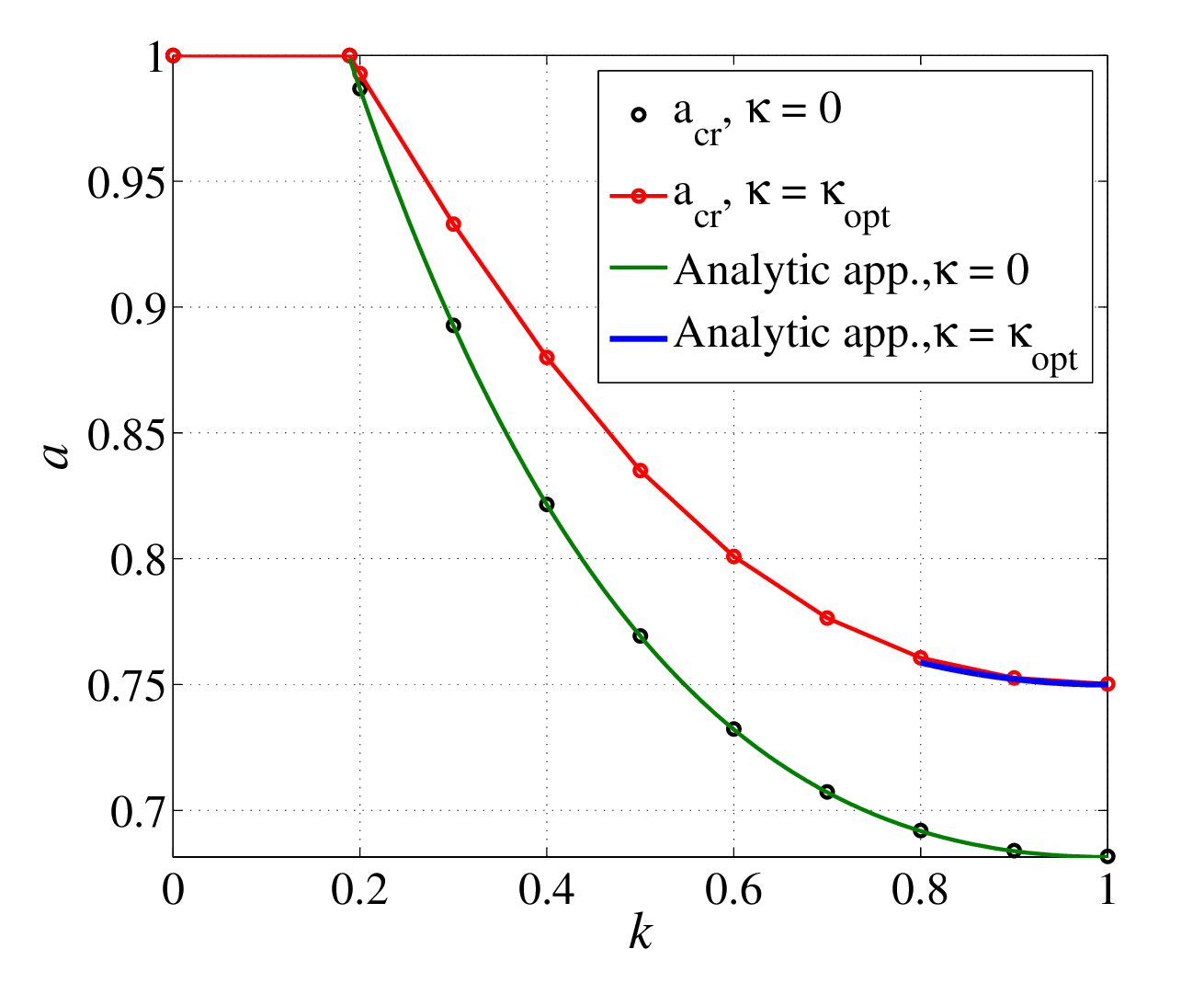}
\caption{\small Maximum forcing amplitude providing linear stability in the limit of zero link stiffness for zero (in black) and optimal (in red) foundation stiffness vs. the coefficient of restitution, with corresponding analytic estimations (in green and blue).}
\label{Figure10}
\end{figure}

\begin{figure}[H]
\includegraphics[scale=0.45]{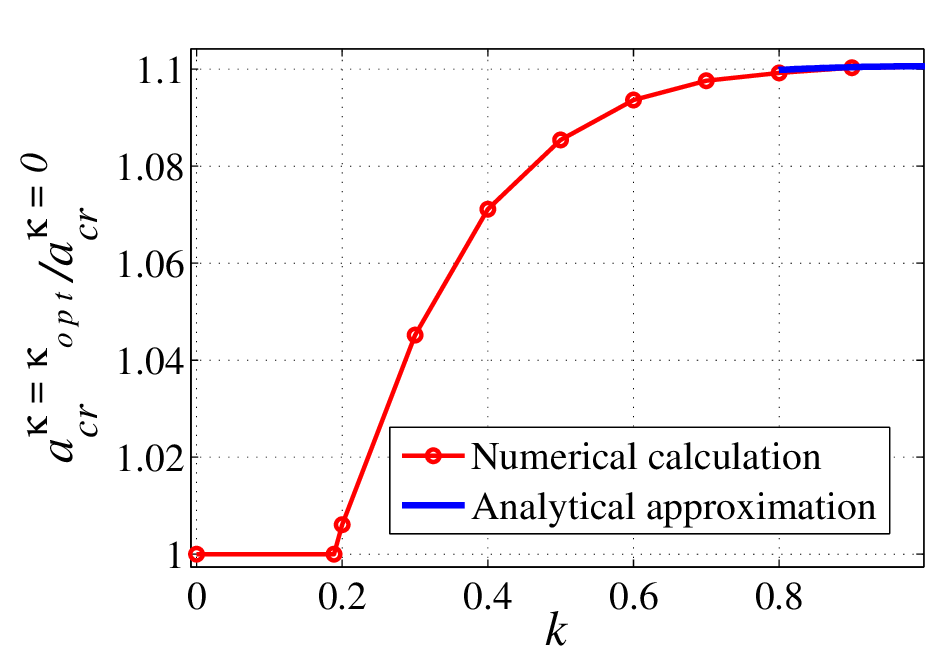}
\caption{\small Maximum amplitude increase factor in the $\gamma \to 0$ limit vs. the coefficient of restitution, corresponding to the introduction of optimal foundation (with the analytic estimate for practical $k$ values).}
\label{Figure11}
\end{figure}

\begin{figure}
\includegraphics[scale=0.45]{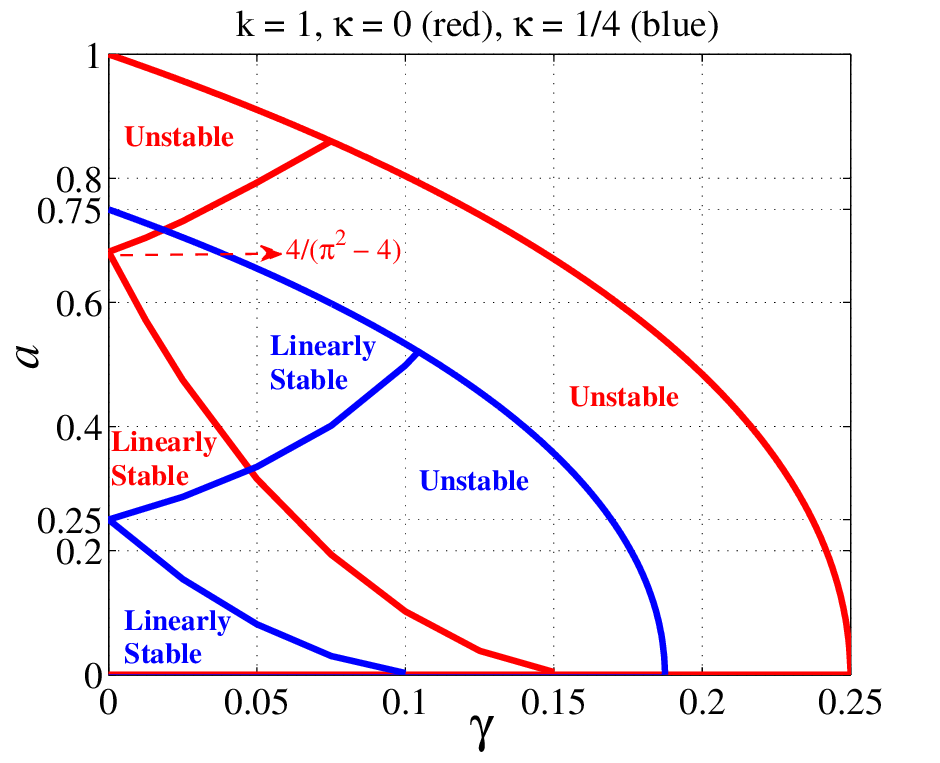}
\caption{\small Existence--stability maps for perfect restitution for zero and optimal foundation stiffness, showing bifurcation points and branches and stable and unstable regions (with analytic results shown on the plot).}
\label{Figure12}
\end{figure}

\begin{figure}[H]
{\includegraphics[width=3.1in,height=2.5in]{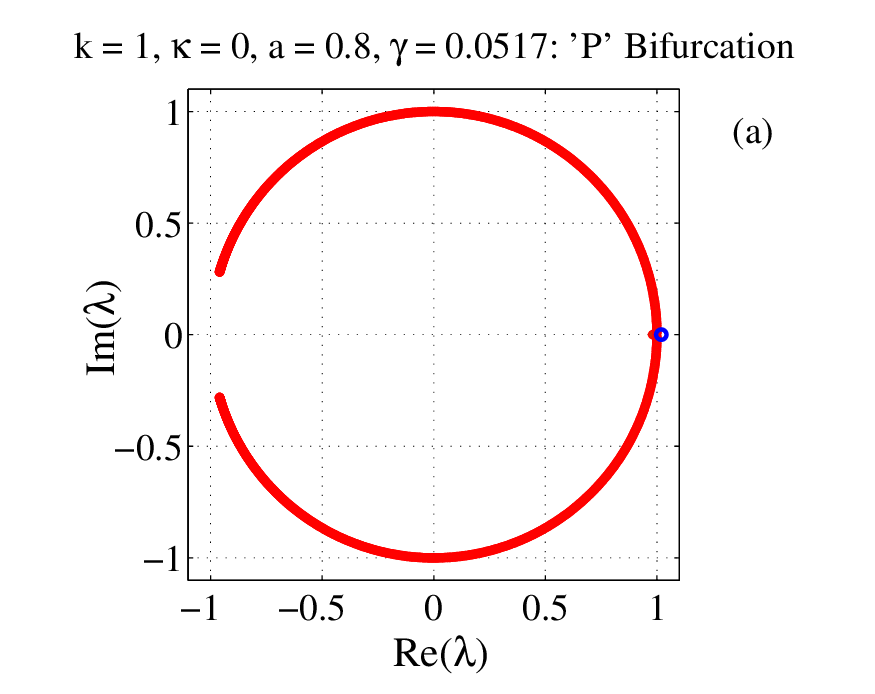}}  \\
{\includegraphics[width=3.1in,height=2.5in]{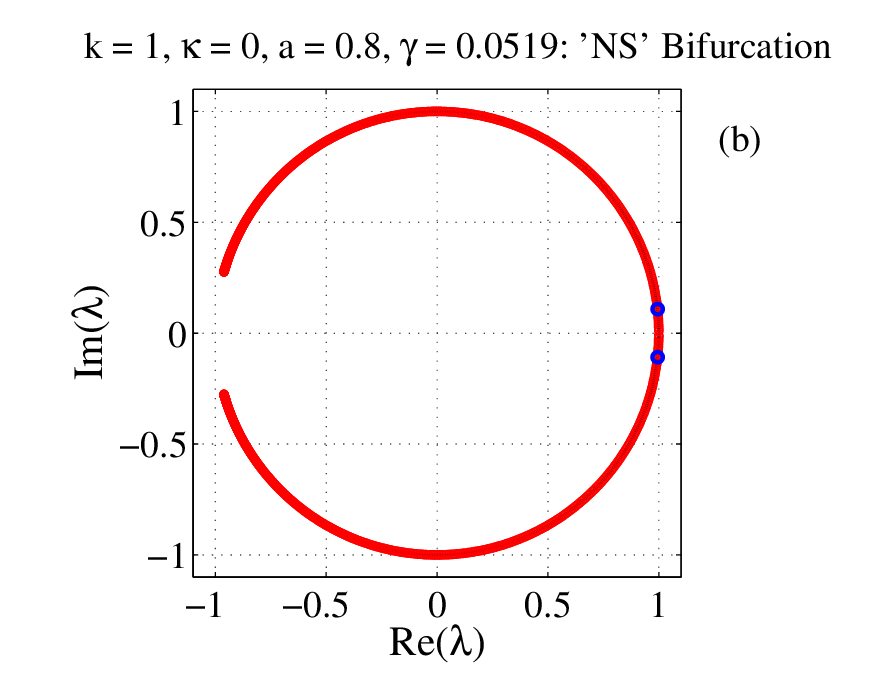}}
\caption{\small Floquet multipliers in the complex plane for a system with $N = 501$ beads (the $N$-convergent value), with critically unstable eigenvalues (in blue) corresponding to pitchfork bifurcation (a) and Neimark-Sacker bifurcation (b), for the $k=1,\kappa=0,a=0.8$ case, as obtained for $\gamma=0.0517$ and $\gamma=0.0519$, respectively.}
\label{Figure13}
\end{figure}

\begin{figure}
{\includegraphics[scale=0.48]{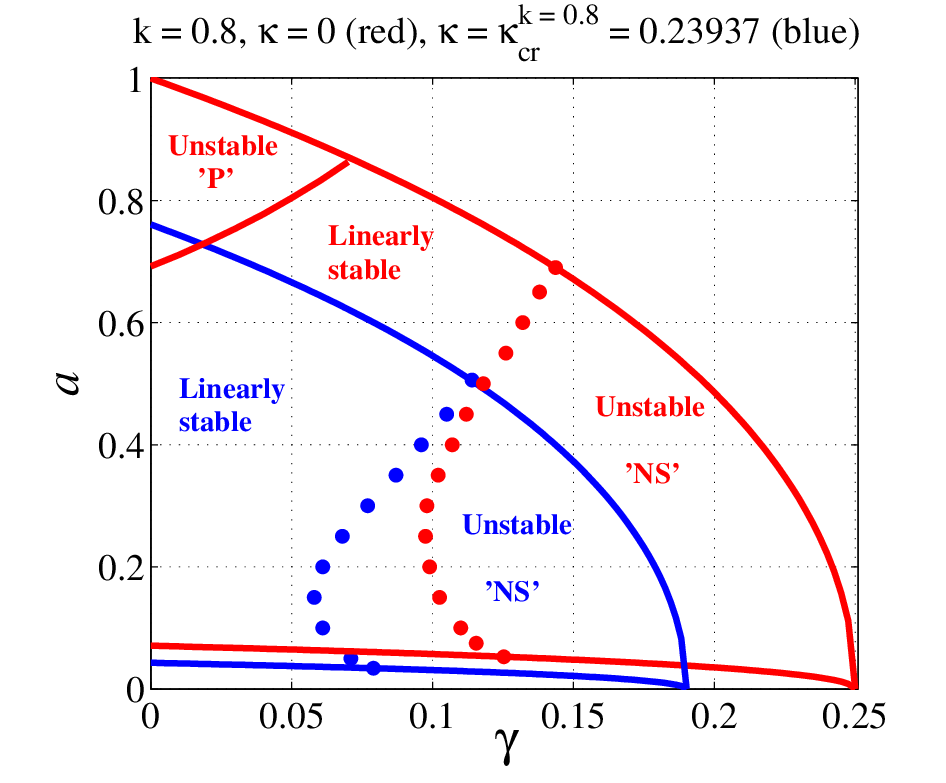}}
\caption{\small Existence-Stability ($a$-$\gamma$) maps for $k=0.8$ for $\kappa=0$ (red) and $\kappa=\kappa_{cr}$ (blue).}
\label{Figure14}
\end{figure}

\begin{figure}[H]
{\includegraphics[width=3.0in,height=2.0in]{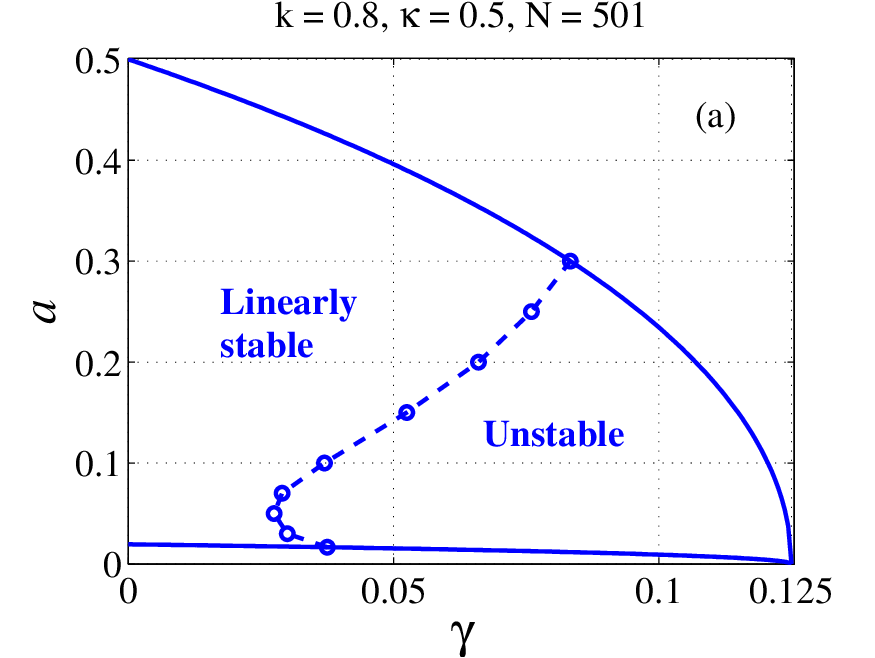}} \\ \\
{\includegraphics[width=3.0in,height=2.0in]{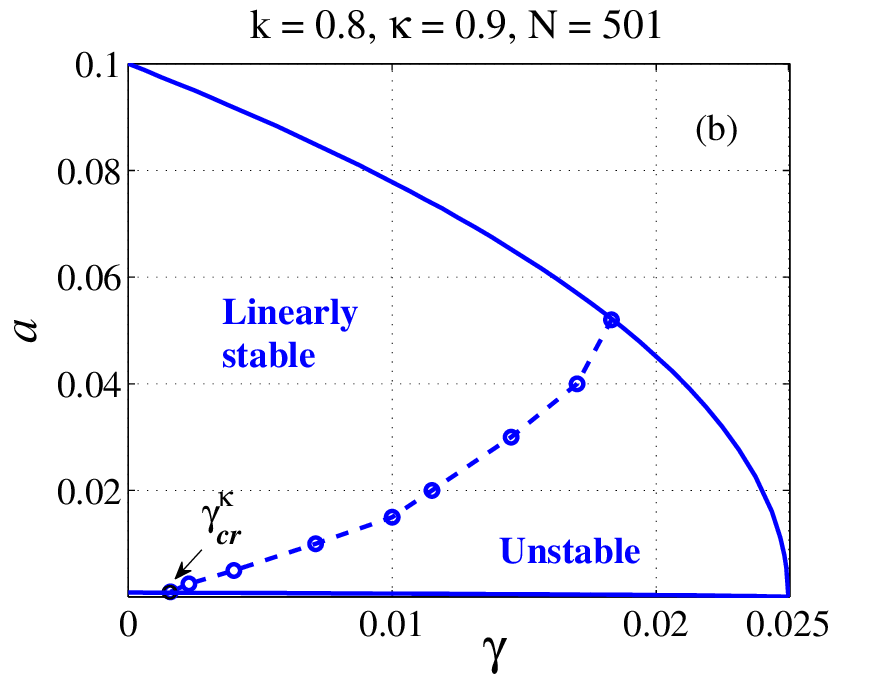}}
\caption{\small Existence--stability maps for $k=0.8$ for two values of stiffness for super-critically stiff foundations, for $\kappa=0.5$ (a) and $\kappa=0.9$ (b), with $N=501$ (as in Figure \ref{Figure14}) beads ($N$-convergent results). Note, also, that  $\gamma_{cr}^{\kappa\to1}\to0$ holds.}
\label{Figure16}
\end{figure}

In this section we present some noteworthy results obtained by numerical solution of the eigenvalue problem discussed in Section \ref{sect5A} for representative values of the parameters in the feasible range. Figures \ref{Figure9}-\ref{Figure11} present several curves related to the $\gamma \to 0$ case, examined in Section \ref{sect5B}, demonstrating the maximum amplitude increase resulting from the introduction of foundation with the right stiffness, as well as the quality of the analytic estimate of this effect,  Figures \ref{Figure12}-\ref{Figure13} exhibit the stability picture in case of perfect restitution, and Figures \ref{Figure14}-\ref{Figure16} show existence--stability maps for more realistic cases, corresponding to various typical values of the foundation stiffness and the coefficient of restitution.

Figure \ref{Figure12} presents two interesting maps. The $\kappa \to 0$ map shows that the linearly stable existence region is pathwise-connected, as is the unstable region, which is also convex. In the unstable region, the two unstable sub-regions, corresponding to pitchfork ('P', top left) and Neimark-Sacker ('NS', bottom right) bifurcations are adjacent and separated by a smooth exchange-of-instability curve. Figure \ref{Figure13} illustrates how a slight increase in the link stiffness in the $\kappa=0$ case changes the type of the instability from individual-bead-instability ('P') to cooperative-behavior- instability ('NS').

The blue map in Figure \ref{Figure12} is not less interesting. On one hand, much as the red map, it shows that the existence domain is divided into three regions, three adjacent sectors originating in a single bifurcation point and dividing the half-plane into three more-or-less equal sectors. However, whereas in the $\kappa=0$ case, there are two adjacent \emph{unstable} regions, in the $\kappa=1/4$ case, there are two \emph{stable} sectors, separated by an \emph{unstable} sector. For parameter-pairs lying in this unstable sector, the system features bifurcation of the Neimark-Sacker type. As one could expect, the addition of elastic foundation  stabilizes the breather for individual bead behavior (eliminating the pitchfork bifurcation-related instability). However, unlike in the $\kappa = 0$ case, cooperative loss of stability ('NS') occurs already in the limit of zero link stiffness (vanishing coupling).

Furthermore, the blue map in Figure \ref{Figure12} shows that the analytically evaluated value of the bifurcation point at $\gamma \to 0$, namely, $a=1/4$, is reproduced numerically, shedding new light on the observation made in \cite{Gendelman2013} regarding non-monotonicity of the stability-limit curve. Indeed, one may view the non-monotonic 'NS' stability-limit curve as a $k$-relaxed version of the phenomenon shown in the blue map in Figure \ref{Figure12}, which is a bifurcation point with two branches dividing, stability-wise, the breather-existence domain into three quasi-convex sectors. Another noteworthy feature seen in the blue map in Figure \ref{Figure12} is the fact that for a value of the link stiffness corresponding to the limit of linear stability for vanishing excitation amplitudes, something like $\gamma \approx 0.1$, nearly the whole amplitude range is unstable, yet there is a small amplitude sub-region, in a small neighborhood above $a=1/2$, for which linear stability holds. This is a somewhat peculiar result: for $\gamma \approx 0.1$, all amplitudes are unstable, yet there is a linearly stable solution for, roughly, a single, finite value, some $a \approx 0.51$. Phenomena of this sort can only occur in nonlinear dynamics. Of course, there remains the question whether the upper-amplitude linearly stable sector shown in the blue map in Figure \ref{Figure12} is stable with respect to finite perturbations as well.

Last, following the more practical perspective, it should be noted that by introducing foundation with critically optimal stiffness in the perfect restitution limit, one does increase the maximum linearly stable amplitude range for weakly-linked breathers by about $10\%$, but for any $\gamma>0$ an unstable region appears for intermediate amplitudes, a phenomenon which is absent in the $\kappa=0$ case. Thus, indeed, some stabilization takes place, but some destabilization can also be said to occur.

Figure \ref{Figure14} is perhaps the most important practical result presented in this paper. Comparison of the red and blue maps shows that for realistic values of the coefficient of restitution (here 0.8), there exists a linearly stable $a$-$\gamma$ stripe, for which (for $k=0.8$) in the range: $\gamma \in (0,0.0183)$ the addition of critically optimally stiff foundation can increase the linearly stable amplitude range by about $10\%$ (in total, averaging over the aforementioned $\gamma$ range), both by decreasing the lower bound and by increasing the upper bound on the existence amplitude. In this sense, if one considers $k=0.8$ (or other values in that neighborhood) to be a practically probable value of the coefficient of restitution, and if only (linearly) stable solutions are considered practical, then, if one is mainly interested in weakly-linked breathers, indeed the addition of linear elastic foundation can increase the amplitude range for which a linearly stable breather would exist, by a non-negligible, and in the same time not a trivially deducible value of about $10\%$.

If one's perspective is not the stabilization of a practical weakly-linked breather but rather the analysis of the effect of the addition of a harmonic spatially uniform potential to a DB, then it can be learned from Figure \ref{Figure14} that this effect is complex, and consists of a decrease in the existence amplitude range and stabilization with respect to pitchfork bifurcations, but also, generally, an increase in the range of instability related to Neimark-Sacker bifurcations. Also, one may argue that the effect of the addition of a uniform harmonic potential to the system is more complex and ambiguous for higher values of the coefficient of restitution. In the examined setting, it appears to bring about global modification of the stability patterns in the parameter space, not adding, however, new bifurcation types.  Regarding the non-monotonicity of the curve bounding the 'NS' bifurcation instability zone, it is illustrated here that the addition of an increasingly stiff quadratic potential can only increase it, as one can learn by comparing the blue maps in Figures \ref{Figure12} and \ref{Figure14} and in Figure \ref{Figure16}.

In conclusion of this section, Figure \ref{Figure16} illustrates the effect of increasing $\kappa$ beyond its critical value (the one for which the parameter-zone corresponding to pitchfork bifurcation first disappears). In addition, it was found that there is a critical physically practical value of $k$, which for the $\kappa=0$ case is $k_{cr}^{\kappa=0}=0.883$, at and above which the linearly stable band within the breather-existence stripe vanishes, and where the 'P' and the 'NS' instability zones become adjacent at $\gamma_{cr}^{\kappa=0}=0.0723$, with $a_{cr,\kappa=0}^{'P'}=0.8655$ and $a_{cr,\kappa=0}^{'NS'}=0.3655$.

\section{Numerical simulations of the DB\lowercase{s} and comparison to analytic solutions}
\label{sect6}

Numerical simulation of the breathers described herein can serve two purposes. The first is validation of the analytic solution, which is relevant for parameters chosen from the stable region. The second is illustration of loss of stability, by performing integration with parameters taken from both stable and unstable regions.

Since the problem was defined with temporal periodicity and thus  assumed as extending temporally from minus infinity to plus infinity, whereas numerical time integration can only be performed for an initial-value problem, clearly, one needs to assume initial conditions for the positions and velocities of all the beads in the breather. In order to be able to compare the numerical solution to the analytic one, as well as in order to illustrate loss of stability, the initial conditions for numerical integration have to be taken from the analytic solution at, say, $t=0$. By integrating the equations of motion from the initial conditions at $t=0$, one can learn whether the chosen parameters lie within the linearly stable region. If round-off errors during numerical integration lead to loss of stability, the corresponding parameter sets can be considered unstable. If round-off errors during integration from initial conditions corresponding to the analytic solution do not lead to loss of stability when using a stable integration scheme, then the corresponding parameter sets could be considered linearly stable.

\subsection{Initial conditions for numerical integration}
\label{sect6A}

Initial conditions for numerical  integration aiming to reproduce the DB solution, can be obtained from Eq. (\ref{eq52}), as follows:
\begin{equation}
\begin{split}
u_n(0)= -\frac{a}{1-\kappa}+\frac{4p}{\pi} \left(-\frac{1}{2\gamma}\right)^{\lvert n \rvert} \times  \\
\sum _{l=0}^{\infty} \frac{ \cos[(2l+1)\phi]}{\sqrt{[(2l+1)^2-(2\gamma+\kappa)]^2-(2\gamma)^2}} \\ \times  \left \lbrace \vphantom{\sqrt{[(1)^2]^2}} (2l+1)^2-(2\gamma+\kappa) \right. \\ \left.-\sqrt{[(2l+1)^2-(2\gamma+\kappa)]^2-(2\gamma)^2} \right \rbrace^{\lvert n \rvert}
\end{split}
\label{eq135}
\end{equation}
\begin{equation}
\begin{split}
\mathpzc{v}_n(0)= \frac{4p}{\pi} \left(-\frac{1}{2\gamma}\right)^{\lvert n \rvert} \times  \\
\sum _{l=0}^{\infty} \frac{ (2l+1)\sin[(2l+1)\phi]}{\sqrt{[(2l+1)^2-(2\gamma+\kappa)]^2-(2\gamma)^2}} \\ \times \left \lbrace \vphantom{\sqrt{[(1)^2]^2}} (2l+1)^2-(2\gamma+\kappa) \right.\\ \left.-\sqrt{[(2l+1)^2-(2\gamma+\kappa)]^2-(2\gamma)^2}\right\rbrace^{\lvert n \rvert}
\end{split}
\label{eq136}
\end{equation}
where use is made of definitions (\ref{eq28}-\ref{eq30}) and (\ref{eq98}), and where $u$ is the displacement and $\mathpzc{v}$ is the velocity.

\subsection{Integration schemes}
\label{sect6B}

In order to perform numerical integration, one has to assume a finite number of beads and introduce some boundary conditions. It was already established in Section \ref{sect4} that the examined system does not allow the existence of phonon-emitting breathers. It was due to this fact that it was possible to calculate the eigenvalues of the monodromy matrix numerically by introducing a cut-off representing periodic boundary conditions without it being associated with a significant energy error. We opt to use the same periodic boundary conditions for numerical integration. Along with a linear impact law for the velocities, coupled to limiting conditions imposed on the displacements, the dynamical system to be integrated becomes:
\begin{equation}
\begin{split}\dot{t}=1 \\ \dot{u}_n= \mathpzc{v}_n\\
\dot{\mathpzc{v}}_n=a\cos{t}-(2\gamma+\kappa)u_n+\gamma G_n(\textbf{u})
\end{split}
\label{eq137}
\end{equation}
where $G_n(\textbf{u})$ is given by:
\begin{equation}
G(\textbf{u})=\left \lbrace \begin{split} u_{n+1}+u_{-n} \ , n=-\frac{N-1}{2}\\ u_{n+1}+u_{n-1} \ , \ |n|<(N-1)/2 \\u_{n-1}+u_{-n} \ , n=\frac{N-1}{2}   \end{split} \right.
\label{eq138}
\end{equation}
$N$ being the total (preferably $-$ and chosen here $-$ odd, for easy enforcement of symmetry) number of degrees of freedom in the system.

In addition to the initial conditions and the ($1^{\text{st}}$-order) time derivatives, one needs an updating scheme, which in our case should include the impact conditions. The simplest way to update the state vector explicitly, is by implementing the backward Euler method:
\begin{equation}
\begin{split}
t^{(j+1)}=t^{(j)}+\Delta{t} \\ \tilde{u}_n^{(j+1)}=u_n^{(j)}+\dot{u}_n^{(j)}\Delta{t} \\  \tilde{\mathpzc{v}}_n^{(j+1)}=\mathpzc{v}_n^{(j+1)}+\dot{\mathpzc{v}}_n^{(j)}\Delta{t}
\end{split}
\label{eq139}
\end{equation}
augmenting it with explicit impact conditions, which may limit the algorithmic stability of the method (in that stability can be inferred from a stable result obtained with the method but instability cannot be inferred from an unstable result $-$ since it may arise from algorithmic rather than physical instability) but does not seem to interfere with the emergence of stable results for parameters taken from a linearly stable region, for $\Delta{t}=10^{-5}$:
\begin{equation}
\begin{split}
u_n^{(j+1)}= \left  \lbrace  \begin{split} \tilde{u}_n^{(j+1)} \ , | \tilde{u}_n^{(j+1)}|<1 \\ u_n^{(j)}, | \tilde{u}_n^{(j+1)}|\ge 1 \end{split} \right.  \\
\mathpzc{v}_n^{(j+1)}= \left  \lbrace  \begin{split} \tilde{\mathpzc{v}}_n^{(j+1)} \ , | \tilde{u}_n^{(j+1)}|<1 \\ -k\mathpzc{v}_n^{(j)}, | \tilde{u}_n^{(j+1)}|\ge 1 \end{split} \right.
\end{split}
\label{eq140}
\end{equation}
where $j$ denotes the current iteration number in the time-stepping loop, and $\Delta{t}$ is the time-step. This explicit 'impact-potential' scheme can be destabilized by the impact conditions, and thus it is adequate when producing a stable solution, but is insufficient when producing a solution that seems unstable, since the instability can be either of physical or of algorithmic origin. Thus, when integrating the equations of motion for parameters considered by Floquet theory to correspond to instability, or whenever the scheme described above produces a solution that deviates from the analytic one for a linearly stable set of parameters (which does not happen in our case), the integration should be repeated with perhaps physically approximate but algorithmically stable a scheme. As mentioned in the introduction, the scheme proposed in \cite{Gendelman2006}, does not represent impact conditions perfectly, but permits controlling the approximation quality, represented by the temporal localization of the impact instances, with exact reproduction of the velocity change, and in a form that allows the use of a formally stable integration scheme.

Employing the model suggested in \cite{Gendelman2006} one produces an alternative version of the last row in (\ref{eq137}), as follows:
\begin{equation}
\begin{split}
\dot{\mathpzc{v}}_n=a\cos{t}-(2\gamma+\kappa)u_n+\gamma G_n(\textbf{u}) \\ -(2\xi+1)u_n^{2\xi}\left[u_n^{2\xi+1}+2\frac{\ln{k}}{\sqrt{\pi^2+(\ln{k})^2}} \mathpzc{v}_n\right] \end{split}
\label{eq141}
\end{equation}
where $\xi \gg 1$ is the localization parameter, which can be shown to reproduce exact impact conditions in the $\xi \to \infty$ limit .

Taking a large enough yet finite $\xi$ (such that further small increase does not noticeably change the solution) and using (\ref{eq137}-\ref{eq139}), with the last row in (\ref{eq137}) replaced by (\ref{eq141}), and the tildes in the left-hand side of (\ref{eq139}) omitted, one gets a stable integration scheme for which, if the time-step is taken small enough and one starts at the initial conditions given in (\ref{eq135}-\ref{eq136}), one should, in principle, reproduce the analytic solution for parameters chosen from the linearly stable region.

Regarding the time-step, one notes the following: in constant time-step schemes, the time-step should be small enough to resolve the highest-frequency waves contributing non-negligibly to the solution. As the analytic solution is given in terms of a Fourier series, the highest non-negligible frequency can be estimated from the convergence of the analytic solution with respect to the frequency. By numerical summation, it is established that in our case, the analytic solution is convergent everywhere in the parameter space if one takes the highest normalized frequency to be $10^5$, which gives a typical temporal period of $2\pi \cdot 10^{-5}$. The basic wave is a sine, which can be adequately discretized by, say, 20 to 100 points in a period. We take $\Delta{t}=10^{-6}$, which means about 63 points to describe one period in the sine function. This value indeed produces a stable scheme in practice.

\subsection{Simulation results}
\label{sect6C}

In this subsection we present the results of numerical integration with the schemes and the initial and boundary conditions as described in the preceding subsections, which illustrate the stabilizing effect of the  introduction of a spatially homogeneous harmonic potential with the (rounded) optimal stiffness constant for a practical restitution coefficient (far enough from unity to correspond to a finite-size attractor in the phase space) and a relatively small link strength, $\gamma$, for a critical value of the excitation amplitude (for which only a system with the critical or nearly-critical foundation stiffness is linearly stable, whereas the $\kappa=0$ case is unstable).

In short, we integrate the equations for: $k=0.8,a=0.73,\gamma=0.01$ and $N=301$ (which is the number associated with the least computational effort, large enough to produce convergence with respect to $N$, for the parameters we chose), once for $\kappa=0$ and once for $\kappa=0.24$. In each case we present the results of the first 5 integration cycles, to illustrate that the integration scheme works fine; and then 5 cycles of integration, starting at cycle 236, to illustrate the deviation of the numerically integrated solution from the analytic one, which occurs in the physically unstable case (and the lack of thereof in the linearly stable case).

The results for the (relatively) late integration times are presented for the 'impact' scheme, although they were obtained also for the 'smooth' scheme, corresponding to the augmented equation of motion in (\ref{eq141}). Comparison of the \emph{stable} results obtained with the different schemes allows one to validate both the algorithmic stability, which is not assured for the 'impact' scheme, and the level of accuracy of the description of impacts, which is not known to be high enough \emph{a priori} for the 'smooth' scheme.

Finally, to give good yet compact representation of the breather, we present four figures for each set of parameters, corresponding to temporal histories of 3 beads, namely, the zeroth, the first and the one positioned farthest from the zeroth bead (the $150^{\text{th}}$ one), and a profile plot of the breather at the end of the $241^{\text{th}}$ cycle. These results, presented in Figures \ref{Figure17}-\ref{Figure20} below, clearly illustrate the $\kappa$-stabilizing effect. The unstable result was reproduced with the 'smooth' scheme with $\xi=300$, for which both schemes gave the same result in the stable case.

\begin{figure}
\begin{center}
{\includegraphics[width=1.8in,height=1.8in]{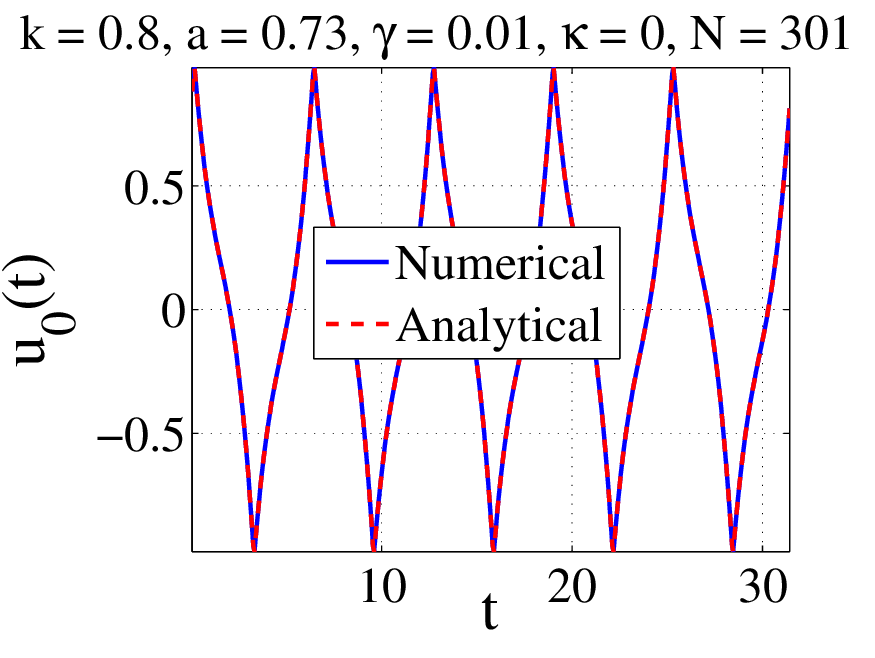}} \\
{\includegraphics[width=1.8in,height=1.8in]{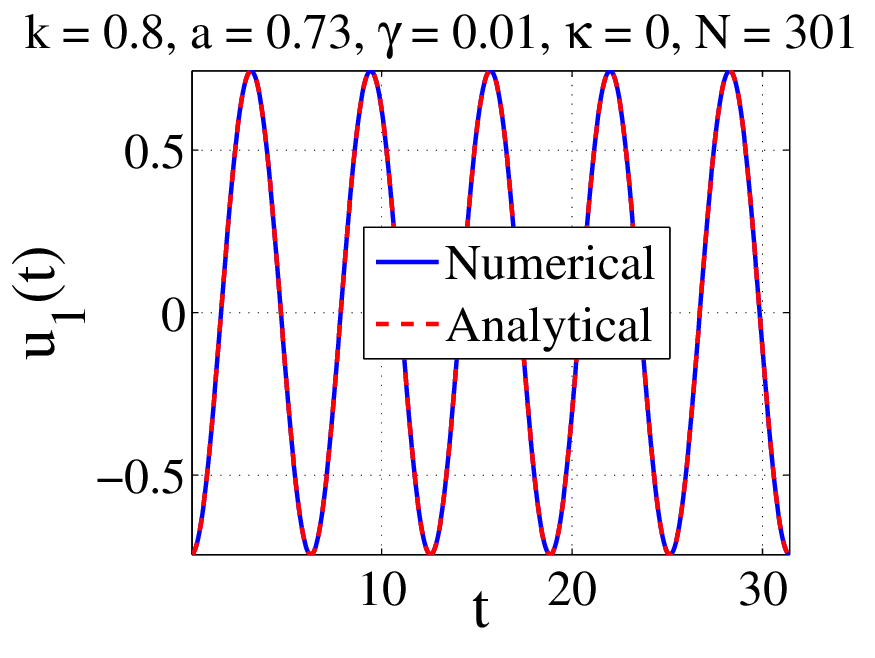}} \\
{\includegraphics[width=1.8in,height=1.8in]{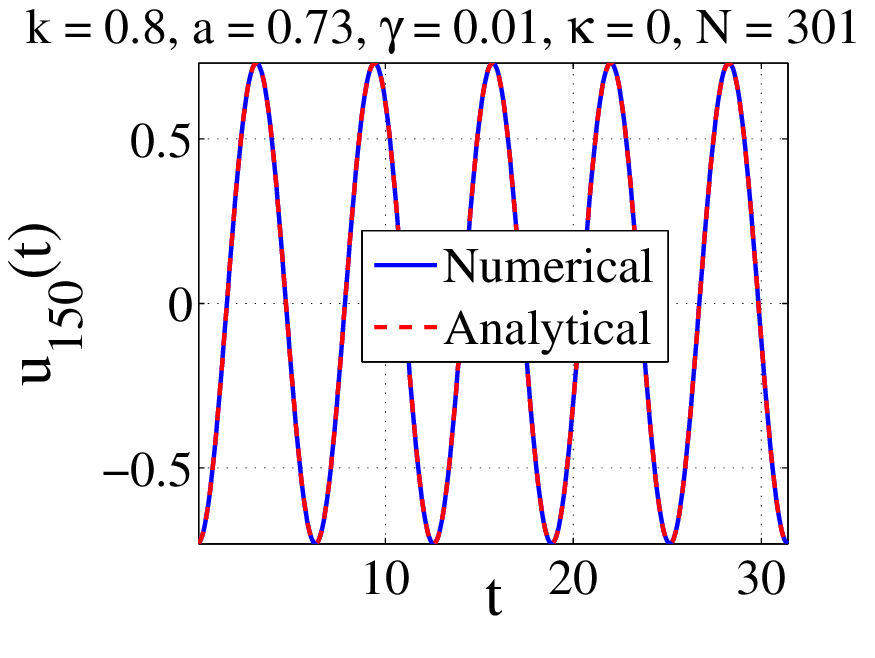}} \\
{\includegraphics[width=1.8in,height=1.8in]{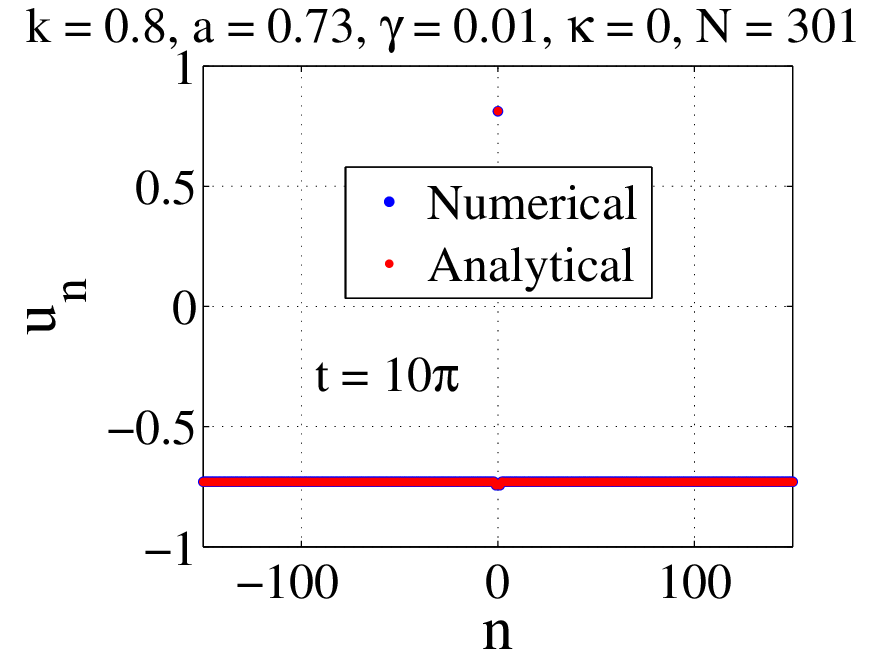}}
\end{center}
\caption{\small Comparison of numerically (blue) and analytically (red) obtained values of $u_0(t)$, $u_1(t)$, $u_{150}(t)$ and $u_n(n)$, at the end of the $5^{\text{th}}$ cycle for $N=301$ and: $k=0.8$, $\gamma=0.01$ and $a=0.73$ for $\kappa=0$, in validation of the analytic solution and the periodic boundary conditions and the scheme itself (for the numerical solution). }
\label{Figure17}
\end{figure}

\begin{figure}
\begin{center}
{\includegraphics[width=1.81in,height=1.81in]{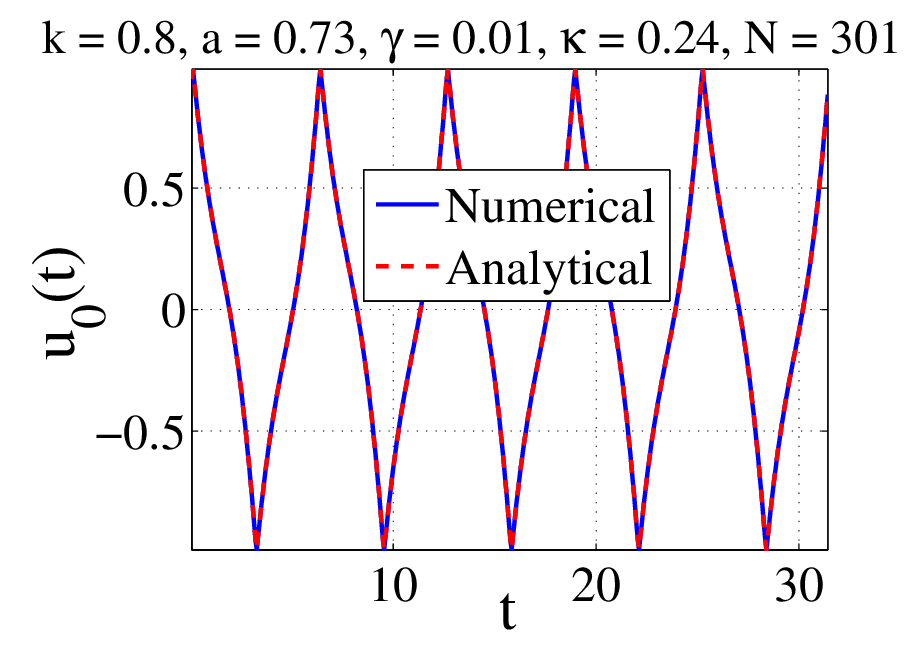}} \\
{\includegraphics[width=1.81in,height=1.81in]{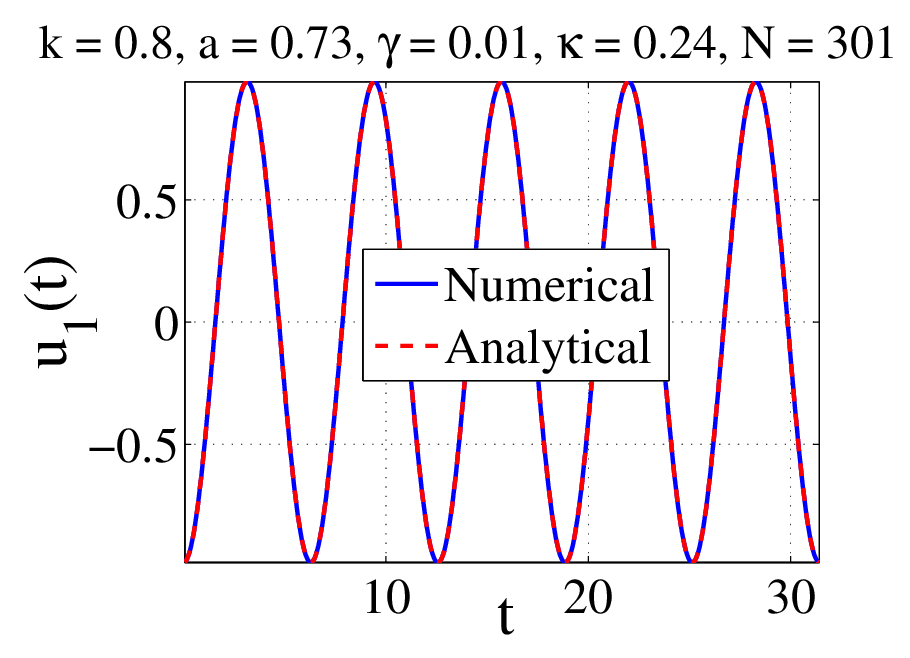}} \\
{\includegraphics[width=1.81in,height=1.81in]{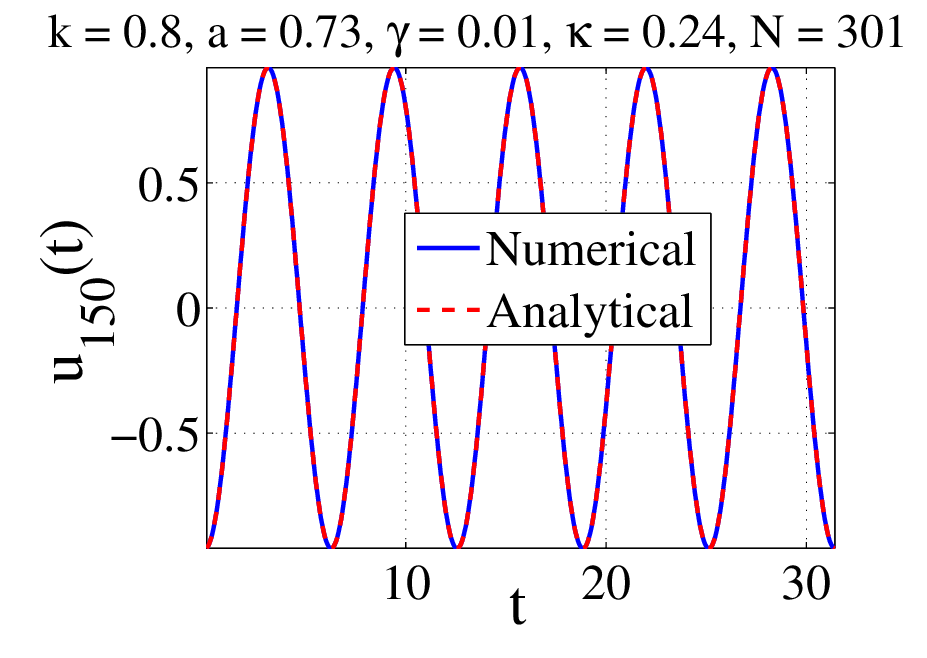}} \\
{\includegraphics[width=1.81in,height=1.81in]{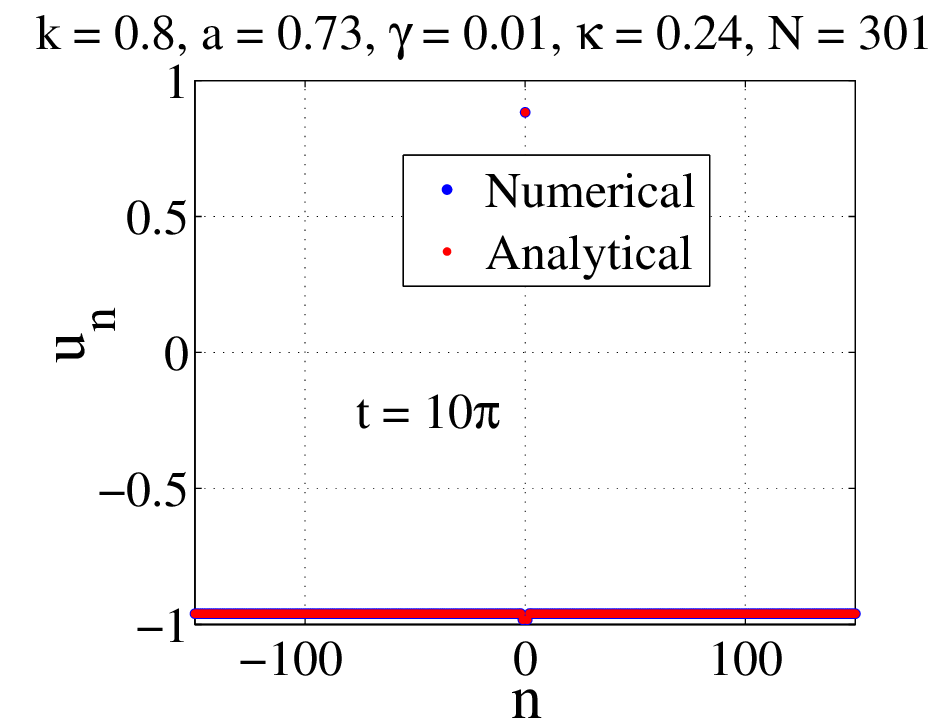}}
\end{center}
\caption{\small  Comparison of numerically (blue) and analytically (red) obtained values of $u_0(t)$, $u_1(t)$, $u_{150}(t)$ and $u_n(n)$, at the end of the $5^{\text{th}}$ cycle for $N=301$ and: $k=0.8$, $\gamma=0.01$ and $a=0.73$ for $\kappa=0.24$, in validation of the analytic solution and the periodic boundary conditions and the scheme itself (for the numerical solution). }
\label{Figure18}
\end{figure}

\begin{figure}
\begin{center}
{\includegraphics[width=1.8in,height=1.8in]{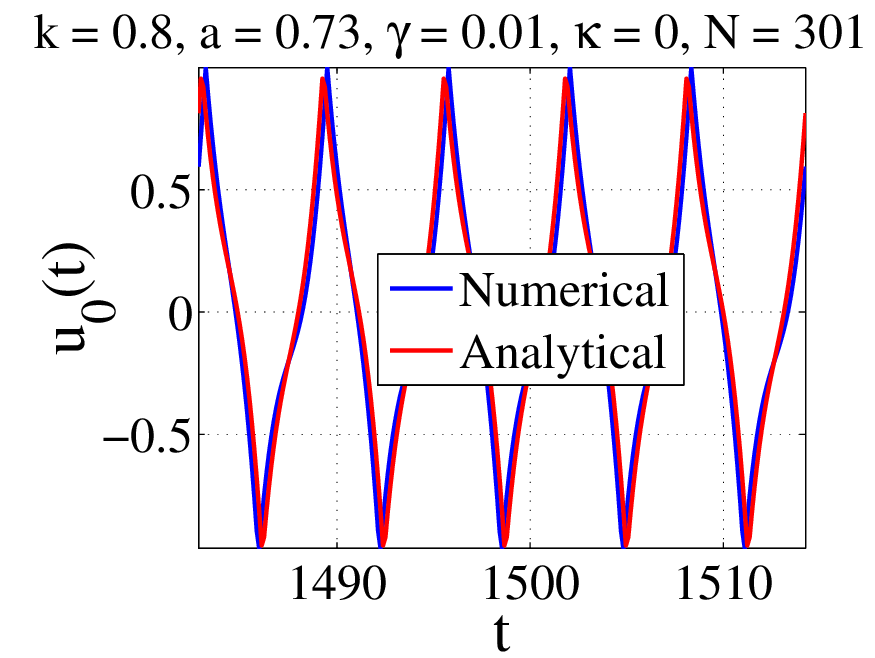}}
{\includegraphics[width=1.8in,height=1.8in]{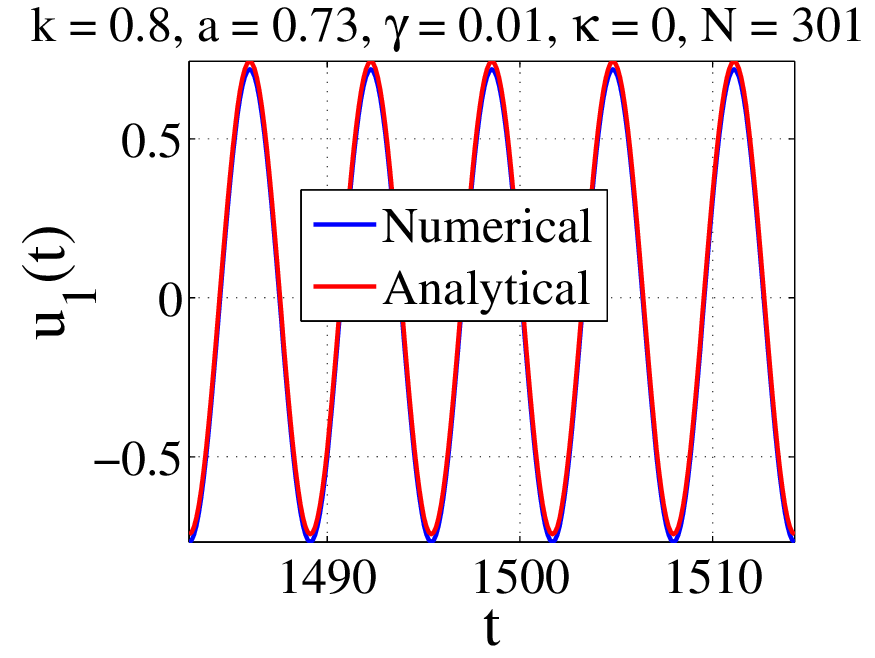}} \\
{\includegraphics[width=1.8in,height=1.8in]{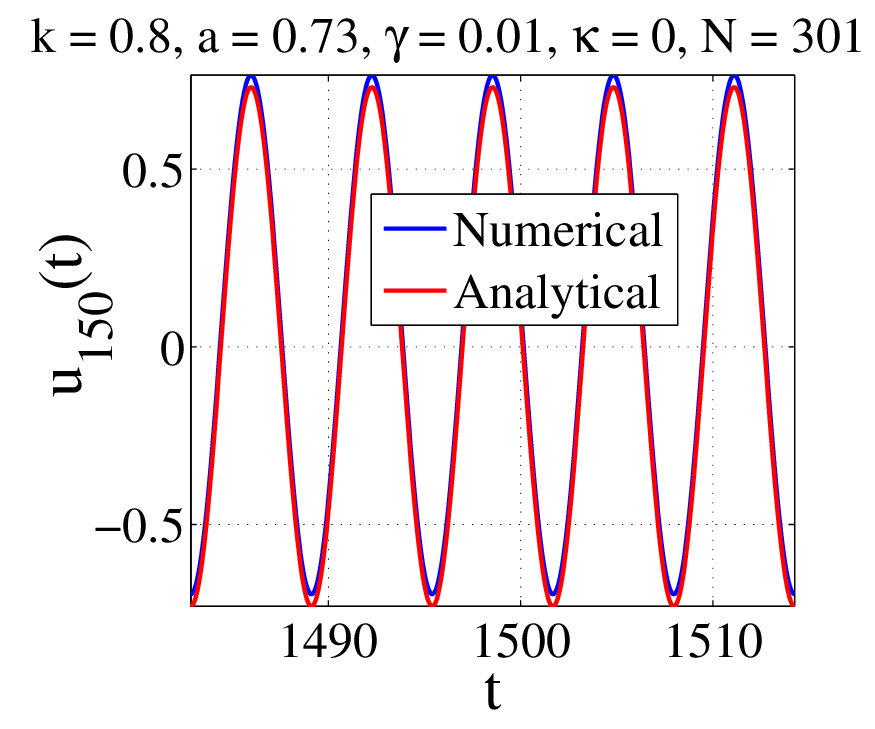}}
{\includegraphics[width=1.8in,height=1.8in]{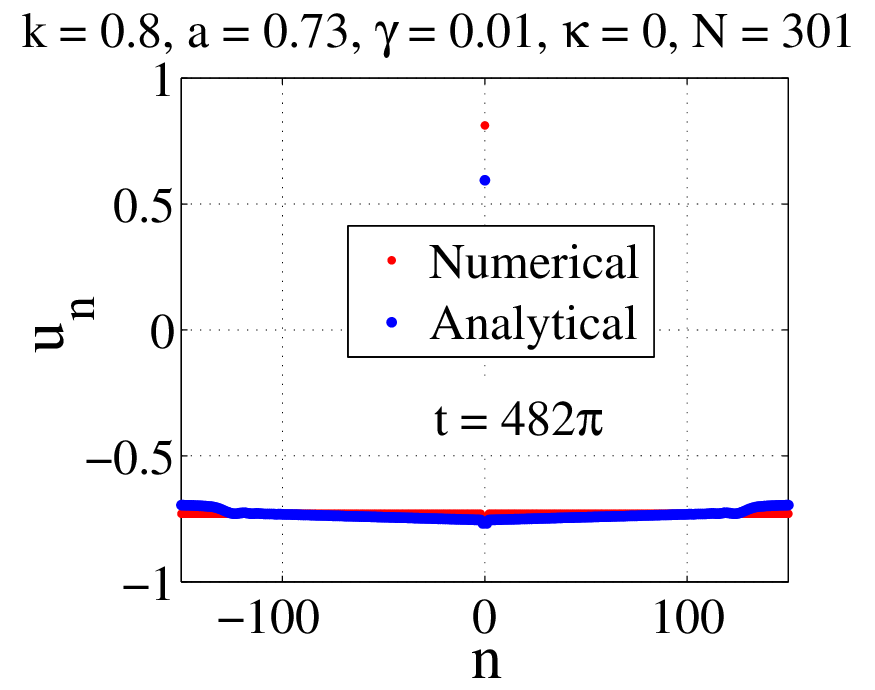}} \\
\end{center}
\caption{\small Comparison of numerically (blue) and analytically (red) obtained displacement histories and profiles for $N=301$ and: $k=0.8$, $\gamma=0.01$ and $a=0.73$ for $\kappa=0$, for large integration times, integrated with the 'impact' scheme. One clearly sees the beginning of the deviation of the numerically integrated result from the analytic one. }
\label{Figure19}
\end{figure}

\begin{figure}
\begin{center}
{\includegraphics[width=1.8in,height=1.8in]{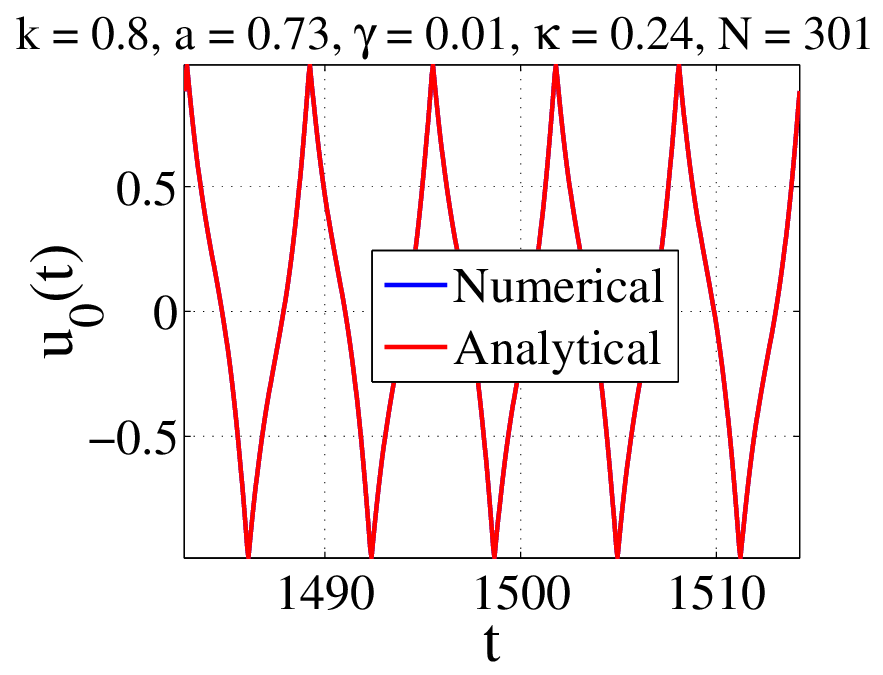}}
{\includegraphics[width=1.8in,height=1.8in]{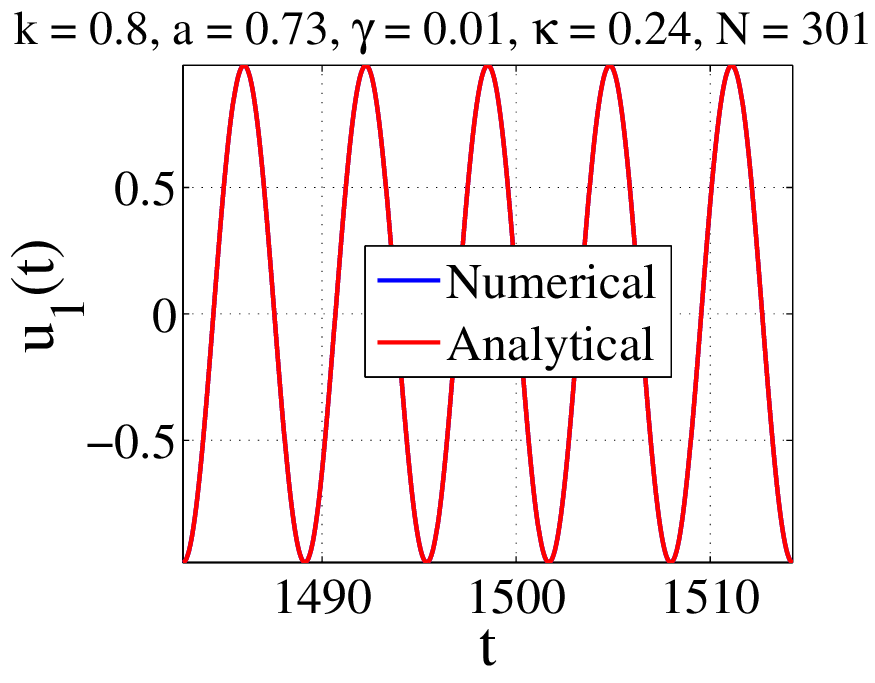}} \\
{\includegraphics[width=1.8in,height=1.8in]{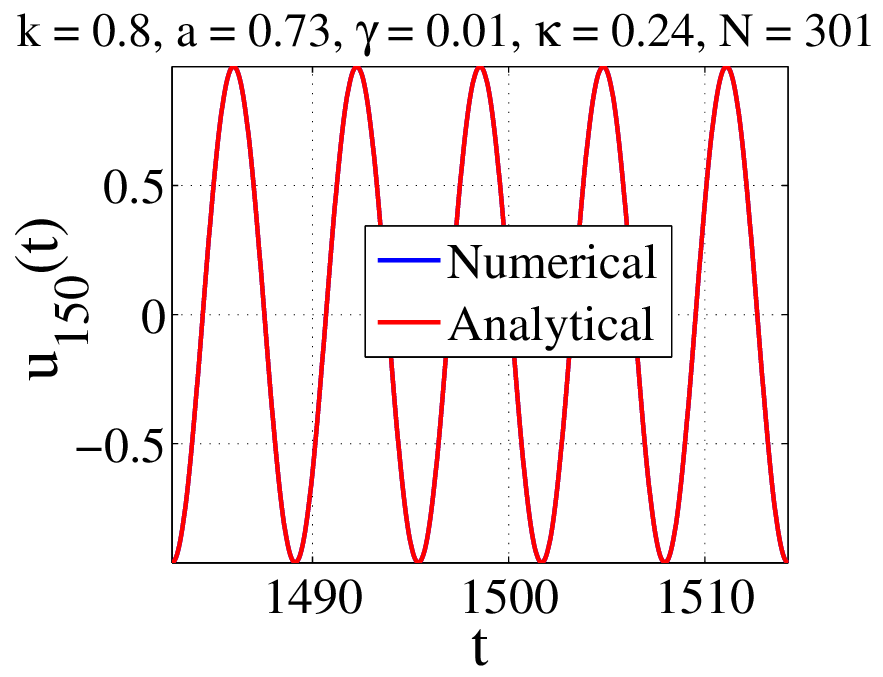}}
{\includegraphics[width=1.8in,height=1.8in]{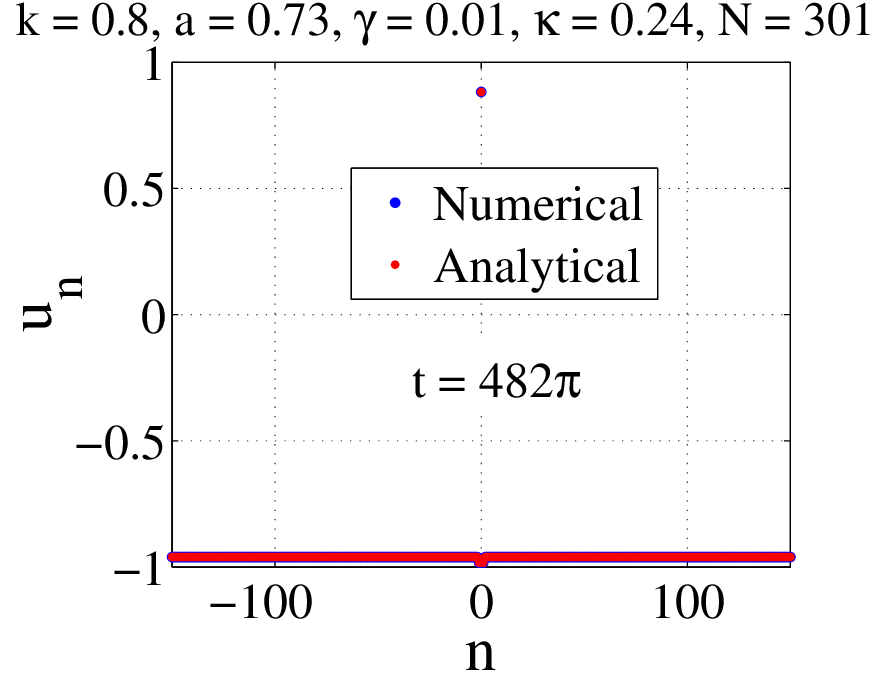}}
\end{center}
\caption{\small Comparison of numerically (blue) and analytically (red) obtained displacement histories and profiles for $N=301,k=0.8$, $\gamma=0.01,a=0.73$ for $\kappa=0.24$, for large integration times, integrated with the 'impact' scheme. One notes that the blue points are completely covered by the red ones, unlike in Figure \ref{Figure19}. This can be viewed as $\kappa$-stabilization. }
\label{Figure20}
\end{figure}

\section{Conclusions}
\label{sect7}

In the present work, we have generalized the result obtained in \cite{Gendelman2013} to account for a uniform harmonic potential added to the harmonically excited infinite linear chain of masses placed between vibro-impact constraints. This forced-damped system was shown to have exact localized solutions corresponding to discrete immobile breathers. The solutions were obtained analytically for arbitrary values of the stiffness of the uniform harmonic potential, the coupling stiffness of the chain, the coefficient of restitution  and the amplitude of the external force. Existence and linear stability characteristics of the solution were examined using a combination of analytic and numerical techniques.

One can note two main results. The first is in the additional steps, with respect to previous work, taken here to address the questions of existence and linear stability of the DB solution. The second is the understanding that the DB can be stabilized for the case of weak coupling, by the introduction of a uniform harmonic on-site  potential. The stabilization is manifested in the form of $10\%$ gain in the excitation amplitude for which the breather remains linearly stable, most of it in the higher amplitude range $-$ that is, the breather can stably exist for approximately $10\%$ higher excitation amplitudes. This result can, in principle, have practical applications. In addition, better understanding of the non-monotonicity of the stability-limiting curve in the amplitude-coupling plane was obtained. It turned out that this non-monotonicity is related to the bifurcation point from which multiple stability-limiting branches originate in the limit of full restitution for critical foundation stiffness.  Furthermore, it was shown that in the limit of full restitution, for a foundation-free chain, two previously reported instability zones, namely the pitchfork bifurcation-related zone and the Neimark-Sacker bifurcation-related zone, become adjacent, leaving but a single  monotonically-bounded linearly stable region in the parameter space.

\section*{Acknowledgments}
The authors are grateful to the Israel Science Foundation (grant 838/13) for financial support.

\renewcommand{\theequation}{A.\arabic{equation}}
\setcounter{equation}{0}  
\setcounter{figure}{0}  

\appendix
\section{Consistency analysis of the exact DB solution}  
\label{AppendixA}

For the first issue mentioned in the end of Section \ref{sect4}, temporal symmetry of the solution with respect to a reference impact instance can be used in conjunction with a sequence of bounds, to show that no additional impacts occur between nominal impacts. The displacement history of the central mass can be obtained by setting $n=0$ in (\ref{eq52}):
\begin{equation}
 \begin{split}
 u_0(t)=- \frac{a\cos t}{1-\kappa} \\ +\frac{4p}{\pi}  \sum _{l=0}^{\infty} \frac{ \cos[(2l+1)(t-\phi)]}{\sqrt{[(2l+1)^2-(2\gamma+\kappa)]^2-(2\gamma)^2}}
\label{eq53}
 \end{split}
   \end{equation}

Temporal symmetry with respect to $t=\phi$ implies: $u_0(t)=u_0(2\phi-t)$. Owing to the fact that $\cos{[\omega(2\phi-t-\phi)]}= \cos{[\omega(\phi-t)]}=\cos{[\omega(t-\phi)]}$, we have:
\begin{equation}
 \begin{split}
  u_0(t)= -\frac{a\cos {(2\phi-t)}}{1-\kappa}\\+\frac{4p}{\pi}  \sum _{l=0}^{\infty} \frac{ \cos[(2l+1)(t-\phi)]}{\sqrt{[(2l+1)^2-(2\gamma+\kappa)]^2-(2\gamma)^2}}
   \label{eq54}
   \end{split}
    \end{equation}

Next, between reference impacts, there holds: $\phi<t<\phi+\pi$ and one thus has:
\begin{equation}
 \begin{split}
 t>\phi \Rightarrow -t<-\phi  \Rightarrow 2\phi-t<2\phi-\phi=\phi  \\ \Rightarrow \cos{(2\phi-t)}>\cos{\phi}  \Rightarrow -\cos{(2\phi-t)} <\\
-\cos{(\phi)}  \Rightarrow u_0(t) < -\frac{a\cos{\phi}}{1-\kappa}\\+ \frac{4p}{\pi}\sum _{l=0}^{\infty} \frac{ 1}{\sqrt{[(2l+1)^2-(2\gamma+\kappa)]^2-(2\gamma)^2}} \\=u_0(\phi)=1
\label{eq55}
\end{split}
 \end{equation}

Similarly,
\begin{equation}
 \begin{split}
 t<\phi+\pi \Rightarrow -t>-\phi-\pi  \Rightarrow 2\phi-t> \\ 2\phi-\phi-\pi= \phi-\pi \Rightarrow \cos{(2\phi-t)}< \\
<\cos{(\phi-\pi)}=\cos{(\pi-\phi)}=-\cos{\phi} \\ \Rightarrow -\cos{(2\phi-t)} >\cos{\phi}
 \Rightarrow u_0(t) > \frac{a\cos{\phi}}{1-\kappa}\\- \frac{4p}{\pi}\sum _{l=0}^{\infty} \frac{ 1}{\sqrt{[(2l+1)^2-(2\gamma+\kappa)]^2-(2\gamma)^2}} \\ =-u_0(\phi)=-1
 \label{eq56}
 \end{split}
  \end{equation}
and therefore: $-1<u_0(\phi<t<\phi+\pi)<1$.

The inequalities in (\ref{eq55}) and (\ref{eq56}) are strong, and this implies that there really are \emph{no} impacts between nominal impacts, at least as for as the central mass is concerned.

As for the displacements of the other masses being always smaller than unity, it is up to the solution of problem (\ref{eq38}), in terms of the upper bound on the excitation amplitude.

Problem (\ref{eq38}) for the upper bound on the excitation amplitude could only be solved numerically, and thus could only work  for $n<N \ll \infty$. For $\lvert n \rvert \to \infty$, due to (\ref{eq39}) and (\ref{eq40}), one would have:
\begin{equation}
\begin{split}
\left \lvert u_{n \to \infty}(t)=-\frac{a\cos {t}}{1-\kappa}+\right. \\   \sum _{l=0}^{\infty}{\lim_{n \to \infty}\left (\frac{A -\sqrt{A^2 -B^2}}{B} \right )}^n  \\ \left. \times \frac{ \frac{4p}{\pi}\cos[(2l+1)(t-\phi)]}{\sqrt{[(2l+1)^2-(2\gamma+\kappa)]^2-(2\gamma)^2}}\right \rvert \\
=\left \lvert-\frac{a\cos {t}}{1-\kappa} \right \rvert <  1 \Rightarrow a<1-\kappa
 \label{eq57}
 \end{split}
  \end{equation}
where $A$ and $B$ in (\ref{eq57}) are as defined in (\ref{eq40}). Thus, infinitely distant masses do not render the problem inconsistent as long as the excitation amplitude is low enough to comply with the constraint in (\ref{eq57}).
We have thus shown the existence of a self-consistent, spatially localized temporally periodic externally excited dissipative system, evident for a non-empty set of parameter values.
The last step in this existence analysis is the investigation of the character of the convergence of the series in (\ref{eq52}).

Regarding the convergence of the series constructing the solution given in (\ref{eq52}), the following can be said. First, one can show that both the displacements and the velocities of all the masses but for the central one are unconditionally convergent. This can be shown as follows. For start, we can express the displacement field as a series: $u_n(t)=\sum_{l=0}^{\infty}U^l_n(t)$. This requires no additional assumptions or actions besides the expression in (\ref{eq52}). Next, expanding $U^l_n(t)$ as a Taylor series in $\gamma$, holding in mind that localized solutions only exist for $\gamma<1/4$, and taking the large $l$ limit, we get:
\begin{equation}
 \begin{split}
   U^l_n(t) \underset{l \to \infty}{\to} -\frac{2}{\pi^2}\frac{a\cos{t}}{1-\kappa}l^{-2} \\ +\left [ \frac{p}{\pi}(-\gamma/4)^{\lvert n \rvert}l^{-2(\lvert n \rvert+1)}+\mathcal{O}(l^{-2(\lvert n  \rvert+2)})\right ] \\ \times \cos{[2l(t-\phi)]}
   \label{eq58}
   \end{split}
   \end{equation}

Summation over a sequence having the limit given in (\ref{eq58}) would always be unconditionally convergent, and indeed, there are no convergence questions in what concerns the displacement field.
However, the velocity field, which can be obtained by direct differentiation with respect to time of the series representation given above, would be: $\dot{u}_n(t)=\sum_{l=0}^{\infty}\dot{U}^l_n(t)$. Differentiation of the asymptotic expansion in (\ref{eq58}) gives an asymptotic expression for an additive term in the series expansion of the velocity field:
\begin{equation}
\begin{split}
\dot{U}^l_n(t) \underset{l \to \infty}{\to} \frac{2}{\pi^2}\frac{a\sin{t}}{1-\kappa}l^{-2}\\ -\left [ \frac{2p}{\pi}(-\gamma/4)^{\lvert n \rvert}l^{-1-2\lvert n \rvert}+\mathcal{O}(l^{-3-2\lvert n  \rvert})\right ] \\ \times \cos{[2l(t-\phi)]}
\label{eq59}
\end{split}
\end{equation}

Summation over a sequence with the limit given in (\ref{eq59}) would always be unconditionally convergent for every integer value of $n$ except for $n=0$, which corresponds to the central mass, where one has: $\dot{U}^l_0(t) \underset{l \to \infty}{\to}\mathcal{O}(l^{-1})$, and $\dot{u}_0(t)=\sum_{l=0}^{\infty}\dot{U}^l_0(t)$ becomes only conditionally convergent.

Hence, the velocity field of the central mass should be treated separately and possibly regularized. Expanding the expression given in (\ref{eq53}) as a Taylor series with respect to $\gamma$ around $\gamma=0$ in the range $\gamma<1/4$ (which is the relevant existence range) gives:
\begin{equation}
\begin{split}
 u_0(t) = -\frac{a\cos{t}}{1-\kappa} +\frac{4p}{\pi}\sum_{j=0}^{\infty}\frac{\Gamma\left(-\frac{1}{2}\right)(-4\gamma)^j}{j!\Gamma\left(-\frac{1}{2}-j\right)} \\ \times\sum_{l=0}^{\infty}{\frac{\cos{[(2l+1)(t-\phi)]}}{[(2l+1)^2-\kappa]^{j+1}}}
 \label{eq60}
 \end{split}
 \end{equation}
where $\Gamma(\cdot{})$ is the generalized factorial function. The obtained expression is a sum of a geometric series with respect to $\kappa$ and can thus be expanded into its power series form. The upper bound on the radius of convergence of the resulting power series for the $l=0$ case would be unity, assuring convergence for $\kappa<1$. We already obtained this anyway as the feasible existence domain guaranteeing that only the central mass impacts, and therefore no additional restrictions arise. Hence, one can write:
\begin{equation}
\begin{split}
 u_0(t) = -\frac{a\cos{t}}{1-\kappa} +\frac{4p}{\pi} \times  \\ \sum_{j=0}^{\infty}\frac{\Gamma\left(-\frac{1}{2}\right)(-4\gamma)^j}{j!\Gamma\left(-\frac{1}{2}-j\right)}   \sum_{l=0}^{\infty}\left [\sum_{r=0}^{\infty}\frac{\kappa^r}{(2l+1)^{2r}}\right ]^{j+1} \\ \times \frac{\cos{[(2l+1)(t-\phi)]}}{(2l+1)^{2(j+1)}}
  \label{eq61}
 \end{split}
 \end{equation}

Separating the terms gives:
\begin{equation}
  \begin{split}
  u_0(t) = -\frac{a\cos{t}}{1-\kappa} \\ +\frac{4p}{\pi}\sum_{l=0}^{\infty} \frac{\cos{[(2l+1)(t-\phi)]}}{(2l+1)^{2}}  \\
+\frac{4p}{\pi} \sum_{l=0}^{\infty} \sum_{r=1}^{\infty}\frac{\kappa^r\cos{[(2l+1)(t-\phi)]}}{(2l+1)^{2(r+1)}} \\
+\frac{4p}{\pi}  \sum_{j=0}^{\infty}\frac{\Gamma\left(-\frac{1}{2}\right)(-4\gamma)^j}{j!\Gamma\left(-\frac{1}{2}-j\right)} \times \\ \sum_{l=0}^{\infty}\left [\sum_{r=0}^{\infty}\frac{\kappa^r}{(2l+1)^{2r}}\right ]^{j+1} \\ \times \frac{\cos{[(2l+1)(t-\phi)]}}{(2l+1)^{2(j+1)}}
\end{split}
\label{eq62}
\end{equation}

So far, everything in (\ref{eq62}) is unconditionally convergent. However, differentiating with respect to time, we get:
\begin{equation}
  \begin{split}
   \dot{u}_0(t) = \frac{a\sin{t}}{1-\kappa}\\ -\frac{4p}{\pi}\sum_{l=0}^{\infty} \frac{\sin{[(2l+1)(t-\phi)]}}{2l+1} \\
-\frac{4p}{\pi}\sum_{l=0}^{\infty} \sum_{r=1}^{\infty}\frac{\kappa^r\sin{[(2l+1)(t-\phi)]}}{(2l+1)^{2r+1}} \\
-\frac{4p}{\pi}\sum_{j=0}^{\infty}\frac{\Gamma\left(-\frac{1}{2}\right)(-4\gamma)^j}{j!\Gamma\left(-\frac{1}{2}-j\right)}  \\ \times \sum_{l=0}^{\infty}\left [\sum_{r=0}^{\infty}\frac{\kappa^r}{(2l+1)^{2r}}\right ]^{j+1} \\ \times \frac{\sin{[(2l+1)(t-\phi)]}}{(2l+1)^{2j+1}}
 \label{eq63}
\end{split}
 \end{equation}

Indeed, as predicted by (\ref{eq59}), the second term in (\ref{eq63}) is only conditionally convergent and thus may be problematic to work with. In order to circumvent this obstacle, we make use of a certain non-continuous function in its formal definition, which has the property that its Fourier sine transform is equal to the series in the second term in (\ref{eq63}), as follows:
\begin{equation}
\begin{split}
 \frac{4}{\pi}\sum_{l=0}^{\infty} \frac{\sin{[(2l+1)(t-\phi)]}}{2l+1} = \\ \operatorname{sgn}(t-\phi)\operatorname{sgn}\left[\pi-2\pi {\left \lbrace \frac{\lvert t-\phi  \rvert}{2\pi} \right  \rbrace _{frac}}\right]
  \label{eq64}
 \end{split}
 \end{equation}
and the exact velocity history of the central mass can then be expressed analytically as:
\begin{equation}
 \begin{split}
 \dot{u}_0(t) = \frac{a\sin{t}}{1-\kappa} -\frac{4p}{\pi}\sum_{j=0}^{\infty}\frac{\Gamma\left(-\frac{1}{2}\right)(-4\gamma)^j}{j!\Gamma\left(-\frac{1}{2}-j\right)} \\ \times \sum_{l=0}^{\infty}\left [\sum_{r=0}^{\infty}\frac{\kappa^r}{(2l+1)^{2r}}\right ]^{j+1} \\ \times \frac{\sin{[(2l+1)(t-\phi)]}}{(2l+1)^{2j+1}} \\
-p\operatorname{sgn}(t-\phi)\operatorname{sgn}\left[\pi-2\pi {\left \lbrace \frac{\lvert t-\phi  \rvert}{2\pi} \right  \rbrace _{frac}}\right]  \\
-\frac{4p}{\pi}\sum_{l=0}^{\infty} \sum_{r=1}^{\infty}\frac{\kappa^r\sin{[(2l+1)(t-\phi)]}}{(2l+1)^{2r+1}} \\
  \label{eq65}
 \end{split}
 \end{equation}

Obviously, the discontinuities incorporated into the solution by the sign function represent velocity discontinuities due to the impacts. The fact that during impact instances the velocity is undetermined, is related to the conditional convergence of the second term in (\ref{eq63}). The expression in (\ref{eq65}), however, unlike the one in (\ref{eq63}), though discontinuous, is to produce no artifacts, should it undergo additional mathematical operations, whether analytic or numerical.

In the limit of zero link stiffness, no  foundation and full restitution, the velocity field of the breather degenerates to the velocity field of a single, $2\pi$-periodically excited bouncing mass, which reads:
\begin{equation}
\begin{split}
   \dot{u}_0(t>0) \underset{q,\kappa,\gamma \to 0}{\to} \frac{a\sin{t}}{1-\kappa}\\ -p\operatorname{sgn}\left[\pi-2\pi {\left \lbrace \frac{t}{2\pi} \right  \rbrace _{frac}}\right]
\end{split}
\label{eq66}
\end{equation}
where $ \lbrace \rbrace_{frac}$ is the fractional-part-of-a-number function.

\renewcommand{\theequation}{B.\arabic{equation}}
\setcounter{equation}{0}  

\section{Approximate analytical treatment of the problem embodied in Inequality (\ref{eq38})}  
\label{AppendixB}

Next, we describe several features of the existence domain of the solution which has already been shown to correspond to a localized breather with no phonon emission, and verified to be self-consistent, unconditionally convergent and regular within, but not in the whole parameter domain: $\kappa<1,\gamma<(1-\kappa)/4,a<1-\kappa$. Additional conditions for the existence of the solution are $a>a_{min}$, as given by (\ref{eq34}), and $a<a_{max}$, arising from the solution of (\ref{eq38}).

The first point worth noting here is that the numerical solution of (\ref{eq38}) is quite tedious a task, being consistent of an iteration loop for the amplitude with two inner minimization loops for $t$ and $n$, and all this inside a loop on $\gamma$, including the computation of truncated infinite series containing the parameters $k$ and $\kappa$. Below we derive a faster method for the computation of the upper bound on the amplitude for which the localized breather solution exists.

Combining (\ref{eq38}) and (\ref{eq52}), we can write an equation for the upper bound on the excitation amplitude as follows:
\begin{equation}
 \hat{a} \cos{t}=A_n(t)p-\text{sgn}[u_n(t)]
  \label{eq67}
  \end{equation}
where
\begin{equation}
 \begin{split}
 \bar{a}=(1-\kappa)  \underset{|n| \in \mathbb{N}, t} \min{\hat{a}} \ , \\  A_n(t) \triangleq \frac{4}{\pi} \left (-\frac{1}{2\gamma} \right )^{|n|}\times \\ \times \sum_{l=0}^{\infty} \frac{ \cos[(2l+1)(t-\phi)]}{\sqrt{[2\gamma+\kappa-(2l+1)^2]^2-(2\gamma)^2}} \times \\ \left \lvert \vphantom{\sqrt{[(1)^2]^2}} (2l+1)^2-(2\gamma+\kappa)\right. \\ \left.-\sqrt{[2\gamma+\kappa-(2l+1)^2]^2-(2\gamma)^2}\right|^{|n|}
  \label{eq68}
 \end{split}
 \end{equation}
and $p$ and $\phi$ are given by (\ref{eq28}-\ref{eq30}).

An approximate explicit expression for the upper bound on the excitation amplitude can be obtained by expanding $\bar{a}$ with respect to the coefficient of restitution $k$ around $k=1$, bearing in mind that a breather is more interesting for larger restitution values, where it is more efficient, or, more conveniently, with respect to $q$ around $q=0$. Taking $q=0$ in the set of equations defined by (\ref{eq28}-\ref{eq30}), (\ref{eq67}) and (\ref{eq68}), one can solve for $\hat{a}_{q=0}$ analytically, getting $p_{q=0}=(1+\hat{a})/\chi_0$ and $\phi_{q=0}=0$ and hence:
\begin{equation}
\hat{a}_{n}^{q=0}(t)=\frac{A_n^{q=0}(t)-\text{sgn}[u_n^{q=0}(t)]\chi_0}{\chi_0\cos{t}-A_n^{q=0}(t)}
\label{eq69}
\end{equation}
where, $\chi_0$ is as in Eq. (\ref{eq30}) and $S_0$ is as given by Eq. (\ref{eq98}). Obviously, for an upper bound we need the minimum with respect to $n$ and $t$ of the expression in (\ref{eq69}), for any feasible $\gamma,\kappa$ pair. Since $A_n^{q=0}(t)$ and $\cos{t}$ are skew-symmetric with respect to $t= \pm \pi/2$, clearly so is $u_n^{q=0}(t)$. Therefore, one would have: $\text{sgn}[u_{n_{min}}^{q=0}(t_{min})]=-\text{sgn}[u_{n_{min}}^{q=0}(\pi-t_{min})]$, and thus minimizing $\hat{a}$ with respect to $t$ in $t \in (-\pi/2,\pi/2)$ would be sufficient, since the complementary half-cycle would contain an $\bar{a}$-critical displacement of the same absolute value of $1$, only opposite in sign. Consequently, we can minimize in $t \in (-\pi/2,\pi/2)$ and with no loss of generality assume that: $\text{sgn}[u_n^{q=0}(t)]=\mathcal{S}_u$ is constant throughout the minimization.

Now, we note that: $-A_{n=2\mathbb{N}}^{q=0}|_{t\in (-\frac{\pi}{2},\frac{\pi}{2})}<0<-A_{n=2\mathbb{N}-1}^{q=0}|_{t\in (-\frac{\pi}{2},\frac{\pi}{2})}$,
and thus:
\begin{equation}
\begin{split}
\underset{t \in (-\frac{\pi}{2},\frac{\pi}{2})}\max{[-A_{|n|\in \mathbb{N}}^{q=0}(t)]} \\ =\underset{n=2\mathbb{N}-1,t \in (-\frac{\pi}{2},\frac{\pi}{2})}\max{[-A_n^{q=0}(t)]}\underset{(\ref{eq39})}{=} \\ \underset{t \in (-\frac{\pi}{2},\frac{\pi}{2})}\max{[-A_{n=1}^{q=0}(t)]}\underset{(\ref{eq68})}{=}-A_{{n=1}}^{q=0}(t=0)
\end{split}
\label{eq71}
\end{equation}

Next, noting that: $\cos{(t=0)}=1$, which means that both $\cos{t}$ and $-A_n^{q=0}(t)$ obtain their maxima at the same time in the considered half-cycle, $t=0$, and realizing that for $X(t)=[const.-X_1(t)]/[X_2(t)+X_1(t)]$, where $X_1(t)$ and $X_2(t)>0$ obtain their maxima at the same point, one has: $\underset{t}\min{X(t)}=[const.-X_1(t_{max})]/[X_2(t_{max})+X_1(t_{max})]$, we get the following result:
\begin{equation}
\begin{split}
\min_{|n| \in \mathbb{N}, t \in (-\frac{\pi}{2},\frac{\pi}{2})}{\hat{a}_n^{q=0}(t)}= \\ \frac{-\mathcal{S}
_u\chi_0-{[-A_{n=1}^{q=0}(t=0)]}}{\chi_0+{[-A_{n=1}^{q=0}(t=0)]}}
\end{split}
\label{eq70}
\end{equation}

Consequently, since $\mathcal{S}_u$ is constant throughout the minimization, we can calculate it at the obtain minimum $(n,t)=(1,0)$, as: $\mathcal{S}_u=\text{sgn}[u_1^{q=0}(t=0)]=-\text{sgn}[1+(1-\kappa)^{-1}\bar{a}_{q=0}] \underset{(*)}{=}-1$, to get:
\begin{equation}
\begin{split}
 \bar{a}_{q=0}=(1-\kappa) \times \\ \frac{\underset{n=1,3,5,...}{\overset{\infty}{\sum}}\frac{1-\frac{n^2-(2\gamma+\kappa)-\sqrt{(2\gamma+\kappa-n^2)^2-4\gamma^2}}{(2\gamma)}}{\sqrt{(2\gamma+\kappa-n^2)^2-4\gamma^2}}}{\underset{m=1,3,5,...}{\overset{\infty}{\sum}}\frac{1+\frac{m^2-(2\gamma+\kappa)-\sqrt{(2\gamma+\kappa-m^2)^2-4\gamma^2}}{(2\gamma)}}{\sqrt{(2\gamma+\kappa-m^2)^2-4\gamma^2}}}
\label{eq72}
\end{split}
\end{equation}
where condition $(*)$ is satisfied since due to (\ref{eq34}) and (\ref{eq39}), (\ref{eq72}) implies that: $0<(1-\kappa)^{-1}\bar{a}_{q=0}<1$.

In the next step we derive a correction for $\bar{a}$ for realistic values of the coefficient of restitution, $k$.

Taking a derivative of (\ref{eq67}) (not differentiating the $\text{sgn}[u_n(t)]$ term for reasons explained below) with respect to some convex, monotonously increasing homogeneous function $\hat{q}(q)$, recalling (\ref{eq28}) and (\ref{eq29}) and rearranging, we get:
\begin{equation}
\begin{split}
\frac{d\hat{a}}{d\hat{q}}=  \frac{A_n(t)p'+p\frac{(pq'+qp')/\hat{a}}{\sqrt{1-q^2p^2/\hat{a}^2}}\frac{\partial{A_n(t)}}{\partial{\phi}}}{\cos{t}-A_n(t)p_{\hat{a}}-\frac{qp\left(\frac{p_{\hat{a}}}{\hat{a}}-\frac{p}{\hat{a}^2}\right)}{\sqrt{1-\frac{q^2p^2}{\hat{a}^2}}}\frac{\partial{A_n(t)}}{\partial{\phi}}}
\end{split}
\label{eq73}
\end{equation}
where $()'$ denotes differentiation with respect to an explicit dependence on a function of $q$ as characterized above, and $()_{\hat{a}}$ denotes differentiation with respect to an explicit dependence on $\hat{a}$. Seeking a (truncated) Taylor series expansion of $\hat{a}(\hat{q})$, we only need the $q \to 0$ limit of $d\hat{a}/d\hat{q}$. Also, in line with the perturbation concept, motivated by the fact that in the interesting scenario one would not have too low a coefficient of restitution, it is assumed that $q$ should be relatively low and thus the pair $(n,t)=(1,0)$ that minimizes $\bar{a}_{q=0}$ should also minimize $\bar{a}(q)$ for: $0<q \ll 1$. Moreover since (due to its smooth $q$-dependence) $u_n^{0<q \ll 1}(t)$ should be only slightly different from $u_n^{q=0}(t)$, clearly one should have: $\text{sgn}[u_n^{0<q \ll 1}(t)]=\mathcal{S}_u$ and hence $d\lbrace\text{sgn}[u_n^{0<q \ll 1}(t)]\rbrace/dq=0$, resulting in Eq. (\ref{eq73}). Furthermore, it is clear from (\ref{eq28}) and (\ref{eq30}) and (\ref{eq68}) that $\frac{\partial{A_n(t=0)}}{\partial{\phi}}|_{q=0}=\frac{\partial{A_n(t=0)}}{\partial{\phi}}|_{\phi=0}=0$.

Next, differentiating (\ref{eq29}) with respect to $\hat{a}$ and taking the $q \to 0$ limit yields: $p_{\hat{a}}^{q=0}=\chi_0^{-1}$.
Hence, substituting $q=0, t=0$ and $n=1$ in (\ref{eq73}) and using the last relation in (\ref{eq71}), we get:
\begin{equation}
\begin{split}
{\hat{a}}'_{q=0}=-p'_{q=0}\chi_0\frac{|A_{{n=1}}^{q=0}(t=0)|}{\chi_0+|A_{{n=1}}^{q=0}(t=0)|}\\ \underset{(\ref{eq69}-\ref{eq72})}{=}  -\frac{1}{2}\chi_0 (1-\hat{a}_{q=0})p'_{q=0}
\end{split}
\label{eq74}
\end{equation}
where $\hat{a}_{q=0}\triangleq(1-\kappa)^{-1}\bar{a}_{q=0}$.

It is clear from (\ref{eq74}) that the $q$-dependence of $\bar{a}$, to first-order approximation, arises from the $q$-dependence of $p$. Thus the $\hat{q}$ in $\hat{a}'=\partial{\hat{a}}/\partial{\hat{q}(q)}$ is the same $\hat{q}$ as in $p'=\partial{p}/\partial{\hat{q}}$. From (\ref{eq29}), and the requirements: $\hat{q}(0)=0$ and $\hat{q}(q) \in C_{\infty}$, we find that to a leading term: $\hat{q}(q)=q^2$ ($\bar{a}$ is linearly independent of $q$, which makes $\bar{a}_{q=0}$ in (\ref{eq72}) a good estimate for high coefficients of restitution). By choosing $\hat{q}(q)=q^2$, we get the same order of approximation with $\hat{a}'_{q=0}$ as we would have got with $\hat{a}''_{q=0}$ had we chosen $\hat{q}(q)=q$, while avoiding second-order differentiation.

Differentiating $p$ in (\ref{eq29}) with respect to the explicit dependence on $q^2$ and then setting $q=0$, results in:
\begin{equation}
p'_{q=0}=-\frac{1}{2}\chi_0^{-3}\frac{(1+\hat{a}_{q=0})^2}{\hat{a}_{q=0}}
\label{eq75}
\end{equation}
\begin{equation}
\hat{a}'_{q=0}=\frac{1}{4}\chi_0^{-2}\frac{(1-\hat{a}_{q=0})(1+\hat{a}_{q=0})^2}{\hat{a}_{q=0}}
\label{eq76}
\end{equation}
\begin{equation}
\bar{a} \approx \bar{a}_{q=0}+(1-\kappa)\hat{a}'_{q=0}q^2
\label{eq77}
\end{equation}

\begin{figure}
\includegraphics[scale=0.33]{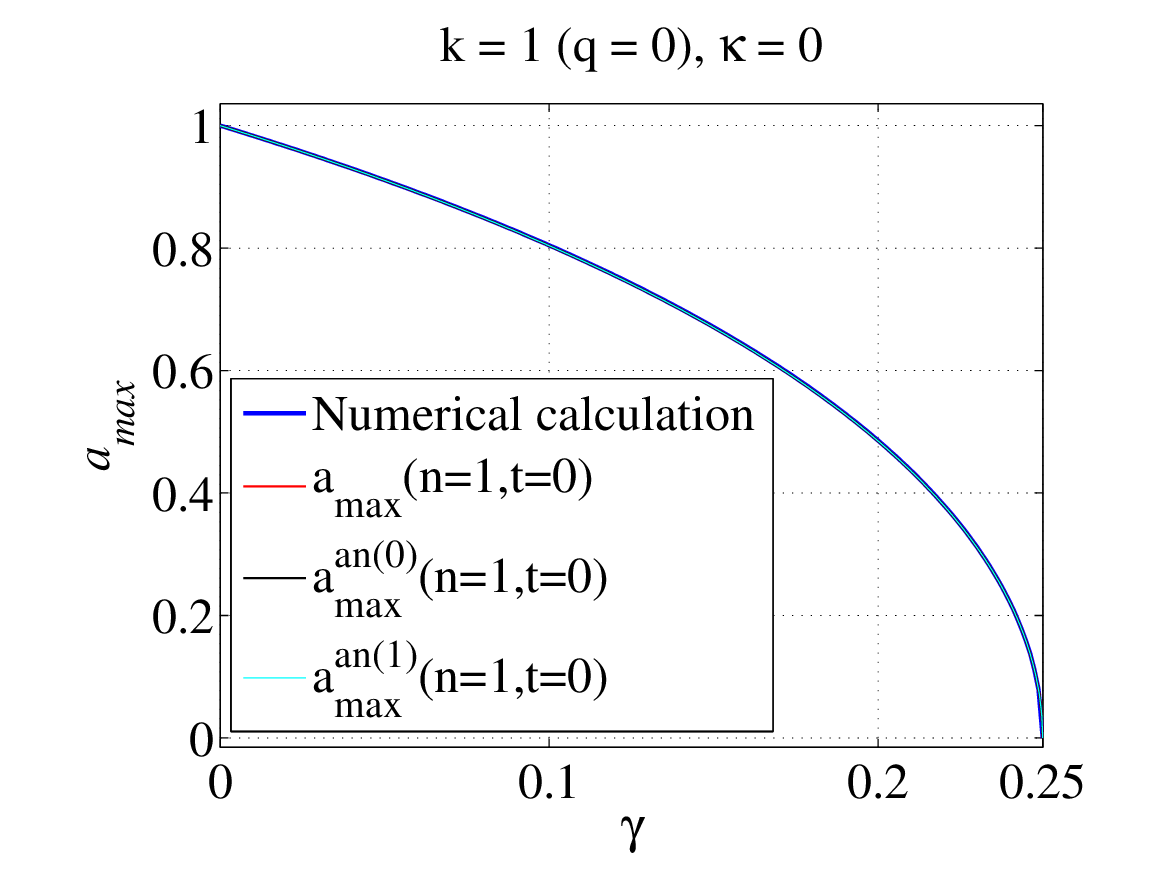}
\caption{\small Comparison of different methods of calculation of maximum amplitudes for perfect restitution and no foundation. Full numerical solution of (\ref{eq38}), the same but assuming $n=1$ and $t=0$, (\ref{eq72}) and (\ref{eq77}).}
\label{Figure2}
\end{figure}

\begin{figure}
\includegraphics[scale=0.33]{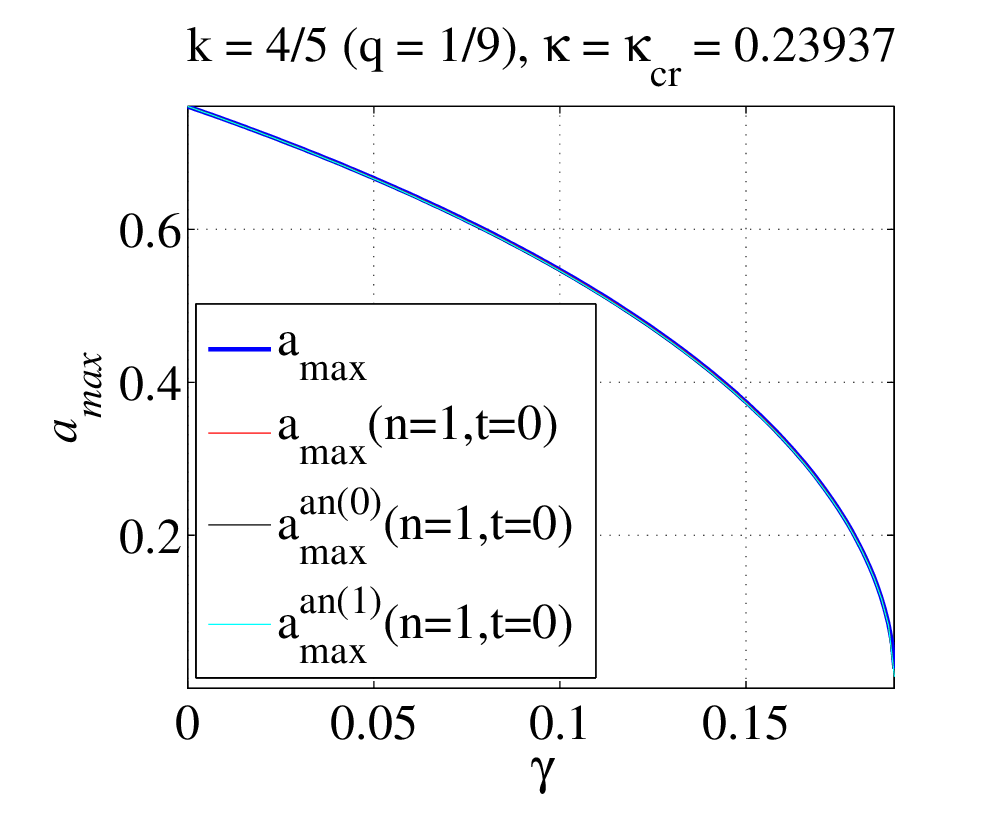}
\caption{\small Different methods of calculation of  $\bar{a}$ for a realistic value of $k$ and the corresponding $\kappa_{cr}$, as emerges from linear stability analysis, as discussed in Section \ref{sect5} (exact solution of (\ref{eq38}), the same with $n=1$ and $t=0$, (\ref{eq72}) and (\ref{eq77})).}
\label{Figure7}
\end{figure}

\begin{figure}
\includegraphics[scale=0.33]{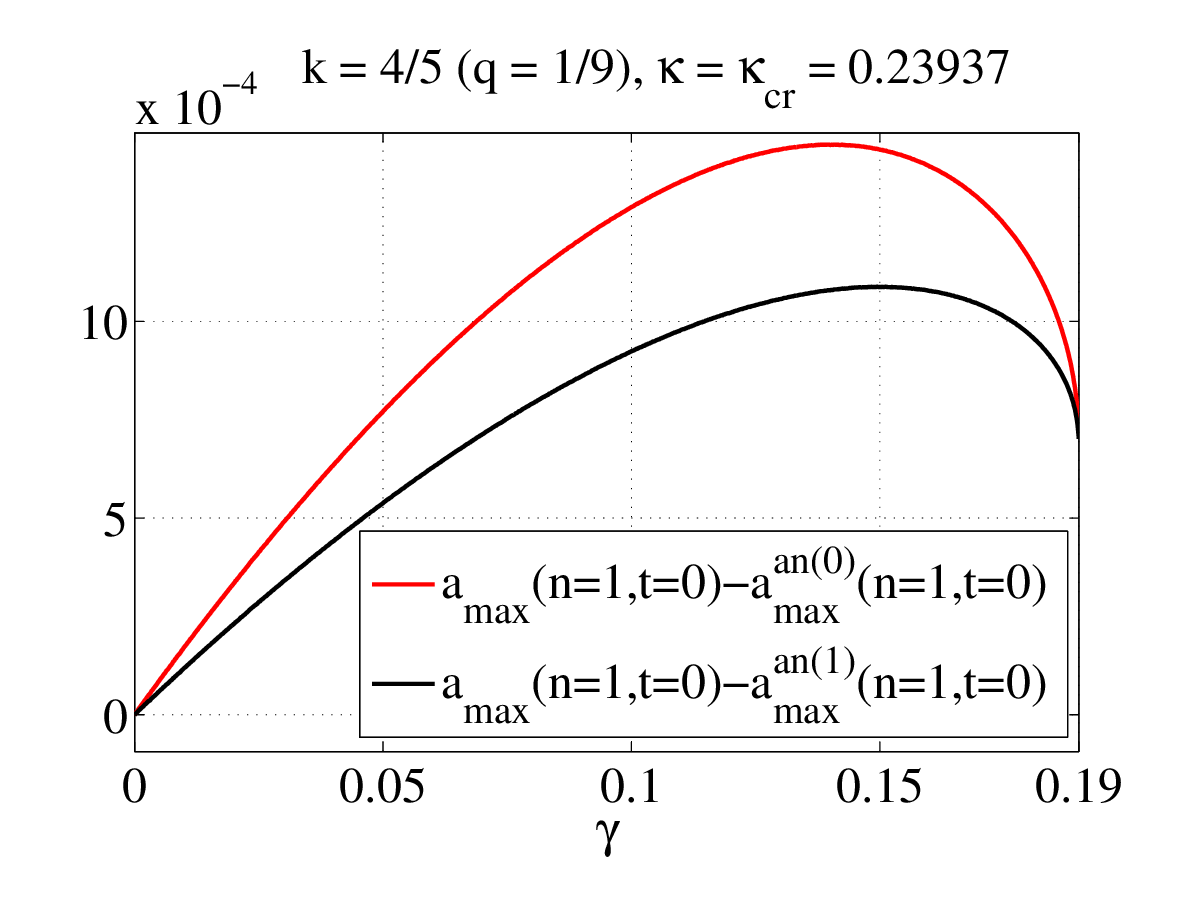}
\caption{\small Comparison of the errors of the zeroth and the first order $q$-expansions of the maximum amplitude for a realistic value of $k$ and the corresponding critical value of $\kappa$, as emerges from linear stability analysis.}
\label{Figure8}
\end{figure}

\begin{figure}[H]
\begin{center}
\includegraphics[scale=0.31]{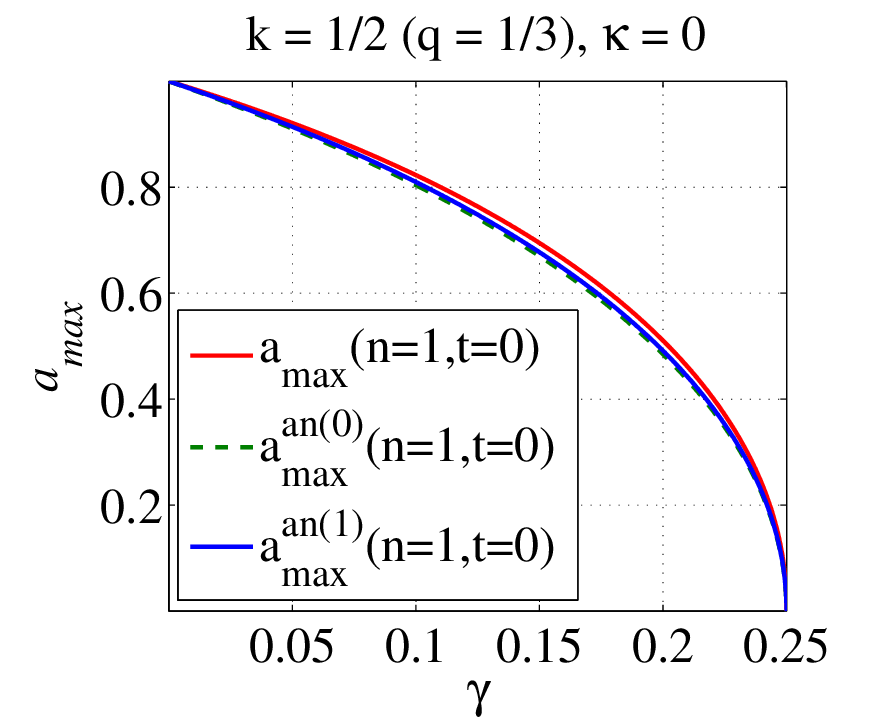}
\includegraphics[scale=0.31]{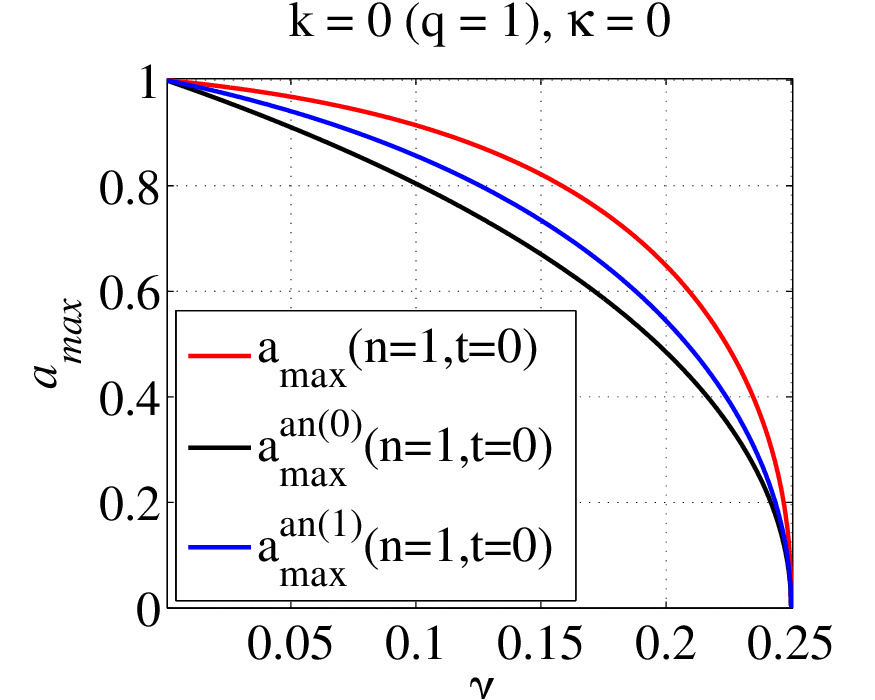}
\includegraphics[scale=0.31]{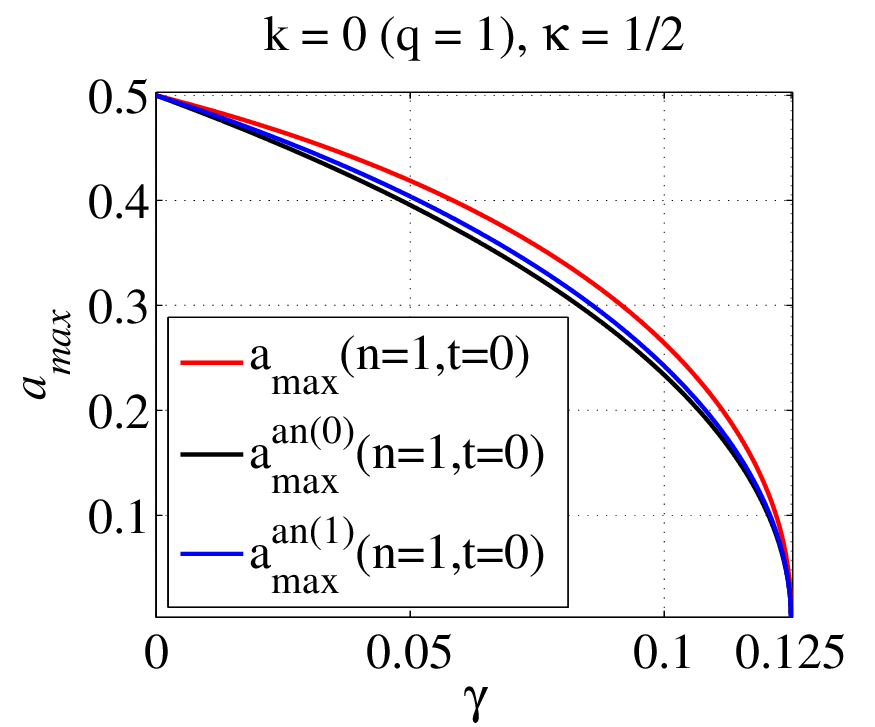}
\includegraphics[scale=0.31]{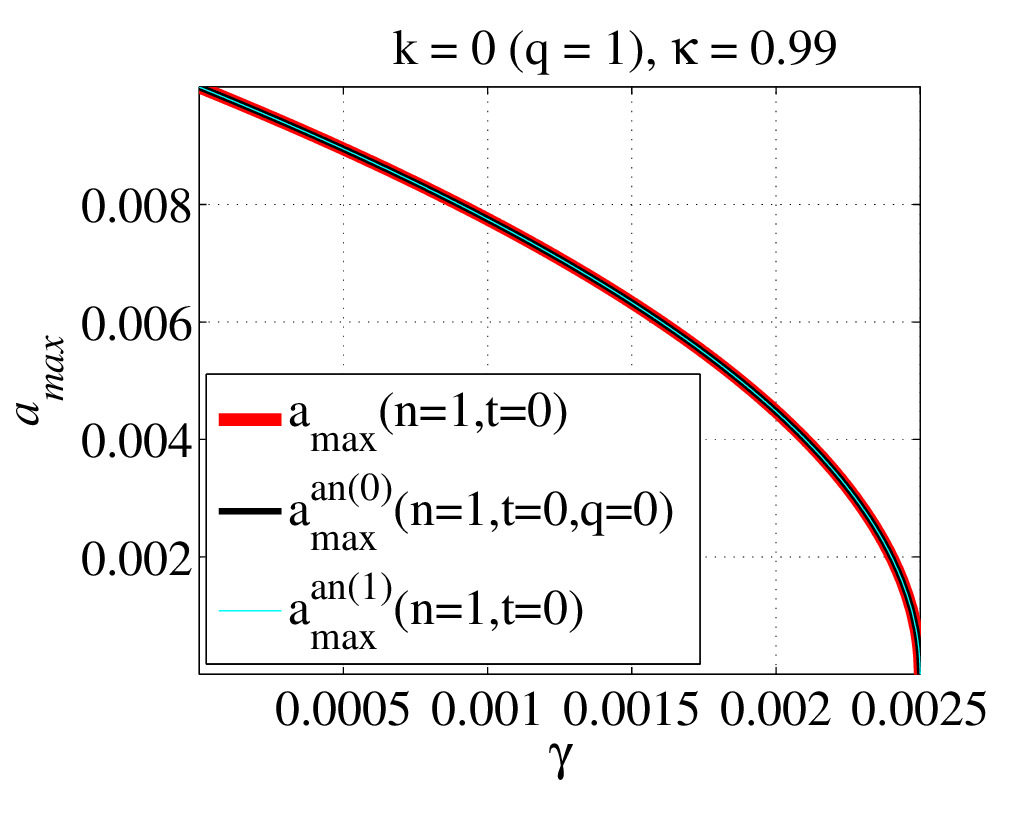}
\end{center}
\caption{\small Comparison of simplified numerical calculation and two different-order analytic approximations of maximum amplitudes for (a) $k=1/2,\kappa=0$, (b) $k \to 0,\kappa=0$, (c) $k=0, \kappa = 1/2$, and (d) $k \to 0, \kappa =0.99$.}
\label{Figure3}
\end{figure}

In Figures \ref{Figure2}-\ref{Figure3}, the quality of the approximation given by (\ref{eq72}), (\ref{eq76}) and (\ref{eq77}) is illustrated in the relevant range of the parameter $\gamma$, for several feasible values of the parameters $k$ and $\kappa$.

The main conclusion arising from  Figures \ref{Figure2}-\ref{Figure3} is that the expansion given in (\ref{eq77}) is worst in the zero-restitution and no-foundation case, where, still, the second term in the expansion gives a considerable correction with respect to the first term. Interestingly, the stiffer the foundation, the better is the agreement between the expansion and the exact solution. Also, clearly, for marginally realistic and higher values of the coefficient of restitution, the expansion in (\ref{eq77}) is good enough for all purposes. Figure \ref{Figure8} shows that even in the case of realistic restitution and the corresponding critical foundation stiffness, the second term in the expansion still noticeably improves the agreement.

Although the $q$-expansion given by  (\ref{eq72}), (\ref{eq76}) and (\ref{eq77}) produces a fairly good estimate for the existence amplitude and certainly can provide a reliable starting point for an iterative numerical solution, it still requires numerical summation of two infinite series. This has two implications: first, numerical summation requires computational effort which may be too expensive if the determination of the maximum existence amplitude is a part of a larger iterative procedure and has to be performed tens to hundreds of times. Second, when calculated numerically, the infinite series in (\ref{eq72}) have to be truncated. Truncation in this case implies non-exact description of the impact of the central mass in the breather.

For finite values of the link stiffness $\gamma$, there is no alternative. However, in the limit of a breather with weak coupling, which may be relevant either for an appropriate physical scenario or, say, in the stability analysis procedure of the sort described in Section \ref{sect5}, one can expand the infinite series with respect to the small parameter  $\gamma$, thus obtaining an estimate for the maximum existence amplitude. Then the remaining series can be summed-up analytically, producing a closed form expression containing only a finite number of elementary functions and being exact in the limit of small $\gamma$ and $q$. In this perspective, we derive below the second-order expansion of the maximum forcing amplitude for which the DB still exists with respect to $\gamma$.

Realizing that $\hat{a}_{q=0}=[S_0-|(\pi/4)A_n^{q=0}(t=0)|]/[S_0+|(\pi/4)A_n^{q=0}(t=0)|]$ and performing second-order Taylor series expansions to obtain:
\begin{equation}
S_0 \underset{\gamma \to 0}{\to} S_{00}(\kappa)+2\gamma S_{01}(\kappa)+6\gamma^2 S_{02}(\kappa)
\label{eq78}
\end{equation}
and
\begin{equation}\begin{split}
|(\pi/4)A_n^{q=0}(t=0)| \underset{\gamma \to 0}{\to} \\ \gamma S_{01}(\kappa)+4\gamma^2 S_{02}(\kappa)
\end{split}
\label{eq79}
\end{equation}
which results in:
\begin{equation}
\begin{split}
\hat{a}_{q=0} \underset{\gamma \to 0}{\to} \\ \frac{S_{00}(\kappa)+\gamma S_{01}(\kappa)+2\gamma^2 S_{02}(\kappa)}{S_{00}(\kappa)+3\gamma S_{01}(\kappa)+10\gamma^2 S_{02}(\kappa)}
\end{split}
\label{eq80}
\end{equation}
we get, by taking the first two derivatives of (\ref{eq80}) with respect to $\gamma$, a second-order expansion of $\hat{a}_{q=0}(\gamma)$, as follows:
\begin{equation}
\begin{split}
\hat{a}_{q=0 \atop \gamma \to 0} =1 -2\frac{S_{01}(\kappa)}{S_{00}(\kappa)}\gamma \\+2\left[3\frac{S^2_{01}(\kappa)}{S^2_{00}(\kappa)}-4\frac{S_{02}(\kappa)}{S_{00}(\kappa)}\right]\gamma^2
\end{split}
\label{eq81}
\end{equation}

Next, substituting, (\ref{eq30}), (\ref{eq78}) and (\ref{eq81}) into (\ref{eq76}) and performing second-order Taylor series expansion with respect to $\gamma$, we get:
\begin{equation}
\begin{split}
\hat{a}'_{q=0 \atop \gamma \to 0} = \frac{\pi^2}{8}\frac{S_{01}(\kappa)}{S^3_{00}(\kappa)}\gamma \\+\frac{\pi^2}{8}\left[4\frac{S_{02}(\kappa)}{S^3_{00}(\kappa)}-7\frac{S^2_{01}(\kappa)}{S^4_{00}(\kappa)}\right]\gamma^2
\end{split}
\label{eq82}
\end{equation}

One notes that for $\gamma=0$, $\bar{a}=1-\kappa$ for every value of $q$, as may be verified by observing Figures \ref{Figure2}-\ref{Figure7}.

Last, to complement the $\gamma$-expansion issue, we note that the functions:
\begin{equation}
S_{0j}(\kappa)=\underset{n=1,3,5,...}{\overset{\infty}{\sum}} \frac{1}{(n^2-\kappa)^{j+1}}
\label{eq83}
\end{equation}
where $j=0,1,2$, involved in the $\gamma$-expansions of $\hat{a}_{q=0}$ and $\hat{a}'_{q=0}$, can be summed-up analytically to give the following closed-form expressions:
\begin{equation}
 S_{00}(\kappa)=\frac{\pi\tan{(\pi\sqrt{\kappa}/2)}}{4\sqrt{\kappa}}
\label{eq84}
\end{equation}
\begin{equation}
\begin{split}
S_{01}(\kappa)=\frac{\pi^2}{16\kappa}[1+\tan^2{(\pi\sqrt{\kappa}/2})] \\ -\frac{\pi\tan{(\pi\sqrt{\kappa}/2)}}{8\kappa^{3/2}}
\end{split}
\label{eq85}
\end{equation}
\begin{equation}
\begin{split}
S_{02}(\kappa)=\frac{\pi^2}{64\kappa^2}[1+\tan^2{(\pi\sqrt{\kappa}/2})] \\ \times [\pi\sqrt{\kappa}\tan{(\pi\sqrt{\kappa}/2})-3] \\ +\frac{3\pi\tan{(\pi\sqrt{\kappa}/2)}}{32\kappa^{5/2}} \end{split}
\label{eq86}
\end{equation}

Finally, the asymptotically exact form of the upper bound on the excitation amplitude is:
\begin{equation}
\bar{a}_{q \to 0 \atop \gamma \to 0}= (1-\kappa)(\hat{a}_{q = 0 \atop \gamma \to 0} +\hat{a}'_{q = 0 \atop \gamma \to 0}q^2)
\label{eq87}
\end{equation}

The expediency of the expression given by the combination of (\ref{eq81}), (\ref{eq82}) and (\ref{eq84}-\ref{eq86}) becomes apparent in Section \ref{sect5}, where derivation of the theoretically optimal foundation stiffness for the breather requires iterative calculations with vanishing values of $\gamma$ and $q$ and a finite value of $\kappa$. There, nothing more than expansion with respect to $q$ and $\gamma$ is required, and exact closed-form $\kappa$-dependence, as well as exact description of the impact conditions are obviously beneficial.

Last, substituting (\ref{eq81},\ref{eq82},\ref{eq84}-\ref{eq86}) into (\ref{eq87}) and performing second-order Taylor series expansion with respect to $\kappa$, one obtains Eq. (\ref{eq89}), a three-variable second-order Taylor series expansion, which should be and indeed seems accurate enough an estimate in the (interesting) range: $q,\gamma,\kappa<1/4$.

\bibliography{NPOVG2014_v5a_ref}

\end{document}